\input{seam.sty}

\def\addots{\mathinner{\mkern1mu\raise1pt\vbox{\kern7pt\hbox{.}}\mkern2mu
\raise4pt\hbox{.}\mkern2mu\raise7pt\hbox{.}\mkern1mu}}

\def\Exp{\mbox{Exp}}

\def\upsilon{\mu}
\def\varsigma{\nu}
\def\face#1#2#3#4{\begin{picture}(3.5,1)(-.6,-.2)
\put(0,-1){\framebox(2,2){\p{}{}}}
\put(-.25,-1.25){\pp{}{#1}}
\put(2.25,-1.25){\pp{}{#2}}
\put(2.25,1.25){\pp{}{#3}}
\put(-.25,1.25){\pp{}{#4}}
\end{picture}}

\begin{document}
\begin{center}
\title{Integrable and Conformal Twisted Boundary Conditions\\
for $sl(2)$ $A$-$D$-$E$ Lattice Models
}
\author{C.H. Otto Chui\footnote{Email: C.Chui@ms.unimelb.edu.au},
Christian Mercat\footnote{Present Address: Technische Universit\"at, Berlin Sfb 288, Strasse des 17. Juni, 136, D-10623 Berlin, Germany; Email: Mercat@Sfb288.Math.TU-Berlin.de}

and Paul A. Pearce\footnote{Email: P.Pearce@ms.unimelb.edu.au}} 
\address{

Department of Mathematics and Statistics, University of Melbourne\\ 
Parkville, Victoria 3010, Australia}
\end{center}
\begin{abstract}
We study integrable realizations of conformal twisted boundary conditions
for $s\ell(2)$ unitary minimal models on a torus.  These conformal field
theories are realized as the continuum scaling limit of critical
$G\!=\!A,D,E$ lattice models with positive spectral parameter \mbox{$u>0$} and Coxeter
number $g$. Integrable seams are constructed by fusing blocks of elementary local face weights. The usual $A$-type fusions are labelled by the Kac labels $(r, s)$ and are associated with the Verlinde fusion algebra.  We introduce a new type of fusion in the two braid limits $u\to\pm i\infty$ associated 
with the graph fusion algebra, and labelled by nodes  $a,b\in G$ respectively. When combined with automorphisms, they lead to general integrable seams labelled by $x=(r,a,b,\kappa)\in(A_{g-2},H,H,{\Bbb Z}_{2})$ where $H$ is the graph $G$ itself for Type I  theories and its parent for Type II theories. Identifying our construction labels with the conformal labels of Petkova and Zuber, we find that the integrable seams are in one-to-one correspondence with the conformal seams. The distinct seams are thus associated with the nodes of the Ocneanu quantum graph. The quantum symmetries and twisted partition functions are checked numerically for $|G|\le 6$. We also show, in the case of $D_{2\ell}$,  that the non-commutativity of the Ocneanu algebra of seams arises because the automorphisms do not commute with the fusions.
\end{abstract}

\section{Introduction}
\setcounter{equation}{0}

There has been much recent
progress~\cite{BePZ, BPPZ98, BPPZ00, BP01, PZ00, Coq00, PZ0011021, PZ01}, %
on understanding integrable boundaries in statistical mechanics, conformal
boundary conditions in rational conformal field theories and the intimate
relations between them on both the cylinder and the torus.  Indeed it appears
that, for certain classes of theories, all of the conformal boundary
conditions on a cylinder can be realized as the continuum scaling limit of
integrable boundary conditions for the associated integrable lattice models.
For $s\ell(2)$ minimal theories, a complete classification has been
given~\cite{BePZ,BPPZ98,BPPZ00} of the conformal boundary conditions on a
cylinder.  These are labelled by nodes $(r,a)$ of a tensor product graph
$A\otimes G$ where the pair of graphs $(A,G)$, with $G$ of \ade type,
coincide precisely with the pairs in the
\ade classification of Cappelli, Itzykson and Zuber~\cite{CIZ}. 
Moreover, the physical content of the boundary conditions on the
cylinder has been ascertained~\cite{BP01,MP} by studying the related
integrable boundary conditions of the associated \ade lattice
models~\cite{Pas} for both positive and negative spectral parameters,
corresponding to \emph{unitary minimal theories} and
\emph{parafermionic theories} respectively.  Recently, the lattice
realization of integrable and conformal boundary conditions for $N=1$
\emph{superconformal theories}, which correspond to the \emph{fused}
$A$ lattice models with positive spectral parameter, has also
been understood~\cite{RP01}.

In this article, we use fusion to complete the task~\cite{CMOP2} of constructing 
integrable realizations of conformal twisted boundary conditions on the torus.
Although the methods are very general we consider $s\ell(2)$ unitary minimal
models for concreteness.  The key idea is that fused blocks of elementary
face weights on the lattice play the role of the local operators in the
theory.  The integrable and conformal boundary conditions on the cylinder are
constructed~\cite{BP01} by acting with these fused blocks on the simple
integrable boundary condition representing the vacuum. 
For the usual $A$-type fusion, associated with the Verlinde fusion algebra, this leads to integrable seams labelled by the Kac labels $(r,s)\in (A_{g-2},A_{g-1})$. 
In this paper we introduce a new type of fusion of $G$-type related to the graph fusion algebra. Integrable seams of this type are labelled by $(r,a)\in (A_{g-2},H)$ where $H$ is the graph $G$ itself for Type I theories and its parent for Type II theories. By the
generalized Yang-Baxter equations, these fused blocks or seams can be
propagated into the bulk without changing the spectrum of the theory.  The
seams so constructed provide integrable and conformal boundary conditions on
the torus.  Fixed boundary conditions $a\in G$ on the edge of the cylinder
are propagated into the bulk by the action of the seam $(1,a)$ on the
distinguished (vacuum) node $1\in G$. Lastly, automorphism seams, which play no role on the 
cylinder, play a crucial role on the torus by providing the extra label
giving rise to the complement of the left and right chiral subalgebras in the
Ocneanu graph. 

In general, for rational conformal field theories on the
torus, we expect the two types of fusions supplemented by the automorphisms to generate
all of the integrable and conformal seams.  In this paper we discuss this assertion in the context of the \ade unitary minimal models.

The paper is organized as follows. In 
Section~\ref{sec:Lattice} we define the \ade series, giving their graphs 
(Sec.~\ref{sec:ADE}), their (proper or improper) graph fusion algebras 
(Sec.~\ref{sec:ADEfusion}), their Ocneanu graphs 
(Sec.~\ref{sec:OcneanuGraph}) and their associated twisted partition functions
(Sec.~\ref{sec:OcneanuGraph} and \ref{sec:Twisted}). The presentation is self contained. 
In Section~\ref{sec:LatticeTwist} we describe the lattice realization of these
twisted boundary conditions. In particular, we define the \ade
lattice models (Sec.~\ref{sec:ADELattice}), the
fusion projectors (Sec.~\ref{sec:FusionProjector}),
the associated fused faces (Sec.~\ref{sec:FusedFace}) and the integrable seams (Sec.~3.4 to 3.6). We
construct the transfer matrices in Section~\ref{sec:Transfer} composed of
regular faces and seams. This is described for the single-row transfer matrix on the torus (Sec.~4.1) and the double-row transfer matrix on the cylinder  (Sec.~\ref{sec:BoundaryWeigths}). The spectra of the transfer matrices and finite-size corrections  are described in Section~\ref{sec:FiniteSize}. The
free energies are computed (Sec.~\ref{sec:SeamFree}) and the numerical conformal parts
are identified with the twisted partition functions (Sec.~\ref{sec:Conf} to 5.6). 

\section{\ade Fusion Graphs and Partition Functions}
\setcounter{equation}{0}
\label{sec:Lattice}

\ade classifications appear in a variety of contexts, namely, graphs, solvable lattice models, 
$\slhat (2)_k$ (Wess-Zumino-Witten) models at level $k$,
and $s\ell (2)$ minimal models.

\subsection{\ade Graphs}
\label{sec:ADE}

The basic \ade objects are graphs.  A simple \emph{graph}
$G$ is given by its \emph{vertices} (or nodes) $a\in G_{0}$ and
\emph{edges} $(a,b)\in G_{1}\subset G_{0}\times G_{0}$.  We are
concerned with unoriented graphs, $(a,b)\in G_{1}\Rightarrow (b,a)\in
G_{1}$.  The \ade graphs, which are the Dynkin diagrams of simply laced Lie algebras, are presented in Table~\ref{tbl:Graphs}.  The number $g$ is
the Coxeter number of the graph $G$ and the exponents $\Exp(G)$ are a
subset (with multiplicities) of the nodes of the $A_{L}$ graph sharing
the same Coxeter number as $G$.
\begin{figure}[htb]
\begin{center}
    \setlength{\unitlength}{.5cm}
\begin{equation*}
\mbox{}\hspace{1.1in}\mbox{}
\begin{array}{ccccc}
    \text{\text{Graph $G$\qquad}}& \mbox{}\quad\qquad\text{$g$}\quad\qquad\mbox{}&\text{$\Exp(G)$}&\text{Type/$H$}&\Gamma\\[10pt]
   \begin{picture}(6,1)
        \put(-3.5,0){\p{}{A_{L}}}
        \put(0,0){\line(1,0){4}}
        \multiput(1,0)(1,0){2}{\pp{}{\bullet}}
        \put(4,0){\pp{}{\bullet}}
        \put(0,0){\pp{}{*}}
        \put(1,0){\pp{}{\square}}
       \put(0,.3){\pp{b}{1}}
        \put(1,.3){\pp{b}{2}}
        \put(2,.3){\pp{b}{3}}
        \put(3,.3){\pp{b}{\cdots}}
        \put(4,.3){\pp{b}{L}}
    \end{picture}  & L+1 & 1, 2, \cdots, L &\mbox{I}&{\Bbb Z_2}\\
     \begin{picture}(6,2)
        \put(-3.5,0){\p{}{D_{\ell +2}\,\mbox{($\ell$ even)}}}
        \put(0,0){\line(1,0){3.5}}
        \put(3.5,0){\line(1,1){1}}
        \put(3.5,0){\line(1,-1){1}}
        \multiput(1,0)(1,0){2}{\pp{}{\bullet}}
        \put(3.5,0){\pp{}{\bullet}}
        \put(4.5,1){\pp{}{\bullet}}
        \put(4.5,-1){\pp{}{\bullet}}
        \put(-.2,-.04){\pp{l}{*}}
        \put(.75,-.04){\pp{l}{\square}}
        \put(0,.3){\pp{b}{1}}
        \put(1,.3){\pp{b}{2}}
        \put(2,.3){\pp{b}{3}}
        \put(2.75,.3){\pp{b}{\cdots}}
        \put(3.5,.3){\pp{b}{\ell}}
        \put(4.5,1){\pp{l}{~\ell+1}}
        \put(4.5,-.7){\pp{l}{~\ell+2}}
    \end{picture}    & 2\ell +2 & 1, 3, \cdots, 2\ell +1, \ell +1 &\mbox{I}&{\Bbb Z_2} \\[10pt]
 \begin{picture}(6,2)
        \put(-3.5,0){\p{}{D_{\ell +2}\,\mbox{($\ell$ odd)}}}
        \put(0,0){\line(1,0){3.5}}
        \put(3.5,0){\line(1,1){1}}
        \put(3.5,0){\line(1,-1){1}}
        \multiput(0,0)(1,0){3}{\pp{}{\bullet}}
        \put(3.5,0){\pp{}{\bullet}}
        \put(4.5,1){\pp{}{\bullet}}
        \put(0,.3){\pp{b}{1}}
        \put(1,.3){\pp{b}{2}}
        \put(2,.3){\pp{b}{3}}
        \put(2.75,.3){\pp{b}{\cdots}}
        \put(3.5,.3){\pp{b}{\ell}}
        \put(4.5,1){\pp{l}{~\ell+1}}
        \put(4.5,-.7){\pp{l}{~\ell+2}}
        \put(4.35,-1.05){\pp{l}{*}}
        \put(3.25,-.05){\pp{l}{\square}}
    \end{picture}    & 2\ell +2 & 1, 3, \cdots, 2\ell +1, \ell +1 &\mbox{II}/A_{2\ell+1} &{\Bbb Z_2}\\
   \begin{picture}(6,2.5)
        \put(-3.5,0){\p{}{E_{6}}}
        \put(0,0){\line(1,0){4}}
        \put(2,0){\line(0,1){1}}
        \multiput(1,0)(1,0){4}{\pp{}{\bullet}}
        \put(2,1){\pp{}{\bullet}}
        \put(0,0){\pp{}{*}}
        \put(1,0){\pp{}{\square}}
        \put(0,.3){\pp{b}{1}}
        \put(1,.3){\pp{b}{2}}
        \put(2,-.3){\pp{t}{3}}
        \put(3,.3){\pp{b}{4}}
        \put(4,.3){\pp{b}{5}}
        \put(2,1.3){\pp{b}{6}}
     \end{picture}  & 12 & 1, 4, 5, 7, 8, 11 &\mbox{I}&{\Bbb Z_2} \\
   \begin{picture}(6,2.5)
        \put(-3.5,0){\p{}{E_{7}}}
        \put(0,0){\line(1,0){5}}
        \put(3,0){\line(0,1){1}}
        \multiput(1,0)(1,0){5}{\pp{}{\bullet}}
         \put(3,1){\pp{}{\bullet}}
        \put(0,.3){\pp{b}{1}}
        \put(1,.3){\pp{b}{2}}
        \put(2,.3){\pp{b}{3}}
        \put(3,-.3){\pp{t}{4}}
        \put(4,.3){\pp{b}{5}}
        \put(5,.3){\pp{b}{6}}
        \put(3,1.3){\pp{b}{7}}
        \put(0,0){\pp{}{*}}
        \put(1,0){\pp{}{\square}}
     \end{picture}  & 18 & 1, 5, 7, 9, 11, 13, 17  &\mbox{II}/D_{10}&1\\
   \begin{picture}(6,2.5)
        \put(-3.5,0){\p{}{E_{8}}}
        \put(0,0){\line(1,0){6}}
        \put(4,0){\line(0,1){1}}
        \multiput(1,0)(1,0){6}{\pp{}{\bullet}}
        \put(4,1){\pp{}{\bullet}}
        \put(0,0){\pp{}{*}}
        \put(1,0){\pp{}{\square}}
        \put(0,.3){\pp{b}{1}}
        \put(1,.3){\pp{b}{2}}
        \put(2,.3){\pp{b}{3}}
        \put(3,.3){\pp{b}{4}}
        \put(4,-.3){\pp{t}{5}}
        \put(5,.3){\pp{b}{6}}
        \put(6,.3){\pp{b}{7}}
        \put(4,1.3){\pp{b}{8}}
     \end{picture}  & 30 & 1, 7, 11, 13, 17, 19, 23, 39&\mbox{I}&1\\
\end{array}
\end{equation*}
\end{center}
\caption{\ade graphs corresponding to the Dynkin diagrams of the classical \ade simply laced 
Lie algebras. The nodes associated with the identity and the fundamental are 
shown by $*$, $\square$ respectively. Also shown are the Coxeter numbers $g$, exponents $\Exp(G)$, the Type I or II, the parent graphs $H\ne G$ and the diagram automorphism group $\Gamma$. The $D_4$ graph is exceptional having the automorphism group $\Bbb S_3$.}
\label{tbl:Graphs}
\end{figure}

A graph $G$ is completely encoded by its
adjacency matrix which we denote by the same letter $G$.  It is a symmetric 
non-negative integer square matrix whose rows and columns are labelled by
the vertices of $G$, defined by $G_{a\, b}=1$ if $a$ and
$b$ are adjacent and $G_{a\, b}=0$ otherwise. 
What is so special about the \ade graphs is that (along with the
tadpole\footnote{The tadpole graph $T_{L}$ is obtained from the graph
$A_{L}$ by adding a loop at the final vertex;  it is not a simple
graph.} series) they are the only ones whose spectra lies in the open
interval $(-2,2)$.
The Perron-Frobenius theorem implies that the largest eigenvalue of
these adjacency matrices is non-degenerate, real and positive and its
eigenvector can be chosen to have non-negative entries.  They are
given explicitly in terms of $q$-deformed integers
$S_{n}=[n]_{q}=\frac{q^{n}-q^{-n}}{q-q^{-1}}$ with $q=\exp
(\pi i/g)$. The largest eigenvalue is $S_{2}=[2]_{q}$ and the
eigenvector $\vec\psi$ is
\begin{eqnarray}
    \vec\psi_{A_{L}} & = & \bigl( [k]_{q}\bigr), {\ss 1\leq k\leq L}
    \notag  \\
   \vec\psi_{D_{\ell+2}} & = & \bigl( [k]_{q}, {\ss 1\leq k\leq 
   \ell},\frac{[\ell]_{q}}{[2]_{q}},\frac{[\ell]_{q}}{[2]_{q}}\bigr)
        \notag \\
    \vec\psi_{E_{6}} & = & \bigl( [1]_{q}, [2]_{q}, [3]_{q}, [2]_{q}, 
    [1]_{q},\frac{[3]_{q}}{[2]_{q}}\bigr)
    \label{eq:ADEsp}  \\
    \vec\psi_{E_{7}} & = & \bigl( [1]_{q}, [2]_{q}, [3]_{q}, [4]_{q}, 
    \frac{[6]_{q}}{[2]_{q}},\frac{[4]_{q}}{[3]_{q}},\frac{[4]_{q}}{[2]_{q}}\bigr)
        \notag \\
    \vec\psi_{E_{8}} & = & \bigl( [1]_{q}, [2]_{q}, [3]_{q}, [4]_{q},  [5]_{q},
    \frac{[7]_{q}}{[2]_{q}},\frac{[5]_{q}}{[3]_{q}},\frac{[5]_{q}}{[2]_{q}}\bigr).
        \notag 
\end{eqnarray}

\subsection{Graph fusion algebras}
\label{sec:ADEfusion}
The integer linear span of the nodes of the graph can be given a
structure of a commutative \emph{graph fusion algebra}.  We first specify two vertices,
the identity vertex $*$ and the fundamental vertex $\square$.  They
are indicated in Table~\ref{tbl:Graphs} and are respectively the
vertices labelled $1$ and $2$ in the $A_{L}$, $D_{2\ell +2}$, $E_{6}$
and $E_{8}$ cases, known as Type I theories, the vertices
$2\ell+1$ and $2\ell-1$ for the $D_{2\ell +1}$ case, and the vertices $1$,
$2$ for the $E_{7}$ case, known as 
Type II theories.  The former models give rise to proper graph
fusion algebras with non negative integer structure constants or Non-negative
Integer Matrix Irreducible Representations (nimreps) which are understood as
adjacency matrices. The latter have some negative structure constants and
do not form proper graph fusion algebras.

The algebra is defined by stating that the edges of the graph $G$ 
encode the action of the fundamental element $\square$:
\begin{equation}
    a \; \square\; = \sum_{b\sim a} b.
    \label{eq:FusionAlg}
\end{equation}
The identity gives one row of the algebra table, the previous formula
gives another, commutativity and associativity determine the rest.  On
the $D_{4}$ example, $\; \square\; = 2$ and associativity gives
\begin{eqnarray}
    (4-3)( \square\;\; \square)&=&
    (4 \; \square - 3 \; \square)\; \square = 0\notag\\
&=&(4-3)(1+3+4)
    \label{eq:assocD4}
\end{eqnarray}
so that $4-3=3\; 3 - 4\; 4$ and $ \square\; 3\; 3= \square$ shows that
$\cases{\!3\; 3 = 4\\ \!4\; 4 = 3}$ and the expansion of $3\; 3\; 3$
implies $3\; 4 = 1$.

The structure constants of this algebra are denoted $\hat N$
\begin{equation}
    a b = \sum_{c\in G} \hat N_{a\, b}{}^{c}\, c.
    \label{eq:hatN}
\end{equation}
The definition of the algebra implies $\hat N_{*}=\text{I}$ and $\hat
N_{\square}=G$ and these matrices themselves form the regular representation of the algebra with the usual matrix product
\begin{equation}
    \hat N_{a}\, \hat N_{b} = \sum_{c\in G} \hat N_{a\, b}{}^{c}\, \hat N_{c}.
    \label{eq:hatNalg}
\end{equation}
As it is a commutative algebra containing the adjacency matrix, its
common set of eigenvectors is given by an orthogonal basis of
eigenvectors of $G$. They are labelled by Coxeter exponents (we have
only given the Perron-Frobenius eigenvector) and the spectral 
decomposition of each matrix onto its eigenvectors give these integers
through a Verlinde-like formula
\begin{equation}
    \hat N_{a\,b}{}^{c}=\sum_{j\in 
    \text{\Exp}(G)}\frac{
    \psi_{a}^{j}\,\psi_{b}^{j}\left(\psi_{c}^{j}\right)^{*}}{\psi_{*}^{j}}.
    \label{eq:VerlindeLike}
\end{equation}

\begin{table}[p]
    \centering
    $$ N_{1}=
    \(\begin{smallmatrix}
        1&0&0&0\\ 0&1&0&0\\0&0&1&0\\0&0&0&1
    \end{smallmatrix}\),\;\;\;\;
    N_{2}=
    \(\begin{smallmatrix}
            0&1&0&0\\ 1&0&1&0\\ 0&1&0&1\\ 0&0&1&0
        \end{smallmatrix}\),\;\;\;\;
    N_{3}=
    \(\begin{smallmatrix}
            0&0&1&0\\ 0&1&0&1\\ 1&0&1&0\\ 0&1&0&0
        \end{smallmatrix}\),\;\;\;\;
    N_{4}=
    \(\begin{smallmatrix}
            0&0&0&1\\ 0&0&1&0\\ 0&1&0&0\\ 1&0&0&0
        \end{smallmatrix}\).\;\;\;\;
$$
$$
\begin{array}{c|c|c|c|ccccccc}
     & 1 & 2 & 3 & 4  \\
     \cline{1-5}
    1 & 1 & 2 & 3 & 4  &&     
    Z_{1}=&\chi_{1}\chi_{1}^{*}&+\chi_{2}\chi_{2}^{*}&+
    \chi_{3}\chi_{3}^{*}&+\chi_{4}\chi_{4}^{*},\\
    \cline{1-5}
    2 & 2 & 1+3 & 2+4 & 3 &&     
    Z_{2}=&\chi_{2}\chi_{1}^{*}&+(\chi_{1}+\chi_{3})\chi_{2}^{*}&+
    (\chi_{2}+\chi_{4}) \chi_{3}^{*}&+\chi_{3}\chi_{4}^{*}, \\
    \cline{1-5}
    3 & 3 & 2+4 & 1+3 & 2  &&     
    Z_{3}=&\chi_{3}\chi_{1}^{*}&+(\chi_{2}+\chi_{4})\chi_{2}^{*}&+
    (\chi_{1}+\chi_{3}) \chi_{3}^{*}&+\chi_{2}\chi_{4}^{*}, \\
    \cline{1-5}
    4 & 4 & 3 & 2 & 1&&     
    Z_{4}=&\chi_{4}\chi_{1}^{*}&+\chi_{3}\chi_{2}^{*}&+
    \chi_{2}\chi_{3}^{*}&+\chi_{1}\chi_{4}^{*}.
\end{array}
$$
    \caption{Fusion matrices, graph fusion algebra
    and  twisted partition functions of $A_{4}$ in terms of affine $s\ell(2)$ characters $\chi_s$.}
    \label{tbl:A4}
\end{table}
\begin{table}[p]
    \centering
    $$ 
\hat N_{1}=
    \(\begin{smallmatrix}
        1&0&0&0\\ 0&1&0&0\\0&0&1&0\\0&0&0&1
    \end{smallmatrix}\),\;\;\;\;
\hat N_{2}=
    \(\begin{smallmatrix}
            0&1&0&0\\ 1&0&1&1\\ 0&1&0&0\\ 0&1&0&0
        \end{smallmatrix}\),\;\;\;\;
\hat N_{3}=
    \(\begin{smallmatrix}
            0&0&1&0\\ 0&1&0&0\\ 0&0&0&1\\ 1&0&0&0
        \end{smallmatrix}\),\;\;\;\;
\hat N_{4}=
    \(\begin{smallmatrix}
            0&0&0&1\\ 0&1&0&0\\ 1&0&0&0\\ 0&0&1&0
        \end{smallmatrix}\).\;\;\;\;
$$
$$
\begin{array}{c|c|c|c|cccccccc}
     &\mathbf{ 1 }& 2 &\mathbf{ 3 }&\mathbf{ 4 } \\
     \cline{1-5}
   \mathbf{ 1 }&\mathbf{ 1 }& 2 &\mathbf{ 3 }&\mathbf{ 4  }&&     
    Z_{1}=&\hat\chi_{1}\hat\chi_{1}^{*}&+
    \hat \chi_{3}\hat\chi_{3}^{*}&+\hat\chi_{4}\hat\chi_{4}^{*},\\
    \cline{1-5}
    2 & 2 &\!\!1\!\!+\!3\!+\!4\!\!& 2 & 2 &&     
    Z_{2\otimes 1}=&\hat\chi_{2}\hat\chi_{1}^{*}&+\hat\chi_{2}\hat\chi_{3}^{*}&+
    \hat\chi_{2} \hat \chi_{4}^{*}&=Z_{1\otimes 2}^*, \\
    \cline{1-5}
   \mathbf { 3 }&\mathbf{ 3 }& 2 &\mathbf{ 4 }&\mathbf{ 1  }&&     
    Z_{3}=&\hat\chi_{3}\hat\chi_{1}^{*}&+
    \hat\chi_{4}\hat\chi_{3}^{*}&+\hat\chi_{1}\hat\chi_{4}^{*}&=Z_4, \\
    \cline{1-5}
    \mathbf{ 4 }&\mathbf{ 4 }& 2 &\mathbf{ 1 }&\mathbf{ 3}&&     
    Z_{4}=&\hat\chi_{4}\hat\chi_{1}^{*}&+
    \hat \chi_{1}\hat\chi_{3}^{*}&+\hat\chi_{3}\hat\chi_{4}^{*}\\[-6pt]
\end{array}
$$
$$ \mbox{}\qquad\  \  \  \ Z_{1'}=Z_{3'}=Z_{4'}=\hat\chi_{2}\hat\chi_{2}^{*} $$
$$\hat\chi_{1}=\chi_{1}+\chi_{5},\;\;\;\;\hat\chi_{2}=\chi_{2}+\chi_{4},\;\;\;\;
\hat\chi_{3}=\hat\chi_{4}=\chi_{3}$$
$$
\begin{array}{c|c|c|c|c|c|c|c|c|}
     &\mathbf{ 1 }& 2 &\mathbf{ 3 }&\mathbf{ 4 }&\mathbf{1}'&2'&\mathbf{3}'&\mathbf{4}' \rule{0pt}{6pt}\\
     \hline
   \mathbf { 1 }& \mathbf { 1 }& 2 & \mathbf { 3 }& \mathbf { 4  }&
 \mathbf {1'}&2'& \mathbf {3'}& \mathbf {4'}\rule{0pt}{6pt}\\
    \cline{1-9}
    2 & 2 &\,1\!\!+\!3\!+\!4\,& 2 & 2 &2'&1'\!+\!3'\!+\!4'&2'&2'\rule{0pt}{6pt}\\
    \cline{1-9}
   \mathbf { 3 }&\mathbf{ 3 }& 2 &\mathbf{ 4 }&\mathbf{1}&\mathbf{3}'&2'&\mathbf{4}'&\mathbf{1}'\rule{0pt}{6pt}\\
    \cline{1-9}
    \mathbf{ 4 }&\mathbf{ 4 }& 2 &\mathbf{ 1 }&\mathbf{ 3}&\mathbf{4}'&2'&\mathbf{1}'&\mathbf{3}'\rule{0pt}{6pt}\\
\hline
    \mathbf{1}'&\mathbf{1}'&2'&\mathbf{4}'&\mathbf{3}'&\mathbf{ 1 }& 2 &\mathbf{ 4 }&\mathbf{ 3  }\rule{0pt}{6pt}\\
    \cline{1-9}
    2'&2'&1'\!+\!3'\!+\!4'&2'&2'& 2 &\,1\!\!+\!3\!+\!4\,& 2 & 2 \rule{0pt}{6pt}\\
    \cline{1-9}
    \mathbf{3}'&\mathbf{3}'&2'&\mathbf{1}'&\mathbf{4}'&\mathbf{ 3 }& 2 &\mathbf{ 1 }&\mathbf{ 4  }\rule{0pt}{6pt}\\
    \cline{1-9}
    \mathbf{4}'&\mathbf{4}'&2'&\mathbf{3}'&\mathbf{1}'&\mathbf{ 4 }& 2 &\mathbf{ 3 }&\mathbf{ 1}\rule{0pt}{6pt}\\
\hline
\end{array}\qquad 
\setlength{\unitlength}{16pt}
\begin{picture}(2,2)
\put(.5,-4){\includegraphics[width=3.0cm]{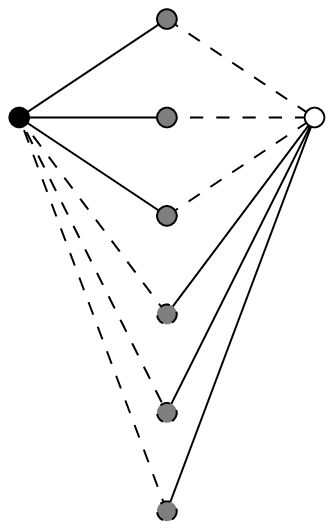}}
\put(3.5,4.1){$1$}
\put(.85,2.3){$2$}
\put(6.0,2.3){$2'$}
\put(3.5,2.8){$3$}
\put(3.5,1.5){$4$}
\put(3.5,0){$1'$}
\put(3.5,-1.3){$3'$}
\put(3.5,-2.6){$4'$}
\put(3.3,-4.4){$\tilde{D}_4$}
\end{picture}
$$
\caption{Fusion matrices, graph fusion algebra, twisted partition
functions, Ocneanu algebra and Ocneanu graph of $D_{4}$. The extended chiral and ambichiral subalgebras are bold.}
    \label{tbl:D4}
\end{table}
\begin{table}[p]
    \centering
    $$ 
\hat N_{1}=
    \(\begin{smallmatrix}
        0&\hbox to 0pt{$\!\!\ss \!-\!1$}&0&1&1\\ \hbox to 0pt{$\!\!\ss \!-\!1$}&0&1&0&0\\0&1&0&0&0\\1&0&0&0&0\\1&0&0&0&0
    \end{smallmatrix}\),\;\;\;\;
\hat N_{2}=
    \(\begin{smallmatrix}
            \hbox to 0pt{$\!\!\ss \!-\!1$}&0&1&0&0\\ 0&0&0&1&1\\ 1&0&1&0&0\\0&1&0&0&0\\0&1&0&0&0
        \end{smallmatrix}\),\;\;\;\;
\hat N_{3}=
    \(\begin{smallmatrix}
            0&1&0&0&0\\ 1&0&1&0&0\\ 0&1&0&1&1\\ 0&0&1&0&1\\ 0&0&1&1&0
        \end{smallmatrix}\),$$ $$
\hat N_{4}=
    \(\begin{smallmatrix}
        1&0&0&0&0\\ 0&1&0&0&0\\0&0&1&0&0\\0&0&0&0&1\\0&0&0&1&0
    \end{smallmatrix}\),\;\;\;\;
\hat N_{5}=
    \(\begin{smallmatrix}
        1&0&0&0&0\\ 0&1&0&0&0\\0&0&1&0&0\\0&0&0&1&0\\0&0&0&0&1
    \end{smallmatrix}\).
$$
\caption{The   $D_{5}$ graph fusion algebra is not a proper graph algebra.}
    \label{tbl:D5}
\end{table}

\begin{table}[htbp]
    \centering
    $$ 
\hat N_{1}=
    \(\begin{smallmatrix}
        1&0&0&0&0&0\\ 0&1&0&0&0&0\\0&0&1&0&0&0\\0&0&0&1&0&0\\
0&0&0&0&1&0\\0&0&0&0&0&1
    \end{smallmatrix}\),\;\;\;\;
\hat N_{2}=
    \(\begin{smallmatrix}
        0&1&0&0&0&0\\ 1&0&1&0&0&0\\0&1&0&1&0&0\\
0&0&1&0&1&1\\0&0&0&1&0&0\\ 0&0&0&1&0&0
        \end{smallmatrix}\),\;\;\;\;
\hat N_{3}=
    \(\begin{smallmatrix}
        0&0&1&0&0&0\\ 0&1&0&1&0&0\\1&0&1&0&1&1\\
0&1&0&2&0&0\\0&0&1&0&0&1\\ 0&0&1&0&1&0
        \end{smallmatrix}\),$$ 
    $$ 
\hat N_{4}=
    \(\begin{smallmatrix}
        0&0&0&1&0&0\\ 0&0&1&0&1&1\\0&1&0&2&0&0\\1&0&2&0&1&1\\
0&1&0&1&0&0\\0&1&0&1&0&0
    \end{smallmatrix}\),\;\;\;\;
\hat N_{5}=
    \(\begin{smallmatrix}
        0&0&0&0&1&0\\ 0&0&0&1&0&0\\0&0&1&0&0&1\\
0&1&0&1&0&0\\1&0&0&0&1&0\\ 0&0&1&0&0&0
        \end{smallmatrix}\),\;\;\;\;
\hat N_{6}=
    \(\begin{smallmatrix}
        0&0&0&0&0&1\\ 0&0&0&1&0&0\\0&0&1&0&1&0\\
0&1&0&1&0&0\\0&0&1&0&0&0\\ 1&0&0&0&0&1
        \end{smallmatrix}\),$$ 

\bigskip
$$
\begin{array}{c|c|c|c|c|c|c}
     &\mathbf{ 1 }& 2 &\mathbf{ 3 }& 4 & \mathbf{ 5 }& \mathbf{ 6 } \\
     \hline
    \mathbf{ 1 }&\mathbf{ 1 }& 2 &\mathbf{ 3 }& 4 & \mathbf{ 5 }& \mathbf{ 6 } 
\\
    \hline
    2 & 2 &\!\!1\!\!+\!3\!\!& \!\!2\!\!+\!4\!\! &\!\!3\!+\!5\!+\!6\!\! &4&4
 \\
    \hline
    \mathbf{ 3 }&\mathbf{ 3 }&  \!\!2\!\!+\!4\!\! &\mathbf{\!\!1\!\!+3\!\!+\!5\!+\!6\!\!}&\!\!2\!+\!4\!+\!4\!\!&\mathbf{\!\!3\!+\!6\!\!}&\mathbf{\!\!3\!+\!5\!\!}\\
    \hline
    4 & 4 &\!\!3\!+\!5\!+\!6\!\!&\!\!1\!\!+\!3\!+\!3\!+\!5\!+\!6\!\!&\!\!2\!+\!4\!\!&\!\!2\!+\!4\!\!
 \\
    \hline
    \mathbf{ 5 }&\mathbf{ 5 }& 4 &\mathbf{\!\!3\!+\!6\!\!}&\!\!2\!+\!4\!\!&\mathbf{\!\!1\!+\!5\!\!}&\mathbf{ 3 }\\
    \hline
    \mathbf{ 6 }&\mathbf{ 6 }& 4 &\mathbf{\!\!3\!+\!5\!\!}&\!\!2\!+\!4\!\!&\mathbf{ 3 }&\mathbf{\!\!1\!+\!6\!\!}
\end{array}
$$
\begin{eqnarray}
  Z_{1}&=&\hat\chi_{1}^{ }\hat\chi_{1}^{*}+
    \hat\chi_{3}^{ }\hat\chi_{3}^{*}+
\hat\chi_{5}^{ }\hat\chi_{5}^{*}+\hat\chi_{6}^{ }\hat\chi_{6}^{*},\notag\\
  Z_{2\otimes 1}&=&\hat\chi_{2}^{ }\hat\chi_{1}^{*}+
    \left(\hat\chi_{2}^{ }+\hat\chi_{4}^{ }\right)\hat\chi_{3}^{*}+
\hat\chi_{4}^{ }\left(\hat\chi_{5}^{*}+\hat\chi_{6}\right)^{*}
= Z_{1\otimes 2}^*,\notag\\
  Z_{3}&=&\hat\chi_{3}^{ }\hat\chi_{1}^{*}+
    \left(\hat\chi_{1}^{ }+\hat\chi_{3}^{ }+\hat\chi_{5}^{ }+\hat\chi_{6}^{ }\right)\hat\chi_{3}^{*}+
\left(\hat\chi_{3}^{ }+\hat\chi_{6}^{ }\right)\hat\chi_{5}^{*}+
\left(\hat\chi_{3}^{ }+\hat\chi_{5}^{ }\right)\hat\chi_{6}^{*},\notag\\
  Z_{4\otimes 1}&=&\hat\chi_{4}^{ }\hat\chi_{1}^{*}+
    \left(\hat\chi_{2}^{ }+2\hat\chi_{4}^{ }\right)\hat\chi_{3}^{*}+
 \left(\hat\chi_{2}^{ }+\hat\chi_{4}^{ }\right)\left(\hat\chi_{5}^{*}
+\hat\chi_{6}\right)^{*}= Z_{1\otimes 4}^*,\notag\\
  Z_{5}&=&\hat\chi_{5}^{ }\hat\chi_{1}^{*}+
    \left(\hat\chi_{3}^{ }+\hat\chi_{6}^{ }\right)\hat\chi_{3}^{*}+
\left(\hat\chi_{1}^{ }+\hat\chi_{5}^{ }\right)\hat\chi_{5}^{*}+
\hat\chi_{3}^{ }\hat\chi_{6}^{*},\notag\\
  Z_{6}&=&\hat\chi_{6}^{ }\hat\chi_{1}^{*}+
    \left(\hat\chi_{3}^{ }+\hat\chi_{5}^{ }\right)\hat\chi_{3}^{*}+
\hat\chi_{3}^{ }\left(\hat\chi_{1}^{ }+
\hat\chi_{6}^{ }\right)\hat\chi_{5}^{*}+
\left(\hat\chi_{1}^{ }+\hat\chi_{6}^{ }\right)\hat\chi_{6}^{*},\notag\\
  Z_{1'}&=&\hat\chi_{2}^{ }\hat\chi_{2}^{*}+
    \hat\chi_{4}^{ }\hat\chi_{4}^{*},\notag\\
  Z_{3'}&=&|\hat\chi_{2}^{ }+\hat\chi_{4}^ {}|^2+|\hat\chi_{4}^{ }|^2,\notag\\
  Z_{5'}&=&Z_{6'}=\hat\chi_{2}^{ }\hat\chi_{4}^{*}+
\hat\chi_{4}^{ }\hat\chi_{2}^{*}+|\hat\chi_{4}^{ }|^2.\notag
\end{eqnarray}
$$\hat\chi_{1}=\chi_{1}+\chi_{9},\;\;\;\;
\hat\chi_{2}=\chi_{2}+\chi_{8},\;\;\;\;
\hat\chi_{3}=\chi_{3}+\chi_{7},$$
$$\hat\chi_{4}=\chi_{4}+\chi_{6},\;\;\;\;
\hat\chi_{5}=\hat\chi_{6}=\chi_{5}.$$
\caption{The graph fusion algebra of $D_{6}$ and its twisted partition
functions in terms of extended characters.  The extended chiral subalgebra $T$
is shown bold (see Fig.~\ref{fig:D2lD2l}) and the extended characters are given in terms
of the $A_{9}$ characters.}
    \label{tbl:D6}
\end{table}


\begin{table}[hptb]
    \centering
    $$ 
\hat N_{1}=
    \(\begin{smallmatrix}
        1&0&0&0&0&0\\ 0&1&0&0&0&0\\0&0&1&0&0&0\\0&0&0&1&0&0\\
0&0&0&0&1&0\\0&0&0&0&0&1
    \end{smallmatrix}\),\;\;\;\;
\hat N_{2}=
    \(\begin{smallmatrix}
        0&1&0&0&0&0\\ 1&0&1&0&0&0\\0&1&0&1&0&1\\
0&0&1&0&1&0\\0&0&0&1&0&0\\ 0&0&1&0&0&0
        \end{smallmatrix}\),\;\;\;\;
\hat N_{3}=
    \(\begin{smallmatrix}
        0&0&1&0&0&0\\0&1&0&1&0&1\\
1&0&2&0&1&0\\0&1&0&1&0&1\\ 0&0&1&0&0&0\\0&1&0&1&0&0
        \end{smallmatrix}\),$$ $$
\hat N_{4}=
    \(\begin{smallmatrix}
        0&0&0&1&0&0\\ 0&0&1&0&1&0\\0&1&0&1&0&1\\
1&0&1&0&0&0\\0&1&0&0&0&0\\ 0&0&1&0&0&0
        \end{smallmatrix}\).\;\;\;\;
\hat N_{5}=
    \(\begin{smallmatrix}
        0&0&0&0&1&0\\0&0&0&1&0&0\\0&0&1&0&0&0\\0&1&0&0&0&0\\
1&0&0&0&0&0\\0&0&0&0&0&1\\
        \end{smallmatrix}\),\;\;\;\;
\hat N_{6}=
    \(\begin{smallmatrix}
        0&0&0&0&0&1\\ 0&0&1&0&0&0\\0&1&0&1&0&0\\
0&0&1&0&0&0\\0&0&0&0&0&1\\ 1&0&0&0&1&0
        \end{smallmatrix}\).
$$
\medskip
$$
\begin{array}{c|c|c|c|c|c|c}
     &\mathbf{ 1 }& 2 & 3 & 4 &\mathbf{ 5 }&\mathbf{ 6 } \\
     \hline
    \mathbf{ 1 }&\mathbf{ 1 }& 2 & 3 & 4&\mathbf{ 5 }&\mathbf{ 6  }\\   
     \hline
    2 & 2 &\!\!1\!\!+\!3\!\!&\!\!2\!+\!4\!+\!6\!\!&\!\!3\!\!+\!5\!\!&4&3\\     
     \hline
    3 & 3 &\!\!2\!+\!4\!+\!6\!\!&\!\!1\!+\!3\!+\!3\!+\!5\!\!&\!\!2\!+\!4\!+\!6\!\!&3&\!\!2\!+\!4\!\!\\     
     \hline
    4 & 4 &\!\!3\!+\!5\!\!&\!\!2\!+\!4\!+\!6\!\!&\!\!1\!\!+\!3\!\!&2&3\\     
     \hline
    \mathbf{ 5 }&\mathbf{ 5 }& 4 & 3 & 2 &\mathbf{ 1 }&\mathbf{ 6 }\\   
     \hline
    \mathbf{ 6 }&\mathbf{ 6 }& 3 &\!\!2\!+\!4\!\!&3&\mathbf{ 6 }&\!\!\mathbf{1\!+\!5}\!\!
\end{array}$$
\medskip
$$
\begin{array}{ccccccc}
    Z_{1}=&\hat\chi_{1}^{ }\hat\chi_{1}^{*}&+
    \hat\chi_{5}^{ }\hat\chi_{5}^{*}&+\hat\chi_{6}^{ }\hat\chi_{6}^{*},\\
   Z_{2\otimes 1}=&\hat\chi_{2}^{ }\hat\chi_{1}^{*}&+\hat\chi_{4}^{ }\hat\chi_{5}^{*}&+
    \hat\chi_{3}^{ }\hat\chi_{6}^{*}&=Z_{1\otimes 2}^*, \\
    Z_{3\otimes 1}=&\hat\chi_{3}^{ }\hat\chi_{1}^{*}&+
    \hat\chi_{3}^{ }\hat\chi_{5}^{*}&+
(\hat\chi_{2}^{ }+\hat\chi_{4}^{ })\hat\chi_{6}^{*}
&=Z_{1\otimes 3}^*,\\
    Z_{4\otimes 1}=&\hat\chi_{4}^{ }\hat\chi_{1}^{*}&+
    \hat \chi_{2}^{ }\hat\chi_{5}^{*}&+\hat\chi_{3}^{ }\hat\chi_{6}^{*}
&=Z_{1\otimes 4}^*,\\
    Z_{5}=&\hat\chi_{5}^{ }\hat\chi_{1}^{*}&+
    \hat\chi_{1}^{ }\hat\chi_{5}^{*}&+\hat\chi_{6}^{ }\hat\chi_{6}^{*},\\
    Z_{6}=&\hat\chi_{6}^{ }\hat\chi_{1}^{*}&+
    \hat\chi_{6}^{ }\hat\chi_{5}^{*}&+
(\hat\chi_{1}^{ }+\hat\chi_{5}^{ })\hat\chi_{6}^{*}.
\end{array}$$
$$\hat\chi_{1}=\chi_{1}+\chi_{7},\;\;\;\;
\hat\chi_{2}=\chi_{2}+\chi_{6}+\chi_{8},\;\;\;\;
\hat\chi_{3}=\chi_{3}+\chi_{5}+\chi_{7}+\chi_{9},$$
$$\hat\chi_{4}=\chi_{4}+\chi_{6}+\chi_{10},\;\;\;\;
\hat\chi_{5}=\chi_{5}+\chi_{11},\;\;\;\;
\hat\chi_{6}=\chi_{4}+\chi_{8}.$$
\caption{The graph fusion algebra of $E_{6}$ and its twisted partition
functions in terms of extended characters.  The ambichiral subalgebra 
is shown bold and the extended characters are given in terms of the $A_{11}$ characters.}
    \label{tbl:E6}
\end{table}

Some algebra tables are given in Tables~\ref{tbl:A4}--\ref{tbl:E6}. 
In the case of the graph $A_{L}$ this reduces to the usual Verlinde
formula, the structure constants are denoted by $N_{i\,j}{}^{k}$ and the
matrix of eigenvectors by $S$.  For an \ade graph $G$ with 
Coxeter number $g=L+1$, another algebra of non-negative integer
matrices with the $A_L$ structure constants  $N_{i\,j}{}^{k}$ is
given by the \emph{fused adjacency matrices} $n_{i}$
defined by the $s\ell(2)$ recurrence relation
\begin{equation}
    n_{1}=\text{I},\qquad n_{2}=G,\qquad n_{i+1}=n_{2}\,n_{i}-n_{i-1}\,
    \ \text{ for }\  2<i< g-1,
    \label{eq:ni}
\end{equation}
which closes with $n_i=0$ for $i>g-2$ and
\begin{equation}
n_{g-2}=\cases{\text{I}, &\mbox{for $D_{2\ell}$, $E_7$, $E_8$}\cr 
\sigma, &\mbox{for $A_L$, $D_{2\ell-1}$, $E_6$}}
\end{equation}
where $\sigma$ is the $\Bbb Z_2$ graph automorphism. Clearly, $E_7$ and $E_8$ do not admit a $\Bbb Z_2$ automorphism. So the fusions contain the $\Bbb Z_2$ graph automorphism in all cases where it exists except for $D_{2\ell}$. 

The matrices $n_i$ also satisfy a Verlinde like property
\begin{equation}
     n_{i\,a}{}^{b}=\sum_{j\in 
    \text{\Exp}(G)}\frac{
    S_{i}^{j}}{S_{*}^{j}}\,\psi_{a}^{j}\left(\psi_{b}^{j}\right)^{*}
    \label{eq:nVerlindeLike}
\end{equation}
with the  algebra structure
\begin{equation}
    n_{i}\,n_{j}=\sum_{k\in A_{L}}N_{i\, j}{}^{k}\,n_{k},\qquad n_i\hat{N}_a=\sum_{b\in G} n_{ia}{}^b \hat{N}_b.
    \label{eq:nAlg}
\end{equation}
The matrices $n_{i}$ are in fact linear combinations of the
graph algebra matrices $\hat N_a$
\begin{equation}
    n_{i}=\sum_{a\in G} n_{i\,1}{}^{a}\hat N_{a}.
    \label{eq:hatNn}
\end{equation}
For $b\in G$, the rectangular 
matrix $V^{b}=(n_{i\,a}{}^{b})_{i\in A_{g-1},a\in G}$ is
called an \emph{intertwiner} because it intertwines the fused
adjacency matrices:
\begin{equation}
    N_{i}\, V^{b}=V^{b}\, n_{i}, \;\; \quad\text{for all}\  i\in A_{L}.
    \label{eq:intertwines}
\end{equation}

One view of these graph algebras is that nodes label
bimodules and edges are homomorphisms between these bimodules.  In the
case of the graph $G$, the edges describe the homomorphisms arising
from tensoring with the fundamental bimodule, the result is isomorphic
to the direct sum of the bimodules which are adjacent to it on the
graph.  For Type I models, one can associate a graph $G_{a}$ to each
vertex in the same manner, by placing $\hat N_{a\, b}{}^{c}$ edges between
the vertex $b$ and the vertex $c$.  For Type II models, this
construction fails.

The graph fusion algebra of a solvable \ade lattice model built on the graph $G$ is not the graph fusion algebra
itself.  Rather, this latter graph encodes the fusion algebra of a
\emph{Wess-Zumino-Witten} $\slhat (2)_{g-2}$ (WZW) theory. The solvable \ade lattice 
model is actually associated with a \emph{minimal model} whose fusion algebra
is given by the \emph{tensor product graph} $A_{g-2}\times G$ where $g$ is the
Coxeter number of $G$.  A vertex of this tensor product graph is of the form
$(r,a)\in A_{g-2}\times G$ and is adjacent with the vertex $(r',b)$ whenever
$r$ and $r'$ are neighbours in $A_{g-2}$ and $a$ and $b$ are neighbours in $G$.  When $G=A_{g-1}$, it is customary to denote such a vertex with
the Kac labels $(r,s)$. Moreover, this tensor product graph is quotiented by the Kac
table symmetry
\begin{eqnarray}
  (r,s)&\sim&(g-1-r, g-s) \text{ ~~if~~ } G=A_{g-1},\\
  (r,a)&\sim&(g-1-r, \sigma(a)) \text{ ~~otherwise.}
\label{eq:KacSymm}
\end{eqnarray}
where $\sigma$ is the identity for $D_{2\ell},E_7,E_8$ and $\Bbb Z_2$ graph automorphism for $D_{2\ell}$ and $E_6$.
The graph fusion algebra for this quotient is $A_{g-2}\otimes G/\sim$ and 
hence it is generated by the two $\slhat (2)$ WZW subalgebras
$1\otimes G$ and $A_{g-2}\otimes 1$.

\begin{figure}[htbp]
\begin{center}
\includegraphics*[0cm,21.9cm][16cm,26.8cm]{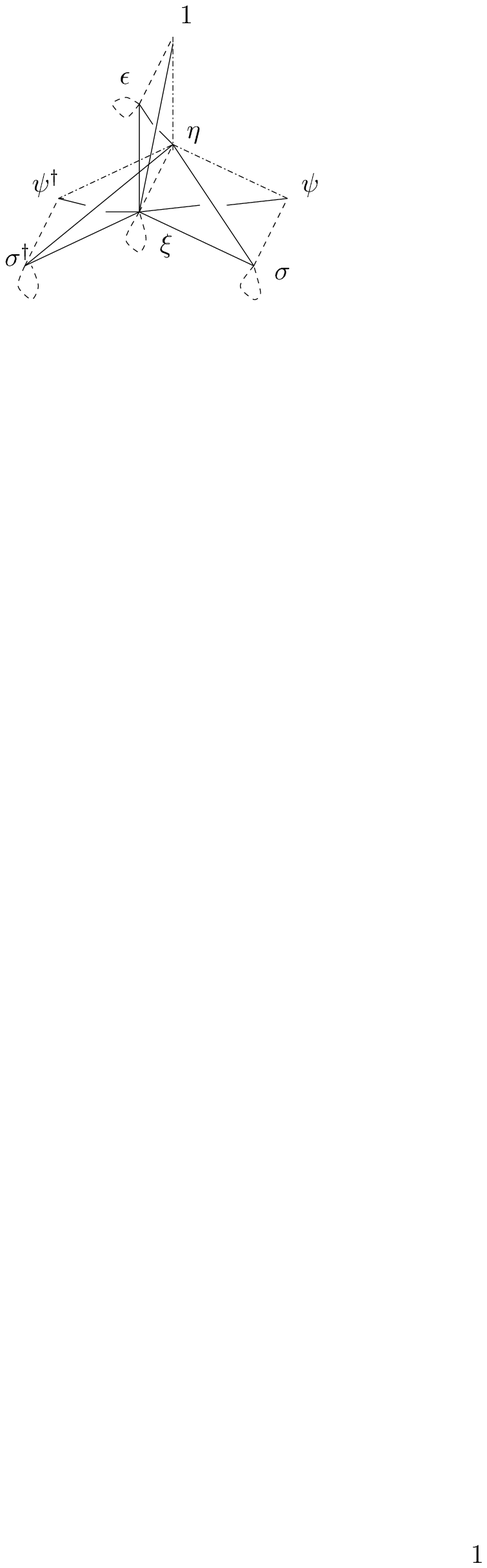}
\end{center}
\caption{The graph $(T_2,D_4)$ coding the fusion algebra of the three-state
Potts model. The solid lines encode the adjacencies of the nodes on the tensor product graph.
The nodes 
$(r,a)=\{(1,1),(1,2),(1,3),(1,4),(2,1),(2,2),(2,3),(2,4)\}$ in $(T_2,D_4)$
are labelled by the associated fields 
$\{1,\eta,\psi,\psi^\dagger,\epsilon,\xi,\sigma,\sigma^\dagger\}$ respectively.}
         \label{fig:T2D4}
\end{figure}

For example, the three-state Potts model is the minimal model associated with
the graph $D_4$. Its fusion algebra is the graph algebra 
$(A_4,D_4)/_\sim = (T_2,D_4)$ pictured in Fig.~\ref{fig:T2D4}. It has eight
vertices and two types of edges, corresponding to the action of the
generators $\epsilon, \eta$, of each subalgebra, $T_2=\{I,\epsilon\}$ and
$D_4=\{I,\eta,\psi,\psi^\dagger\}$ where we use the labelling of Fig.~\ref{fig:T2D4}. 
You read off that $\xi
\eta=\epsilon+\sigma+\sigma^\dagger$ and $\xi \epsilon=\xi+\eta$. 
Assuming associativity, you can then infer for example that $\xi\xi=
I+\epsilon+\psi+\psi^\dagger+\sigma+\sigma^\dagger$ and work out
the rest of the algebra table, which is also given by the tensor product
of the $\hat N$ matrices for $D_4$ (listed in Table~\ref{tbl:D4}) with
the fusion matrices of $T_2$, namely
\begin{equation}N_1=
\begin{pmatrix}
1&0\\ 0&1
\end{pmatrix}, \qquad N_2=
\begin{pmatrix}
0&1\\ 1&1
\end{pmatrix}.
\end{equation}

\subsection{Ocneanu graph algebras and twisted partition functions}
\label{sec:OcneanuGraph}
In this section we review the properties of Ocneanu (quantum double) graphs and their fusion algebras~\cite{Ocn,Ocn2,PZ0011021,PZ01,CoqTwist01}.  The related 
\emph{Double Triangle Algebra} (DTA) governs many aspects of both the
statistical mechanics models and their associated conformal field theories. It is
called the algebra of \emph{quantum symmetries} of the problem.  We are
interested here in 
the algebra of
twisted boundary conditions on the torus, called the \emph{twisted 
fusion algebra}~\cite{PZ0011021,PZ01}. 
We discuss the twisted
boundary conditions and associated twisted partition functions briefly summarizing the work of Petkova and Zuber~\cite{PZ0011021,PZ01}. We then present an alternative approach using tensor products~\cite{Ocn,Ocn2,CoqTwist01}.

The twisted boundary conditions are encoded by the Ocneanu~\cite{Ocn} graph fusion algebra $\widetilde G$ whose structure constants are denoted by $\tilde N_{x\, y}{}^z$
for $x, y, z\in \widetilde G$. The matrices $\tilde{N}_y=\{\tilde{N}_{xy}{}^z\}$ satisfy
\begin{equation}
\tilde{N}_x\tilde{N}_y=\sum_{z\in\tilde{G}}\tilde{N}_{xy}{}^z\tilde{N}_z.
\label{twistAlg}
\end{equation}
Petkova and Zuber give explicit expressions for these structure constants for the $\ade$ graphs. 
The seam index 
\begin{equation}
x=\cases{(a,b,\kappa)\in(H,H,\Bbb Z_2),&\mbox{WZW}\cr
(r,a,b,\kappa)\in(A_{g-2},H,H,\Bbb Z_2),&\mbox{minimal}}
\end{equation}
labels conformal twisted boundary conditions or seams. 
The index $\kappa=1,2$ labels the automorphisms $\eta=\sigma^{\kappa-1}=I,\sigma$.
The seams $x$ are not all distinct due to quantum symmetry
\begin{equation}
x=(r,a,b,\kappa)\equiv x'=(r',a',b',\kappa')\quad\Leftrightarrow\quad 
Z_{(r,a,b,\kappa)}=Z_{(r',a',b',\kappa')}
\end{equation}
that is, seams giving rise to the same twisted partition functions are considered equivalent. In some cases, such as $D_4$, it is necessary to use unspecialized characters to see the full quantum symmetry. 
For the WZW models, it suffices to take $x$ of the form
\begin{equation}
x=\cases{s\in A_{g-1},&G=A_{g-1}\cr
(a,\kappa)\in(D_{2\ell},\Bbb Z_2),&G=D_{2\ell}\cr
s\in A_{4\ell-1},&G=D_{2\ell+1}\cr
(a,b)\in(G,G), &G=E_{6,8}\cr
(a,b)\in(D_{10},D_{10}),&G=E_7}
\end{equation}
and similarly for the minimal models with $r\in A_{g-2}$ added.

The modular invariant torus partition functions of the \ade minimal models are
\begin{eqnarray}
Z(q)=\sum_{(r,s),(r',s')}
Z_{(r,s),(r',s')}\chi_{r,s}(q)\chi_{r',s'}(\overline{q}),\qquad
Z_{(r,s),(r',s')}=\displaystyle\delta_{r,r'}
\sum_{a\in T}\,n_{s 1}{}^a\,n_{s' 1}{}^{\zeta(a)}
\end{eqnarray}
where $n_s=n_s(H)$ are the fusion matrices of the Type I parent $H$ of $G$, 
\begin{eqnarray}
T=T_1&=&\cases{\{1,2,\ldots,L\},&G=A_L\cr
\{1,3,5,\ldots,2\ell-1,2\ell\},&G=D_{2\ell}\cr
\{1,2,3,\ldots,4\ell-1\},&G=D_{2\ell+1}\cr
\{1,5,6\},&G=E_6\cr 
\{1,3,5,7,9,10\},&G=E_7\cr
\{1,7\},&G=E_8}
\end{eqnarray}
and the involutive twist $\zeta$ is the identity for Type I theories but for Type II theories has the action
\begin{equation}
\zeta=\cases{s\mapsto 4\ell-s,\ s=2,4,\ldots,2\ell-2,&G=D_{2\ell+1}\cr
\{1,3,5,7,9,10\}\mapsto \{1,9,5,7,3,10\},&G=E_7
}
\end{equation}
The twisted partition functions are given by the toric matrices $P_{ab}^{(\kappa)}$
\begin{equation}
Z_{(r,a,b,\kappa)}(q)=\!\!\sum_{(r',s'),(r'',s'')}
N_{r\,r'}^{(A_{g-2})}{}^{r''}\;[P_{ab}^{(\kappa)}]_{s' s''}\;
\chi_{r',s'}(q)\,\chi_{r'',s''}(\overline{q}),\quad 
[P_{ab}^{(\kappa)}]_{s' s''}=\sum_{c\in T_\kappa} n_{s' a}{}^c\,n_{s''
b}{}^{\zeta(c)}
\label{PZtwistedPFs}
\end{equation}
where
\begin{eqnarray}
T_2&=&\cases{
\{2,4,\ldots,2\ell-2\},&G=D_{2\ell}\cr
\;T_1,&\mbox{otherwise}}
\end{eqnarray}
except for the special seams denoted $x=(r,X)$ of $E_7$~\cite{PZ01}, which are given by $Z_{(r,X)}(q)=Z_{(r,6,2,1)}(q)-Z_{(r,4,2,1)}(q)$.

The Ocneanu graph fusion algebra is a double graph algebra 
combining left and right copies of the graph fusion algebra $H$ connected through a left-right symmetric
subalgebra called the \emph{ambichiral subalgebra}.
For graphs $G$ of Type I,  $H$ is $G$ itself. For graphs $G$
of Type II, it involves the parent graph algebra $H$ and a twist $\zeta$.
Although the graph algebras of Type II theories are not proper graph
algebras, the parent graph $H$ is always of Type I.
The Ocneanu graph has two types of edges (plain and dashed), corresponding to the action of
the left and right copies of the generator of $H$.

\begin{figure}[htbp]
\begin{center}
\vspace{-.4in}
\setlength{\unitlength}{20pt}
\begin{picture}(6,6)
\put(.5,-4){\includegraphics[width=3.75cm]{fig/D4D4.eps}}
\put(3.5,4.1){$1$}
\put(.85,2.3){$2$}
\put(6.0,2.3){$2'$}
\put(3.5,2.8){$3$}
\put(3.5,1.5){$4$}
\put(3.5,0){$1'$}
\put(3.5,-1.3){$3'$}
\put(3.5,-2.6){$4'$}
\end{picture}
\vspace{.75in}
\end{center}
\caption{The Ocneanu graph $\widetilde{D}_4$. The nodes can be labelled by $x=(a,b)$ or $(a,\kappa)$. 
We have set $a=a\otimes 1$, $2'=1\otimes 2$ with $2\otimes 2$ decomposed into $1',3',4'$. The nodes \mbox{$\{1,2,3,4,1',2',3',4'\}$} are alternatively labelled by $(a,\kappa)=\{(1,1), (2,1), (3,1), (4,1), (1,2), (2,2), (3,2), (4,2)\}$ \mbox{respectively}.}
         \label{fig:D4D4}
%
\begin{center}\input{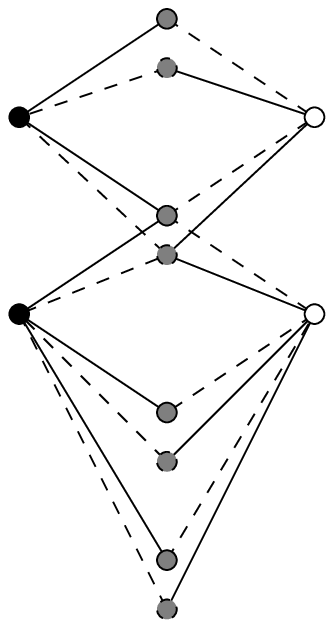}
\end{center}
\caption{The Ocneanu graph $\widetilde{D}_{2\ell}$ for $\ell=3$.}
         \label{fig:D2lD2l}
\end{figure} 

An alternative construction of the Ocneanu graph fusion algebra, which emphasizes the chiral and ambichiral structure, can be given~\cite{Ocn,Ocn2,CoqTwist01} 
in terms of tensor products over the subalgebra $T$.
We start with an example and consider $G=D_{4}$. The nodes $T=\{1, 3, 4\}$ form a
$\mathbb{Z}_{3}$ subalgebra of the graph fusion algebra $G$ corresponding to the extended chiral algebra. This gives rise to the ambichiral subalgebra of $\tilde{G}$. One can then construct the tensor product algebra over $T$
\begin{equation}
    D_{4}\otimes_{\mathbb{Z}_{3}} D_{4}
    \label{eq:D4D4}
\end{equation}
The  tensor product $\otimes_T$ over $T$ means that $a\otimes t\,b\equiv a\,t\otimes b$ for all $a,b\in G$ and $t\in T$. 
In $D_4$ for example $2\otimes 3=2\, 3\otimes
1=2\otimes 1$.  So this algebra has six distinct elements, 
$1\otimes 1$, 
$1\otimes 2 = 3\otimes 2 = 4\otimes 2$, 
$1\otimes 3 = 3\otimes 1$, 
$1\otimes 4 = 4\otimes 1$ , 
$2\otimes 1 = 2\otimes 3 = 2\otimes 4$, 
and $2\otimes 2$.  It clearly has two
generators, a left and a right, $\square_{L}=2\otimes 1$ and
$\square_{R}=1\otimes 2$.  They generate the \emph{left} respectively \emph{right
chiral subalgebras}.  One would like to encode this algebra in a graph as
previously but there is an obstruction:
\begin{equation}
            (2\otimes 2)(1\otimes 2)=2\otimes(1+3+4)
=2\otimes 1+2\otimes  1+2\otimes 1
    \label{eq:2212} 
\end{equation}
so while there is only one edge from $2\otimes 1$ to $2\otimes 2=(2\otimes
1)(1\otimes 2)$, there would be three in the opposite direction.  This
problem is solved by splitting the node $2\otimes 2$ into three different
ones, $1',3',4'$, using a non central extension of the algebra by an algebra
of $2\times 2$-matrices~\cite{CoqTwist01}.  The detail of this extension is
not needed to compute twisted partition functions but we note that it leads to non-commutativity.  
One then obtains the graph $\widetilde{D}_4$ presented in Fig.~\ref{fig:D4D4}.

The same procedure works for $G=H=D_{2\ell}$ (see Fig.~\ref{fig:D2lD2l}). It has a $J_{\ell +1}$
subalgebra generated by the odd vertices (the forked extremities are both
taken as odd) over which the tensor square is taken and extended by an
algebra of $2\times 2$-matrices. The result is encoded in a graph
$\widetilde{D}_{2\ell}$, depicted in Fig.~\ref{fig:D2lD2l}, where the
ambichiral $1,3,\ldots,2\ell -3, 2\ell -1, 2\ell$ nodes are duplicated by
$1',3',\ldots,(2\ell -3)', (2\ell -1)', (2\ell)'$, and where the even
vertices have a left and a right counterpart $2\otimes 1, 4\otimes 1,\ldots,
2(\ell-1)\otimes 1$ and $1\otimes 2, 1\otimes 4,\ldots, 1\otimes 2(\ell-1)$.
The algebra structure can be worked out beginning with $1'$, which satisfies
\begin{eqnarray}
1'(1\otimes 2)&=&2\otimes 1,\\
1'(2\otimes 1)&=&1\otimes 2,\notag
\end{eqnarray}
while $2\otimes 2=1'+3'$.

The Type I exceptional cases $E_{6}$ and $E_{8}$ are simpler as there
is no need to extend the algebra.  The subalgebras in these cases are
generated by $T=\{1,5,6\}$ and $\{1,7\}$ respectively.  The $A_{L}$ case is even simpler
as we tensorise over the full graph algebra, yielding the same algebra back again
$A_{L}\otimes_{A_{L}}A_{L}\simeq A_{L}$ so all the elements are ambichiral.

\begin{figure}[hbtp]
\begin{center}\input{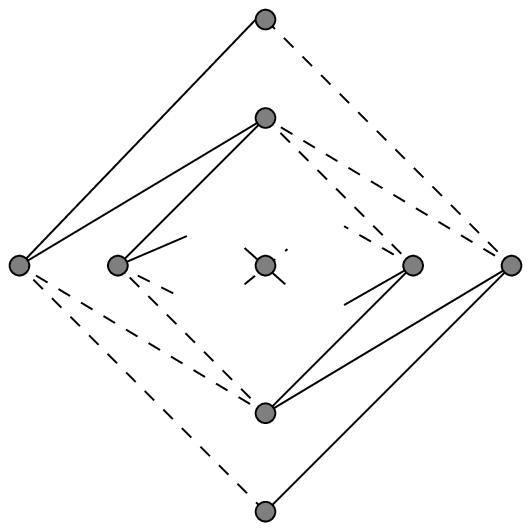}
\end{center}
\caption{The Ocneanu graph $\widetilde{D}_{2\ell+1}$.}
         \label{fig:D2lp1}
\end{figure}

For the Type II models, $G=D_{2\ell +1}, E_{7}$, the
Ocneanu graph algebra is defined through the square tensor of the
parent theory $H=A_{4\ell -1}, D_{10}$ twisted by the involution $\zeta$. 
For $G=D_{2\ell +1}$, the twisted fusion algebra is defined by $A_{4\ell\mi1} \otimes_{\zeta}
A_{4\ell\mi1}$ where $a\otimes b=a\zeta(b)\otimes 1=1\otimes \zeta(a) b$. It is
the graph algebra of the Ocneanu graph $\tilde{D}_{2\ell +1}$ defined by
${4\ell \mi1}$ vertices forming two $A_{4\ell \mi1}$ graphs with plain and
dashed edges, sharing the same numbering for the odd vertices (forming
$\Exp(D_{2\ell+1})$) but where the even ones are flipped, as pictured
in Fig.~\ref{fig:D2lp1}.
For $G=E_{7}$, the parent theory is $H=D_{10}$ and the automorphism is
given by interchanging the nodes $3$ and $9$, so that the twist
fusion algebra is $D_{10}\otimes_{T,\zeta}D_{10}$ where $T$ is the $D_{10}$
ambichiral subalgebra (its odd vertices, counting forked vertices as both odd).
The left (right) chiral subalgebra is generated by the left 
(right) generator and the ambichiral subalgebra. They are both
isomorphic to the primitive graph algebra $H$ and, as in the Type I case, the
ambichiral subalgebra is their intersection.

To summarize, the vertices of the Ocneanu graphs are given by distinct pairs
$(a,b)\in H$ ($H$ is the graph $G$ or its parent graph), coding a left and a
right element, or equivalently for the $D_{2\ell}$ models, a pair $(a,\kappa)$ with $\kappa\in\{1,2\}$.
The twist fusion algebras just described are for the $\slhat (2)_{g-2}$ WZW
theories.  The Ocneanu graph fusion algebras associated with \ade minimal models 
are more involved. These are given by two copies of the graph algebra
$(A_{g-2},H)/\sim$ whose vertices are pairs $(r,a)$
with $r\in A_{h-2}$ and $a\in H$, subject to the Kac table symmetry
$(r,s)\sim (g-1-r, g-s)$ if $H=A_{g-1}$ and $(r,a)\sim (g-1-r,a)$ 
otherwise. The vertices are, in general, labelled by $x=(r,a,b,\kappa)\in
(A_{g-2},H,H,\Bbb{Z}_2)$.
In the $(A_4,D_4)$ case for example, the Ocneanu graph is simply the tensor
product of $\tilde{D}_4$, the Ocneanu graph of $D_4$, with the tadpole $T_2$. It has three types
of edges, associated with the action of the left, the right and the tadpole
generators.

\subsection{Twisted partition functions and extended characters} 
\label{sec:Twisted}
Given an \ade graph $G$, we 
associate to each node $a\in G$, an 
\emph{extended character} $\hat \chi_{a}(q)$
which is a generating function in the modular parameter $q$.  The ordinary
characters $\chi_{i}(q)$ are the $A$-type characters. The extended characters
are 
\begin{equation}
    \hat\chi_{a}(q)=\sum_{i\in A_{g-1} }n_{i\,1}{}^{a}\chi_{i}(q)
    \label{eq:hatChi}
\end{equation}

Explicitly, in the $D_{4}$ case,
\begin{equation}
    \hat\chi_{1}=\chi_{1}+\chi_{5},\;\;\hat\chi_{2}=\chi_{2}+\chi_{4},
\;\;    \hat\chi_{3}=\hat\chi_{4}=\chi_{3}.
    \label{eq:extendedCharD4}
\end{equation}
We use the extended characters mostly for Type I graphs, that is, either $G$ itself or its parent $H$ if $G$ is of Type II.

The twisted partition
functions, which are sesquilinear combination of characters, can now be written in terms of extended characters 
in the following way~\cite{PZ0011021,Coq00,PZ01,CoqTwist01}.
First, the modular invariant partition function corresponding to the unit vertex of the Ocneanu graph 
is associated with a sum over the subalgebra $T$
\begin{equation}
    Z(q):=\sum_{a\in T}\hat \chi_{a}(q)\left( \hat
    \chi_{\zeta(a)}(q) \right)^{*}.
    \label{eq:Z1}
\end{equation}
where the automorphism $\zeta$ is the identity for the Type I models.
For Type II theories,
the sum over $T$ is a sum over (a subset of)
the parent graph. In the $D_{2\ell +1}$ case for example, we sum over $A_{4\ell
-1}$ (they are all ambichiral). Moreover, the vertex $a$, associated with
$\hat\chi_{a}(q)$, is paired by the twist $\zeta$ with $\hat\chi_{\zeta(a)}(q)^{*}$.
It is also a sesquilinear form in terms of ordinary characters and its
diagonal part gives the exponents $\Exp(G)$ of Table~\ref{tbl:Graphs}.
This partition function has the property of \emph{modular invariance}
in the modular parameter $q=\exp(2\pi i\tau)$, that is to say, $Z(1+\tau)=Z(\tau)$ and
$Z(-1/\tau)=Z(\tau)$.  The other twisted partition functions \emph{are not}
modular invariant.

The twisted partition function of other elements
$a\otimes_T b$ of the Ocneanu graph, with or without an automorphism seam $\eta=\sigma^{\kappa-1}$, are obtained by the action of this
element on the terms of $Z(q)$
\begin{equation}
    Z_{a\otimes_T b}^{(\kappa)}(q):=
    \sum_{c\in T_\kappa}\hat \chi_{a\,c}(q)
     \left( \hat \chi_{\zeta(c)\,b}(q) \right)^{*}
=    \sum_{c\in T_\kappa, d,e\in H}{\hat N}_{a\, c}{}^d{\hat N}_{b\, \zeta(c)}{}^e
\hat \chi_{d}(q)
     \left( \hat \chi_{e}(q) \right)^{*}
    \label{eq:Zab}
\end{equation} 
where $a\, c$ and $c\, b$ denote the product in the graph fusion algebra, that is
to say, it is a linear combination of elements of the algebra and the
characters are extended by linearity in their indices. The graph $H=G$ 
for Type I theories but is the parent graph for Type II theories. 
Petkova and Zuber~\cite{PZ01} indicated that the twisted partition functions could be written in terms of the extended characters but did not write down an explicit formula. 
Notice that this implies that the $\hat N$ structure constants involved in the
calculation are all positive integers since no use is made of the graph 
fusion algebra for Type II models, only for the parent model which is always of Type I.
For example, the action of the left generator changes the left term in $\hat
\chi_{a}(q)\left(\hat \chi_{\zeta(a)}(q) \right)^{*}$ to a sum over the 
neighbours of $a$ in $H$
\begin{eqnarray}
    Z_{\square_{L}}(q)=Z_{2\otimes 1}(q)= \sum_{a\in T}\hat
    \chi_{2\,a}(q)\left( \hat \chi_{\zeta(a)}(q) \right)^{*} \notag=
\sum_{a\in T,\; b\,{\sim}_{\ss H}\, a}
  \hat \chi_{b}(q)\left( \hat \chi_{\zeta(a)}(q) \right)^{*}.
    \label{eq:Z2}
\end{eqnarray}
Notice that $Z_{a\otimes_T b}=\left(Z_{b\otimes_T a}\right)^{*}$ and, in
particular for the Type I theories, the ambichiral twisted partition
functions are real.  A complete list of these twisted partition
functions in terms of ordinary characters is given in~\cite{PZ01,CoqTwist01}.  
For the cases $D_{6}$ and $E_{6}$ 
we give a list of them in terms of extended characters in
Tables~\ref{tbl:D6} and \ref{tbl:E6}.


The twisted partition functions just described are for the $\slhat(2)_{g-2}$,
or WZW models. The twisted partition functions for the minimal models (\ref{PZtwistedPFs}) involve
pairs $(r,a)\in A_{g-2}\times H/\sim$ of indices in place of single indices
$a\in H$. Here, $g$ is the Coxeter number of $H$ and $\sim$ is the Kac table
symmetry. The ambichiral subalgebra is the product of the two ambichiral
subalgebras for each graph.
In the case $(A_4,D_4)/\sim=(T_2,D_4)$, for example, the modular 
invariant partition function is
\begin{eqnarray}
  \label{eq:Z1A4D4}
  Z_1=\sum_{x\in T}\hat\chi_x\hat\chi_x^*=\!\!\sum_{r\in\{1,3\},\,
    a\in\{1,3,4\}\subset D_4}\!\!
\hat\chi_{r,a}^2
=\hat\chi_{1,1}^2+\hat\chi_{3,1}^2+
\hat\chi_{1,3}^2+\hat\chi_{3,3}^2+
\hat\chi_{1,4}^2+\hat\chi_{3,4}^2
\end{eqnarray}
where we have kept the labels $1=4$ and $2=3$ for $r\in
T_2=A_4/{\mathbb{Z}_2}$. The other $15$ twisted partition functions are
obtained by action of the twisted fusion algebra. For example,
\begin{eqnarray}
  \label{eq:Z2A4D4}
  Z_{(1,2)\otimes 1}&=&
\hat\chi_{1,2}(\hat\chi_{1,1}+\hat\chi_{1,3}+\hat\chi_{1,4})^*+
\hat\chi_{3,2}(\hat\chi_{3,1}+\hat\chi_{3,3}+\hat\chi_{3,4})^*\\
  Z_{(3,1)\otimes 1}&=&
\hat\chi_{3,1}\hat\chi_{1,1}^*+(\hat\chi_{1,1}+\hat\chi_{3,1})\hat\chi_{3,1}^*+
\hat\chi_{3,3}\hat\chi_{1,3}^*+(\hat\chi_{1,3}+\hat\chi_{3,3})\hat\chi_{3,3}^*
\notag\\&&\;\;
+\hat\chi_{3,4}\hat\chi_{1,4}^*+(\hat\chi_{1,4}+\hat\chi_{3,4})\hat\chi_{3,4}^*
\end{eqnarray}
and so on as listed in Table~\ref{tab:A4D4}.

\begin{table}[htbp]
  \centering
  \begin{eqnarray}
Z_1&=& \hat\chi_{1,1}^{ }\hat\chi_{1,1}^*+ \hat\chi_{3,1}^{ }\hat\chi_{3,1}^*+ 
   \hat\chi_{1,3}^{ }\hat\chi_{1,3}^*+ \hat\chi_{3,3}^{ }\hat\chi_{3,3}^*+ 
   \hat\chi_{1,4}^{ }\hat\chi_{1,4}^*+ \hat\chi_{3,4}^{ }\hat\chi_{3,4}^*
\notag \\
Z_{\eta\otimes 1}&=&
   \hat\chi_{1,2}(\hat\chi_{1,1}^{ }+\hat\chi_{1,3}^{ }+\hat\chi_{1,4}^{ })^*
+ \hat\chi_{3,2}(\hat\chi_{3,1}^{ }+\hat\chi_{3,3}^{ }+\hat\chi_{3,4}^{ })^*
=Z_{1\otimes\eta}^*\notag \\ 
Z_{\psi}&=&
\hat\chi_{1,3}^{ }\hat\chi_{1,1}^*+ \hat\chi_{3,3}^{ }\hat\chi_{3,1}^*+ 
   \hat\chi_{1,4}^{ }\hat\chi_{1,3}^*+ \hat\chi_{3,4}^{ }\hat\chi_{3,3}^*+ 
   \hat\chi_{1,1}^{ }\hat\chi_{1,4}^*+ \hat\chi_{3,1}^{ }\hat\chi_{3,4}^*
=Z_{\psi^\dagger}\notag \\
Z_{\epsilon}&=& \hat\chi_{3,1}^{ }\hat\chi_{1,1}^*+ 
(\hat\chi_{1,1}^{ }+\hat\chi_{3,1}^{ })\hat\chi_{3,1}^*+ 
   \hat\chi_{3,3}^{ }\hat\chi_{1,3}^*+
   (\hat\chi_{3,3}^{ }+\hat\chi_{1,3}^{ })\hat\chi_{3,3}^*+\notag \\ && \;\; 
   \hat\chi_{3,4}^{ }\hat\chi_{1,4}^*+ 
(\hat\chi_{3,4}^{ }+\hat\chi_{1,4}^{ })\hat\chi_{3,4}^*\notag \\
Z_{\xi\otimes 1}&=&
   \hat\chi_{3,2}(\hat\chi_{1,1}^{ }+\hat\chi_{1,3}^{ }+\hat\chi_{1,4}^{ })^* + 
(\hat\chi_{1,2}^{ }+\hat\chi_{3,2}^{ })
(\hat\chi_{3,1}^{ }+\hat\chi_{3,3}^{ }+\hat\chi_{3,4}^{ })^*
=Z_{1\otimes\xi}^*\notag \\
Z_{\sigma}&=&
\hat\chi_{1,3}^{ }\hat\chi_{1,1}^*+ \hat\chi_{3,3}^{ }\hat\chi_{3,1}^*+ 
   \hat\chi_{1,4}^{ }\hat\chi_{1,3}^*+ \hat\chi_{3,4}^{ }\hat\chi_{3,3}^*+ 
   \hat\chi_{1,1}^{ }\hat\chi_{1,4}^*+ \hat\chi_{3,1}^{ }\hat\chi_{3,4}^*
\notag \\
&=&Z_{\sigma^\dagger}\notag \\
Z_{1'}&=&|\hat\chi_{1,2}^{ }|^2+|\hat\chi_{3,2}^{ }|^2
=Z_{\psi'}=Z_{{\psi^\dagger}'}
\notag \\ 
Z_{\epsilon'}&=&\hat\chi_{1,2}^{ }\hat\chi_{3,2}^*+
\hat\chi_{3,2}^{ }\hat\chi_{1,2}^*+|\hat\chi_{3,2}^{ }|^2=Z_{\sigma'}
=Z_{{\sigma^\dagger}'}\notag
  \end{eqnarray}
$$
\hat\chi_{r,1}:=\chi_{r,1}+\chi_{r,5},\;\;\;\;
\hat\chi_{r,2}:=\chi_{r,2}+\chi_{r,4},\;\;\;\;
\hat\chi_{r,3}=\hat\chi_{r,4}:=\chi_{r,3}.$$
  \caption{The $(A_4,D_4)$ twisted partition functions of the 3-state Potts model. The extended characters are given in terms of the Virasoro minimal characters $\chi_{r,s}$.}
  \label{tab:A4D4}
\end{table}

\section{Lattice Realization of Twisted Boundary Conditions}
\setcounter{equation}{0}
\label{sec:LatticeTwist}

\subsection{\ade lattice models}
\label{sec:ADELattice}

A solvable~\cite{BaxterBook} \ade lattice model~\cite{ABF,Pas} is associated with
a graph $G$, of $A$, $D$ or $E$ type.  We place spins on the sites of 
the square lattice, where
the spin states are taken to be the nodes of the graph $G$ and
neighbouring sites on the lattice must be neighbouring nodes on
the graph.  The probability distribution of spins is defined by the
critical (unfused) Boltzmann weight of each face (or plaquette) of
spins, depending on a spectral parameter $u$.  For four spins
$a,b,c,d\in G$ such that $(a,b),(b,c),(c,d),(d,a)$ are pairs
of neighbours in $G$, the Boltzmann weight is
\begin{equation}
    W^{11}\W{d&c\\a&b}{|u}=
    \setlength{\unitlength}{.4cm}
\begin{picture}(3.5,1)(-.6,-.2)
\put(0,-1){\framebox(2,2){\p{}{u}}}
\put(.1,-.9){\pp{bl}{\searrow}}
\put(-.25,-1.25){\pp{}{a}}
\put(2.25,-1.25){\pp{}{b}}
\put(2.25,1.25){\pp{}{c}}
\put(-.25,1.25){\pp{}{d}}
\end{picture}
=s(\lambda-u)\,\delta_{ac}+
    s(u)\,\sqrt{\frac{\psi_{a}\psi_{c}}{\psi_{b}\psi_{d}}}\;\delta_{bd}
           \label{eq:W11}
\end{equation}
 and zero otherwise.  Here, $g$ is the Coxeter number of $G$,
 $\lambda=\frac{\pi}{g}$, $s(u)=\frac{\sin (u)}{\sin (\lambda)}$ and
 $\psi_{a}$ is the entry, associated with the node $a$, of the 
Perron-Frobenius eigenvector of the adjacency matrix $G$.

These Boltzmann weights can be represented~\cite{BP01} by a local
face operator $X_{j}(u)$
\setlength{\unitlength}{14pt}
\begin{equation}
X_{j}(u)=
\vtop to 1.6\unitlength{}
\begin{picture}(2.5,1)(-.5,.9)
    \multiput(0,1)(1,-1){2}{\line(1,1){1}}
    \multiput(0,1)(1,1){2}{\line(1,-1){1}}
\put(1,1){\pp{}{u}}
\put(1,.3){\pp{}{\to}}
\multiput(0,-0.5)(0,0.25){13}{\pp{}{.}}
\multiput(1,-.5)(0,0.25){2}{\pp{}{.}}  
\multiput(1,2.25)(0,0.25){2}{\pp{}{.}}
\multiput(2,-0.5)(0,0.25){13}{\pp{}{.}}
\put(0,-.6){\pp{t}{j\mi 1}}
\put(1,-.6){\pp{t}{j}}
\put(2,-.6){\pp{t}{j\plus 1}}
\end{picture}
\;=\; s(\lambda-u)\text{I}+
    s(u)e_{j}          
\label{eq:X11}
\end{equation}

\medskip\noindent
 in the Temperley-Lieb algebra $\mathcal{T}(N,\, \lambda)$
\begin{eqnarray}
    e_{j}^{2} & =& s(2\lambda)\, e_{j}\notag \\
    e_{j}\, e_{k}\, e_{j} & =&  e_{j},\qquad\qquad |j-k|=1  \label{eq:TLJ}\\
    e_{j}\, e_{k} & = &e_{k}\, e_{j},\qquad\quad |j-k|>1 \notag
\end{eqnarray}
where $e_{j}=X_{j}(\lambda)$ is a Temperley-Lieb generator and $j$ is
an integer labelling the $N$ positions along a row of the lattice.  

\subsection{Fusion projectors}
\label{sec:FusionProjector}
Each of the \ade models gives rise to a hierarchy of \emph{fused} models
whose Boltzmann weights we are going to describe.  They are associated
with \emph{blocks} of faces where the internal spins are summed
over in a particular way.


 We first define recursively the fusion operators $P^r_j$, for
 $r=1,2,\ldots,g$ as follows
\begin{equation}
\begin{array}{l}
P^1_j\;=\;P^2_j\;=\;I \\
{}\\
P^r_j\;=\;
\frac{1}{S_{r-1}}\;P^{r-1}_{j+1}\;X_j(-(r\mi2)\lambda)\:P^{r-1}_{j+1}\,,
\quad r\geq 3\, ,\end{array}
\label{eq:FusionOperator}
\end{equation}
where $S_{k}=s(k \lambda)$ and $j$ is appropriately restricted~\cite{BP01}.  
Thus, $P^r_j$ can be
expressed as a function of $e_j,e_{j+1},\ldots, e_{j+(r-3)}$.  In
particular,
\begin{equation}
P_{j}^{3}=\frac{1}{S_{2}}\;
\setlength{\unitlength}{14pt}
\begin{picture}(2,1)(0,.7)
    \multiput(0,1)(1,-1){2}{\line(1,1){1}}
    \multiput(0,1)(1,1){2}{\line(1,-1){1}}
\put(1,1){\pp{}{-\lambda}}
\put(1,.3){\pp{}{\to}}
\end{picture}
= I - \frac{1}{S_{2}}\;
\begin{picture}(2,1)(0,.7)
    \multiput(0,1)(1,-1){2}{\line(1,1){1}}
    \multiput(0,1)(1,1){2}{\line(1,-1){1}}
\put(1,1){\pp{}{+\lambda}}
\put(1,.3){\pp{}{\to}}
\end{picture} \; .
        \label{eq:DefP}
\end{equation}
We shall represent the fusion operators diagrammatically as
\setlength{\unitlength}{5mm}
\begin{equation}
\raisebox{-1.75\unitlength}[1.75\unitlength][1.75\unitlength]{
\textB{1}{3.5}{0.5}{2}{}{P^r_j}
\textB{1.8}{3.5}{0.9}{2}{}{=}
\begin{picture}(6,3.5)
\multiput(0,2)(5,-1){2}{\line(1,1){1}}
\multiput(0,2)(5,1){2}{\line(1,-1){1}}
\multiput(1,1)(0,2){2}{\line(1,0){4}}
\multiput(0,0.5)(0,0.25){13}{\pp{}{.}}
\multiput(1,0.5)(0,0.25){2}{\pp{}{.}}
\multiput(1,3.25)(0,0.25){2}{\pp{}{.}}
\multiput(5,0.5)(0,0.25){2}{\pp{}{.}}
\multiput(5,3.25)(0,0.25){2}{\pp{}{.}}
\multiput(6,0.5)(0,0.25){13}{\pp{}{.}}
\put(-0.05,0){\pp{b}{j-1}}\put(1.1,0){\pp{b}{j}}
\put(5.3,0){\pp{br}{j+r-3}}\put(5.7,0){\pp{bl}{j+r-2}}
\put(6.4,1.7){\p{}{.}}\end{picture} }
\end{equation}
It is easy to show that this operator is in fact a projector. 
Moreover,
\begin{gather}
    P_{j'}^{r'}P_{j}^{r}=P_{j}^{r}P_{j'}^{r'}=P_{j}^{r},\qquad
    \text{ for } 0\leq j'-j\leq r-r'.
    \label{eq:PP}\\
\setlength{\unitlength}{10pt}
    \begin{picture}(8,3.5)(0,1.8)
\multiput(0,2)(7,-1){2}{\line(1,1){1}}
\multiput(0,2)(7,1){2}{\line(1,-1){1}}
\multiput(1,1)(0,2){2}{\line(1,0){6}}
\multiput(1.5,2)(1,-4){2}{
        \begin{picture}(4,3.5)
\multiput(0,2)(3,-1){2}{\line(1,1){1}}
\multiput(0,2)(3,1){2}{\line(1,-1){1}}
\multiput(1,1)(0,2){2}{\line(1,0){2}}
        \end{picture}
        }
\multiput(0,-1.5)(0,0.25){28}{\pp{}{.}}
\multiput(8,-1.5)(0,0.25){28}{\pp{}{.}}
\end{picture}
\;=\;
    \begin{picture}(8,3.5)(0,1.8)
\multiput(0,2)(7,-1){2}{\line(1,1){1}}
\multiput(0,2)(7,1){2}{\line(1,-1){1}}
\multiput(1,1)(0,2){2}{\line(1,0){6}}
\multiput(0,.5)(0,0.25){13}{\pp{}{.}}
\multiput(8,.5)(0,0.25){13}{\pp{}{.}}
\end{picture}
\vtop to 3\unitlength { } \; .
\notag
\end{gather}
In particular, the local face operator 
$
\setlength{\unitlength}{14pt}
\vtop to 1.1\unitlength{}
\begin{picture}(2,1)(0,.7)
    \multiput(0,1)(1,-1){2}{\line(1,1){1}}
    \multiput(0,1)(1,1){2}{\line(1,-1){1}}
\put(1,1){\pp{}{+\lambda}}
\put(1,.3){\pp{}{\to}}
\end{picture} \;=S_{2}e_{j}$  
is a projector orthogonal to all the $P^{r}_{j'}$ for $0\leq j-j'\leq 
r-3$. This fact 
is a defining property of the orthogonal projector $P^{r}_{j}$
\begin{equation}
    \text{Im }P^{r}_{j}=\bigcap_{k=j}^{j+r-3}\text{Ker }e_{k}.
    \label{eq:Ker}
\end{equation}
 Clearly, we can decompose the projector  $P^{r}_{j}$ onto the space of paths with given
 end points: $P^{r}(a,b)$ is the fusion projector acting on paths from
 $a$ to $b$ in $r\mi 1$ steps.  Its rank is given by the fused
 adjacency matrix entries
\begin{equation}
    \text{Rank}\left(P^{r}(a,b)\right)=n_{r\; a}{}^{b}.
    \label{eq:RankP}
\end{equation}

The $+1$ eigenvectors of $P^{r}(a,b)$ are thus indexed by an integer
$\gamma=1,2,\ldots,n_{r\; a}{}^{b}$ referred to as the \emph{bond
variable}.  We denote these eigenvectors by $\vec U^{r}_{\gamma}(a,b)$ and call them 
\emph{fusion vectors} or \emph{essential paths}.  
Explicitly, in the
representation~\eqref{eq:W11} of the Temperley-Lieb algebra
$\mathcal{T}(r-2,\, \lambda)$, these generators act on the paths from
$a$ to $b$ in $r-1$ steps as
\begin{gather}
    e_{k}(a_{0},a_{1},\ldots ,a_{k-1},a_{k},a_{k+1},\ldots 
    ,a_{r-2},b)=\qquad\qquad\qquad\qquad
    \label{eq:ek}
    \\ \qquad\qquad
    \delta_{a_{k\mi 1},a_{k\plus 1}}
    \sum_{c\sim a_{k\mi 1}}
    \frac{\psi_{a_{k}}^{\half}\psi_{c}^{\half}}
    {\psi_{a_{k\mi 1}}^{\half}\psi_{a_{k\plus 1}}^{\half}}
(a_{0},a_{1},\ldots ,a_{k-1},c ,a_{k+1},\ldots ,a_{r-2},b).
\notag
\end{gather}
In the $D$ and $E$ cases multiplicities occur and there is some freedom 
in the choice of fusion vectors corresponding to a unitary change of basis.  
In the $A_{L}$ case, however, there is a unique fusion path. 
As an example, there are two paths on $A_{L}$, going from the
 node $2$ to itself in $2$ steps, namely $(2,1,2)$ and $(2,3,2)$.  As
 they both backtrack, the fusion vector $\vec U^{3}_{1}(2,2)$ is
 unique, proportional to their difference
 $\psi_{3}^{\half}(2,1,2)-\psi_{1}^{\half}(2,3,2)$ and the fused 
adjacency matrix entry is $n_{32}{}^2=1$. 
 In the $D_{4}$ case, there is the path $(2,4,2)$ as well, so that
 there are two linearly independent fusion vectors, the previous one
 and $(2,3,2)-(2,4,2)$ or any similar linear combination (notice that
 $\psi_{3}=\psi_{4}$).  The fused adjacency matrix has a two as
 the corresponding entry $n_{32}{}^2=2$.  The general form of the
 unique $A_{L}$ fusion vector at fusion level $s\leq L$ between the
 vertex $2$ and the vertex $s-1$ is given by the following formula of
 cancelling alternating backtracking paths, generalizing the one
 just described for $s=3$
\begin{eqnarray}
        \vec U_{1}^{s}(2,s-1)=\Bigl(
    \psi_{1}^{\mi\half}\psi_{2}^{\mi\half}(2,1,2,3,\ldots,s\mi1)
    -\psi_{2}^{\mi\half}\psi_{3}^{\mi\half}(2,3,2,3,\ldots ,s\mi1)\notag\\
   \mbox{}
    +\psi_{3}^{\mi\half}\psi_{4}^{\mi\half}(2,3,4,3,\ldots ,s\mi1)
+\ldots 
    +(-1)^{s}\psi_{s\mi1}^{\mi\half}\psi_{s}^{\mi\half}(2,\ldots ,s\mi1,s,s\mi1)
    \Bigr)
\label{eq:Us2sm1}
\end{eqnarray}
and similarly for $D_{\ell}$ with fusion level $s<\ell -1$.  But at
fusion level $s=\ell -1$, the fork gives rise to a two dimensional
space of fusion vectors. One choice of orthonormal basis is given (with $\nu$ 
the appropriate normalisation constant) by
\begin{eqnarray}
    \vec U_{1}^{\ell -1}(2,\ell-2)=\frac{1}{\nu}\Bigl(
    \psi_{1}^{\mi\half}\psi_{2}^{\mi\half}(2,1,2,3,\ldots,\ell\mi2)
    -\psi_{2}^{\mi\half}\psi_{3}^{\mi\half}(2,3,2,3,\ldots ,\ell\mi2)
    \notag\\ \mbox{}
    +\psi_{3}^{\mi\half}\psi_{4}^{\mi\half}(2,3,4,3,\ldots ,\ell\mi2)
+\ldots\hspace{1.4in}\mbox{} \\ \mbox{}
    +\frac{1}{2}(-1)^{\ell\mi1}\psi_{\ell\mi2}^{\mi\half}\psi_{\ell\mi1}^{\mi\half}
    (2,\ldots ,\ell-2,\ell\mi1,\ell-2)
    +\frac{1}{2}(-1)^{\ell\mi1}\psi_{\ell-2}^{\mi\half}\psi_{s}^{\mi\half}(2,\ldots 
    ,\ell\mi2,\ell,\ell\mi2)
    \Bigr),    \label{eq:Ulm12lm2}
\notag\\
    \vec U_{2}^{\ell -1}(2,\ell-2)=\frac{1}{\sqrt{2}}\Bigl((2,\ldots ,\ell-2,\ell\mi1,\ell-2)
     -(2,\ldots 
    ,\ell\mi2,\ell,\ell\mi2)\Bigr).\qquad\ \ 
\notag
\end{eqnarray}

\subsection{Fused face operators}
\label{sec:FusedFace}
The fusion projectors allow us to define the $(p,q)$-fused face operators
consisting of $q$ rows of $p$ local face
operators with relative shifts in the spectral parameter by $\pm \lambda$ from
one face to the next
\setlength{\unitlength}{14pt}
\begin{equation}
\textB{3.5}{0}{0.5}{2}{}{X^{pq}_{j}(u)=}
\begin{picture}(5,6)
    \multiput(0,2)(2,-2){2}{\line(1,1){3}}
    \multiput(0,2)(3,3){2}{\line(1,-1){2}}
\put(2.5,2.5){\pp{}{X^{pq}(u)}}
\put(2,.3){\pp{}{\to}}
\multiput(0,-0.5)(0,0.25){25}{\pp{}{.}}
\multiput(2,-.5)(0,0.25){2}{\pp{}{.}}  
\multiput(3,5.25)(0,0.25){2}{\pp{}{.}}
\multiput(5,-0.5)(0,0.25){25}{\pp{}{.}}
\put(0,-.6){\pp{t}{j-1}}
\put(2,-.6){\pp{t}{j+ q\mi 1}}
\put(3,5.75){\pp{b}{j+ p\mi 1}}
\put(5,-.6){\pp{t}{j+ p+ q\mi 2}}\end{picture}
\textB{2}{0}{1}{2}{}{=}
\begin{picture}(7,6)(-2,0)
    \multiput(0,2)(.5,-.5){2}{\line(1,1){3}}
    \multiput(2,0)(-.5,.5){2}{\line(1,1){3}}
    \multiput(0,2)(.5,.5){2}{\line(1,-1){2}}
    \multiput(3,5)(-.5,-.5){2}{\line(1,-1){2}}
\put(4.5,3){\pp{}{u}}
\put(2.,5.5){\vector(1,-1){1}}
\put(2.,5.25){\pp{br}{u+(q\mi 1)\lambda}}
\put(.5,3.75){\vector(0,-1){1.75}}
\put(-.5,4.5){\pp{}{u+(q\mi p)\lambda}}
\put(3,-1){\vector(-2,3){1}}
\put(3.1,-1){\pp{tl}{u-(p\mi 1)\lambda}}
\put(4.2,3.3){\vector(-1,1){.85}}
\put(.8,2.3){\vector(1,1){1.85}}
\put(1.7,0.8){\vector(-1,1){.85}}
\put(2.3,.8){\vector(1,1){1.85}}
\multiput(0,-0.5)(0,0.25){16}{\pp{}{.}}
\multiput(2,-.5)(0,0.25){2}{\pp{}{.}}  
\multiput(3,5.25)(0,0.25){2}{\pp{}{.}}
\multiput(5,-.5)(0,0.25){25}{\pp{}{.}}
\put(0,0){\pp{lb}{\diagup}}
\put(0,0){\pp{lt}{\diagdown}}
\multiput(.5,.5)(0,-1){2}{\line(1,0){1}}
\put(1,0){\pp{}{P_{j}^{q\plus 1}}}
\put(2,0){\pp{tr}{\diagup}}
\put(2,0){
\put(0,0){\pp{lt}{\diagdown}}
\multiput(.5,.5)(0,-1){2}{\line(1,0){2}}
\put(1.5,0){\pp{}{P_{\!j\plus q}^{p\plus 1}}}
\put(3,0){\pp{tr}{\diagup}}
\put(3,0){\pp{rb}{\diagdown}}
}
\end{picture}
\textB{2}{0}{1}{2}{}{.}
\vtop to 1.5\unitlength {}
           \label{eq:Xpq}
\end{equation}
The position of the projectors and spectral parameters can be 
 altered by \emph{pushing-through}
\setlength{\unitlength}{14pt}
\begin{equation}
    \vtop to 1.5\unitlength{}
\textB{2.25}{0}{0.5}{2}{}{X^{pq}_{j}(u)=}
\begin{picture}(6,5.5)(-1,0)
    \multiput(0,2)(.5,-.5){2}{\line(1,1){3}}
    \multiput(2,0)(-.5,.5){2}{\line(1,1){3}}
    \multiput(0,2)(.5,.5){2}{\line(1,-1){2}}
    \multiput(3,5)(-.5,-.5){2}{\line(1,-1){2}}
\put(3,4.5){\pp{}{u}}
\put(.5,3.5){\vector(0,-1){1.5}}
\put(0,3.6){\pp{b}{u-(p\mi 1)\lambda}}
\put(1,-1){\vector(2,3){1}}
\put(1,-1){\pp{t}{u+(q\mi p)\lambda}}
\put(3.3,4.2){\vector(1,-1){.85}}
\put(.8,2.3){\vector(1,1){1.85}}
\put(.8,1.7){\vector(1,-1){.85}}
\put(2.3,.8){\vector(1,1){1.85}}
\multiput(0,-0.5)(0,0.25){16}{\pp{}{.}}
\multiput(2,-.5)(0,0.25){2}{\pp{}{.}}  
\multiput(3,5.25)(0,0.25){4}{\pp{}{.}}
\multiput(5,-.5)(0,0.25){27}{\pp{}{.}}
\put(3,5){
\put(0,0){\pp{lb}{\diagup}}
\multiput(.5,.5)(0,-1){2}{\line(1,0){1}}
\put(1,0){\pp{}{P_{\!j\plus p}^{q\plus 1}}}
\put(2,0){\pp{tr}{\diagup}}
\put(2,0){\pp{rb}{\diagdown}}
}
\put(2,0){
\put(0,0){\pp{lt}{\diagdown}}
\multiput(.5,.5)(0,-1){2}{\line(1,0){2}}
\put(1.5,0){\pp{}{P_{\!j\plus q}^{p\plus 1}}}
\put(3,0){\pp{tr}{\diagup}}
\put(3,0){\pp{rb}{\diagdown}}
}
\end{picture}
\textB{2}{0}{1}{2}{}{=}
\begin{picture}(4.5,5.5)(0,0)
    \multiput(0,2)(.5,-.5){2}{\line(1,1){3}}
    \multiput(2,0)(-.5,.5){2}{\line(1,1){3}}
    \multiput(0,2)(.5,.5){2}{\line(1,-1){2}}
    \multiput(3,5)(-.5,-.5){2}{\line(1,-1){2}}
\put(.5,2){\pp{}{u}}
\put(3,-1){\vector(-2,3){1}}
\put(3.1,-1){\pp{t}{u+(q\mi 1)\lambda}}
\put(3.3,4.2){\vector(1,-1){.85}}
\put(2.7,4.2){\vector(-1,-1){1.85}}
\put(.8,1.7){\vector(1,-1){.85}}
\put(4.2,2.7){\vector(-1,-1){1.85}}
\multiput(0,-0.5)(0,0.25){27}{\pp{}{.}}
\multiput(2,-.5)(0,0.25){2}{\pp{}{.}}  
\multiput(3,5.25)(0,0.25){4}{\pp{}{.}}
\multiput(5,-.5)(0,0.25){27}{\pp{}{.}}
\put(3,5){
\put(0,0){\pp{lb}{\diagup}}
\multiput(.5,.5)(0,-1){2}{\line(1,0){1}}
\put(1,0){\pp{}{P_{\!j\plus p}^{q\plus 1}}}
\put(2,0){\pp{tr}{\diagup}}
\put(2,0){\pp{rb}{\diagdown}}
}
\put(0,5){
\put(0,0){\pp{lb}{\diagup}}
\put(0,0){\pp{lt}{\diagdown}}
\multiput(.5,.5)(0,-1){2}{\line(1,0){2}}
\put(1.5,0){\pp{}{P_{j}^{p\plus 1}}}
\put(3,0){\pp{rb}{\diagdown}}
}
\end{picture}
\textB{2.5}{0}{1.5}{2}{}{=}
\begin{picture}(5,5.5)(0,0)
    \multiput(0,2)(.5,-.5){2}{\line(1,1){3}}
    \multiput(2,0)(-.5,.5){2}{\line(1,1){3}}
    \multiput(0,2)(.5,.5){2}{\line(1,-1){2}}
    \multiput(3,5)(-.5,-.5){2}{\line(1,-1){2}}
\put(2,.5){\pp{}{u}}
\put(4.2,3.3){\vector(-1,1){.85}}
\put(2.7,4.2){\vector(-1,-1){1.85}}
\put(1.7,0.8){\vector(-1,1){.85}}
\put(4.2,2.7){\vector(-1,-1){1.85}}
\multiput(0,-1)(0,0.25){29}{\pp{}{.}}
\multiput(2,-1)(0,0.25){4}{\pp{}{.}}  
\multiput(3,5.25)(0,0.25){4}{\pp{}{.}}
\multiput(5,-1)(0,0.25){29}{\pp{}{.}}
\put(0,0){
\put(0,0){\pp{lb}{\diagup}}
\put(0,0){\pp{lt}{\diagdown}}
\multiput(.5,.5)(0,-1){2}{\line(1,0){1}}
\put(1,0){\pp{}{P_{j}^{q\plus 1}}}
\put(2,0){\pp{tr}{\diagup}}
}
\put(0,5){
\put(0,0){\pp{lb}{\diagup}}
\put(0,0){\pp{lt}{\diagdown}}
\multiput(.5,.5)(0,-1){2}{\line(1,0){2}}
\put(1.5,0){\pp{}{P_{j}^{p\plus 1}}}
\put(3,0){\pp{rb}{\diagdown}}
}
\end{picture}\,
\textB{2}{0}{1}{2}{}{.}
           \label{eq:PushThru}
\end{equation}
These properties imply several others, namely the \emph{Transposition 
Symmetry}
\begin{equation}
    X^{pq}_{j}(u)^{T}=X^{qp}_{j}(u+(q\mi p)\lambda)
    \label{eq:Transpose}
\end{equation}
the \emph{Generalized
Yang-Baxter Equation} (GYBE)
\newsavebox{\Xpq} 
\newsavebox{\Xqp}
\newsavebox{\Xqq}
\begin{equation}
\setlength{\unitlength}{17pt}
   \savebox{\Xpq}(3,3){
   \begin{picture}(3,3)(0,2)
    \multiput(0,2)(2,-2){2}{\line(1,1){1}}
    \multiput(0,2)(1,1){2}{\line(1,-1){2}}
\put(1.5,1.5){\pp{}{X^{pq'}(u)}}
\put(2,.3){\pp{}{\to}}
\end{picture}}
   \savebox{\Xqp}(4,4){
   \begin{picture}(4,4)(0,1)
    \multiput(0,1)(1,-1){2}{\line(1,1){3}}
    \multiput(0,1)(3,3){2}{\line(1,-1){1}}
\put(2,2){\pp{}{X^{qp}(v)}}
\put(1,.3){\pp{}{\to}}
\end{picture}}
   \savebox{\Xqq}(4,5){
   \begin{picture}(4,5)(0,2)
    \multiput(0,2)(2,-2){2}{\line(1,1){3}}
    \multiput(0,2)(3,3){2}{\line(1,-1){2}}
\put(2.5,2.5){\pp{}{X^{qq'}(u+v)}}
\put(2,.3){\pp{}{\to}}
\end{picture}}
\begin{picture}(8,8.5)(0,-.5)
\multiput(0,-.5)(0,0.25){33}{\pp{l}{.}}
\multiput(2,-.5)(0,0.25){2}{\pp{l}{.}}
\multiput(3,-.5)(0,0.25){6}{\pp{l}{.}}
\multiput(3,7.25)(0,0.25){2}{\pp{l}{.}}
\multiput(4,6.25)(0,0.25){6}{\pp{l}{.}}
\multiput(6,-.5)(0,0.25){33}{\pp{l}{.}}
    \put(0,4){\usebox{\Xqp}}
    \put(1,3){\usebox{\Xqq}}
    \put(0,2){\usebox{\Xpq}}
    \put(7,3){=}
\end{picture}
\begin{picture}(6,8.5)(0,-.5)
\multiput(0,-.5)(0,0.25){33}{\pp{l}{.}}
\multiput(2,-.5)(0,0.25){6}{\pp{l}{.}}
\multiput(3,-.5)(0,0.25){2}{\pp{l}{.}}
\multiput(3,6.25)(0,0.25){6}{\pp{l}{.}}
\multiput(4,7.25)(0,0.25){2}{\pp{l}{.}}
\multiput(6,-.5)(0,0.25){33}{\pp{l}{.}}
    \put(3,6){\usebox{\Xpq}}
    \put(0,3){\usebox{\Xqq}}
    \put(2,1){\usebox{\Xqp}}
\end{picture}
                \label{eq:GYBE}
\end{equation}
the \emph{Inversion Relation}
\begin{gather}
\setlength{\unitlength}{17pt}
  \mbox{}\hspace{-.7in} X^{pq}_{j}(u)
    X^{qp}_{j}(-u)
\;=\;
\begin{picture}(7.8,2)(-.5,1)
    \multiput(0,1)(1,-1){2}{\line(1,1){2}}
    \multiput(0,1)(2,2){2}{\line(1,-1){1}}
\put(1.5,1.5){\pp{}{X^{pq}(u)}}
\put(.3,1){\pp{}{\downarrow}}
  \put(3,2){\begin{picture}(4,4)(0,1)
    \multiput(0,1)(1,1){2}{\line(1,-1){2}}
    \multiput(0,1)(2,-2){2}{\line(1,1){1}}
\put(1.5,.5){\pp{}{X^{qp}(-u)}}
\put(.3,1){\pp{}{\downarrow}}
\end{picture}}
\multiput(-.5,0)(0.25,0){30}{\pp{}{.}}
\multiput(-.5,3)(0.25,0){30}{\pp{}{.}}
\multiput(-.5,1)(0.25,0){2}{\pp{}{.}}
\multiput(6.25,1)(0.25,0){2}{\pp{}{.}}
\end{picture}
=\;s_{1}^{p\,q}(u)\,s_{1}^{p\,q}(-u)\:P_{j}^{q+1}P_{j+q}^{p+1}
                \label{eq:InversionRelation}
\end{gather}
where $s^{p\,q}_{i}(u)=\displaystyle
\prod_{j=0}^{p-1}\prod_{k=0}^{q-1}s(u+(i\mi j\plus k)\lambda)$ (we will 
also use the notation $s^{p}_{i}$ for $q=1$)
and the \emph{Abelian Property}
\begin{gather}
\setlength{\unitlength}{17pt}
  X^{pq}_{j}(u+(p\mi 1)\lambda)
    X^{qp}_{j}(v+(q\mi 1)\lambda)
=  X^{pq}_{j}(v+(p\mi 1)\lambda)
    X^{qp}_{j}(u+(q\mi 1)\lambda)
\,.
                \label{eq:Abelian}
                \notag
\end{gather}

The braid limits of the fused face operators are
\begin{equation}
X_j^{pq}(i\epsilon\infty)
:=\lim_{u\rightarrow i\epsilon\infty}\frac{e^{-i\epsilon\frac{(g+1)pq}{2}\lambda}}
{s_0^{1\,q}(u)\ s_{-1}^{p-1\,q}(u)}\ X_j^{pq}(u)\, ,
\qquad\quad \epsilon=\pm1
\end{equation}
It follows that
\begin{equation}
X_j^{pq}(+i\infty)=X_j^{pq}(-i\infty)^*
\label{eq:braidpm}
\end{equation}
and the inversion relation becomes
\begin{equation}
X^{pq}_j(i\epsilon\infty)\,X^{pq}_j(i\epsilon\infty)^{\dag}=P_j^{q+1}\,P_{j+q}^{p+1}
\label{eq:braid-inversion}
\end{equation}
so that these operators are unitary.

The fused face operators, contracted against the fusion vectors, yield the
$(p,q)$-fused Boltzmann weights which depend not only on the spins at
the four corners but also on the bond variables on the edges
\begin{gather}
    \setlength{\unitlength}{.8cm}
   \mbox{}\hspace{-.7in}
    W^{pq}\W{
    d&\gamma& c\\
    \delta&&\beta\\
    a&\alpha&b
    }{|u}=
    \begin{picture}(2,0)(0,.85)
\put(0.5,0.5){\framebox(1,1){$
u
$}}
\put(.5,.52){\pp{bl}{\sss \searrow}}
\put(0.6,0.62){\pp{bl}{\sss pq}}
\put(0.45,0.45){\pp{tr}{a}}
\put(1,.5){\pp{}{\alpha}}
\put(1.55,0.45){\pp{tl}{b}}
\put(1.5,1){\pp{}{\beta}}
\put(1.55,1.55){\pp{bl}{c}}
\put(1,1.5){\pp{}{\gamma}}
\put(0.45,1.55){\pp{br}{d}}
\put(0.5,1){\pp{}{\delta}}
\end{picture}
=
    \frac{1}{s^{p\, q\mi 1}_{0}(u)}
    \setlength{\unitlength}{.7cm}
\vtop to 2\unitlength{}
\begin{picture}(6,1.75)(-2.5,-.25)
\put(0,-1){\framebox(3,2){\p{}{X^{pq}(u)}}}
\put(.1,-.9){\pp{bl}{\searrow}}
\put(-.25,-1.25){\pp{}{a}}
\put(1.5,-1.4){\pp{}{\vec U^{p-1}_{\alpha}(a,b)}}
\put(3.25,-1.25){\pp{}{b}}
\put(3.2,0){\pp{l}{{\vec U^{q-1}_{\beta}(c,b)}^{\dag}}}
\put(3.25,1.25){\pp{}{c}}
\put(1.5,1.4){\pp{}{{{\vec U^{p-1}_{\gamma}(d,c)}}^{\dag}}}
\put(-.25,1.25){\pp{}{d}}
\put(-.2,0){\pp{r}{\vec U^{q-1}_{\delta}(d,a)}}
\end{picture}           \label{eq:Wpq}
\end{gather}

\vspace{-.15in}\noindent
where the function $s^{p\, q\mi1}_{0}(u)$ 
eliminates some scalar factors common to all the spin configurations 
which appear in the process of fusion.
In the $A_{L}$ case the bond variables are redundant.
The fused Boltzmann weights satisfy the reflection symmetry 
\begin{gather}
    W^{pq}\W{
    d&\gamma& c\\
    \delta&&\beta\\
    a&\alpha&b
    }{|u}
=
\frac{s^{q\, p\mi1}_{q\mi p}(u)}{s^{p\, q-1}_{0}(u)}
    \;
    W^{qp}\W{
    d&\delta& a\\
    \gamma&&\alpha\\
    c&\beta&b
    }{|u+(q\mi p)\lambda}
    \label{eq:WSymmetry}
\end{gather}
and
\emph{Crossing Symmetry}
\begin{gather}
    W^{pq}\W{
    d&\gamma& c\\
    \delta&&\beta\\
    a&\alpha&b
    }{|u}
=\sqrt{\frac{\psi_{a}\psi_{c}}{\psi_{b}\psi_{d}}}
     \;
   \frac{s^{q\, p\mi 1}_{0}(\lambda\mi u)}
    {s^{p\, q\mi 1}_{0}(u)}
    \;
    W^{qp}\W{
    a&\delta& d\\
    \alpha&&\gamma\\
    b&\beta&c
    }{|\lambda\mi u}.
    \label{eq:CrossingSymmetry}
\end{gather}
We use these fused Boltzmann weights to construct commuting 
transfer matrices with seams.

\subsection{Integrable seams on the torus}
\label{sec:Seam}
Simple integrable seams are modified faces.  Surprisingly, they produce some new twisted boundary conditions even for the Ising model~\cite{UG1,UG2}. 
They come in four different types,
$r$, $s$, $a$ and $\eta$-type where $r\in A_{g-2}$, $s\in A_{g-1}$, $a\in H$ and $\eta\in\Gamma$.  
Here $H=G$ for Type I theories and is the parent for Type II theories.  
A composite integrable seam $x$ consists of several simple 
seams glued together with four spins $c,d,e,f$ at the corners. The integrable seams 
$x=(r,a,b,\sigma^{\kappa-1})$ give rise to the conformal seams $(r,a,b,\kappa)$ in 
the continuum scaling limit. 

First, let us consider the case $r=1$ equivalent to  the WZW factor in the minimal coset.
The adjacency matrix
of an integrable seam $x=(a,b,\sigma^{\kappa-1})$  is given by
\begin{equation}
\tilde{n}_x=\cases{\hat{N}_a\,\hat{N}_b\,\sigma^{k-1},&\mbox{$G$ of Type I}\cr
n_a^{(G\,H)}\,n_b^{(G\,H)}\,\sigma^{k-1},&\mbox{$G$ of Type II and $x\ne X$}\cr 
n_2^{(G\,H)}(n_6^{(G\,H)}-n_4^{(G\,H)}),&\mbox{$G=E_7$ and $x=X$}}
\end{equation}
where $n_a^{(G\,H)}$ with $a\in H$ are the intertwiners~\cite{BPPZ00} of $G$ relative to $H$ 
\begin{equation}
n_{a\,b}^{(G\,H)}{}^c =
\sum_{m\in \mbox{$\scriptsize\Exp$}(G)} \frac{\psi^{(H)}_a{}^m}{\psi^{(H)}_1{}^m}
\psi^{(G)}_b{}^m \psi^{(G)}_c{}^{m\,*}
\label{intGrelH}
\end{equation}
and $X$ is the special node~\cite{PZ01} of the $E_7$ Ocneanu graph. 
The matrices $(\tilde n_x)_c{}^d$ and $(\tilde n_x)_f{}^e$ encode the allowed adjacency between spins $c,d$ at the bottom and $f,e$ at the top of a composite seam. 
Although our interpretation of these matrices is different,  these matrices coincide exactly with the $\tilde n_x$ matrices of Petkova and Zuber. Our definition, however, is intrinsic to the
seam
so we do not need to invoke boundary conditions on the cylinder.  The matrices $\tilde n_x$ form a (non-faithful) representation
of the \mbox{Ocneanu} graph fusion algebra
\begin{equation}
\tilde{n}_x\tilde{n}_y=\sum_z\tilde{N}_{xy}{}^z\tilde{n}_z
\end{equation}
where the explicit expressions for $\tilde{N}$ are given in \cite{PZ01}.

For the unitary minimal models $(A_{g-2},G)$, the Ocneanu graphs are the quotient of the tensor product graph, $A_{g-2}\otimes
\tilde G/\sim$, where $\sim$ is given by the Kac table symmetry \eqref{eq:KacSymm}. There are $({g-2})|\tilde{G}|/2$  
distinct nodes on this Ocneanu graph  and the corresponding graph structure constants
$\tilde{N}_{(\rho,y)}=\{\tilde{N}_{(\rho',y')\, (\rho,y)}{}^{(\rho'',y'')}\}$
are given by
\begin{equation}
\tilde{N}_{(\rho',y')\, (\rho,y)}{}^{(\rho'',y'')}=Q_{(\rho,y)}{}^{(r,x)}Q_{(\rho',y')}{}^{(r',x')}Q_{(\rho',y')}{}^{(r'',x'')}
N_{r\,r'}^{(A_{g-2})}{}^{r''}\, (\tilde{N}_x)_{x'}{}^{x''}
\end{equation}
 where $(\rho,y)=(\rho,a,b,\kappa)$, 
\begin{equation}
Q_{(\rho,y)}{}^{(r,x)}=
\delta_{\rho,r}\,\delta_{y, x} +
\delta_{\rho,g-1-r}\,\delta_{y, \eta(x)}
\end{equation}
where the indices $(\rho,y)$ range over
\begin{equation}
\left\{
\begin{array}{ll}
 1\leq \rho\leq \frac{g-2}{2},\quad 1\leq y \leq |\tilde{G}|, & \quad\mbox{$g$ even},\\
 1\leq \rho\leq g-2, \quad 1\leq y \leq \frac{g-1}{2}, &\quad\mbox{$G=A_{g-1}$, $g$ odd,}
\end{array} \right.
\end{equation}
and the matrices $Q$ act on each index of the tensor product $ N_{r}^{(A_{g-2})}\otimes\tilde{N}_x$
to quotient out the Kac table symmetry. The matrices 
\begin{equation}
\tilde{n}_{(\rho,y)}=Q_{(\rho,y)}{}^{(r,x)}\,N_{r}^{(A_{g-2})}\otimes\tilde{N}_x
\end{equation}
now form a (non-faithful) representation of the Ocneanu graph fusion algebra for minimal models.

We construct composite seams of type $x=(r,s,\eta)$ and first define $W^{q}_{(r,1,1)}$, the $r$-type seam for the
$(p,q)$-fused model.  It is a usual $(r-1,q)$-fused face (it doesn't
depend on the horizontal fusion level $p$) with an extra parameter
$\xi$ acting as a shift in the spectral parameter, and another choice
for the removal of common scalar factors
\begin{equation}
  W^{q}_{(r,1,1)}\W{d&\gamma&c\\ \delta&&\beta\\ a&\alpha & 
  b}{|u,\xi}=\!\!
  \setlength{\unitlength}{10mm}
\begin{picture}(1.5,0)(.2,.85)
\multiput(0.5,0.5)(1,0){2}{\line(0,1){1}}
\multiput(0.5,0.5)(0,1){2}{\line(1,0){1}}
\put(.5,.52){\pp{bl}{\sss \searrow}}
\put(0.55,0.57){\pp{bl}{\sss q}}
\put(0.45,0.45){\pp{tr}{a}}
\put(1,.5){\pp{}{\alpha}}
\put(1.55,0.45){\pp{tl}{b}}
\put(1.5,1){\pp{}{\beta}}
\put(1.55,1.55){\pp{bl}{c}}
\put(1,1.5){\pp{}{\gamma}}
\put(0.45,1.55){\pp{br}{d}}
\put(0.5,1){\pp{}{\delta}}
\put(1,1){\pp{}{{}^{r}\!(u,\xi)}}
\end{picture}
                = \,
  \frac{s_{0}^{r\mi 1\, q\mi 1}(u+\xi)}{s_{-1}^{r\mi 2\, q}(u+\xi)}
  \; W^{(r\mi 1)\, q}\W{
    d&\gamma& c\\
    \delta&&\beta\\
    a&\alpha&b
    }{|u+\xi}.
    \label{eq:rSeam}
\end{equation}
An $s$-type seam with $s\in A_{g-1}$ is the normalized \emph{braid limit} of an $r$-type
seam, it does not depend on the spectral parameter
\begin{equation}
  W^{q}_{(1,s,1)}\W{d&\gamma&c\\ \delta&&\beta\\ a&\alpha & 
  b}{.}=\!
  \setlength{\unitlength}{8mm}
\begin{picture}(2,0)(0,.85)
\multiput(0.5,0.5)(1,0){2}{\line(0,1){1}}
\multiput(0.5,0.5)(0,1){2}{\line(1,0){1}}
\put(.5,.52){\pp{bl}{\sss \searrow}}
\put(0.55,0.57){\pp{bl}{\sss q}}
\put(0.45,0.45){\pp{tr}{a}}
\put(1,.5){\pp{}{\alpha}}
\put(1.55,0.45){\pp{tl}{b}}
\put(1.5,1){\pp{}{\beta}}
\put(1.55,1.55){\pp{bl}{c}}
\put(1,1.5){\pp{}{\gamma}}
\put(0.45,1.55){\pp{br}{d}}
\put(0.5,1){\pp{}{\delta}}
\put(1,1){\pp{}{(1,s)}}
\end{picture}
                        \! = \;
  \lim_{\xi\to i \infty}\;\frac{e^{-i\frac{(g+1)(s-1)q}{2}\lambda}}
  {s_{0}^{1q}(u+\xi)}\;
  W^{q}_{(s,1)}\W{d&\gamma&c\\ \delta&&\beta\\ a&\alpha & b}{|u,\xi}.
    \label{eq:sSeam}
\end{equation}
In general the $s$-type weights are complex. The complex conjugate gives the weights in the 
other braid limit $\xi\to -i\infty$.
By the reflection and crossing symmetries \eqref{eq:WSymmetry} and \eqref{eq:CrossingSymmetry}
\begin{equation}
  W^{q}_{(1,s,1)}\W{d&\gamma&c\\ \delta&&\beta\\ a&\alpha & b}{.}=
  \sqrt{\frac{\psi_a\psi_c}{\psi_b\psi_d}}\
     W^{q}_{(1,s,1)}\W{a&\alpha & b\\ \delta&&\beta\\d&\gamma&c}{.}^{*}
    \label{eq:ReflectSeam}
\end{equation}
These braid-limit face weights provide us with a representation of the 
braid group.

It is known~\cite{Zuber86,LienartEtAl} that discrete symmetries play an important role in twisted boundary conditions. In fact, there is an integrable seam corresponding to each discrete symmetry. 
Specifically, the graph automorphisms $\eta\in\Gamma$, satisfying
$G_{a,b}=G_{\eta(a),\eta(b)}$, leave the face weights invariant
\begin{equation}
W^{pq}\W{
d&\gamma&c\\ 
\delta&&\beta\\ 
a&\alpha & b}{|u}\;=\;
\setlength{\unitlength}{8mm}
\begin{picture}(2,0)(0,.85)
\multiput(0.5,0.5)(1,0){2}{\line(0,1){1}}
\multiput(0.5,0.5)(0,1){2}{\line(1,0){1}}
\put(.5,.52){\pp{bl}{\sss \searrow}}
\put(0.6,0.62){\pp{bl}{\sss pq}}
\put(0.45,0.45){\pp{tr}{a}}
\put(1,.5){\pp{}{\alpha}}
\put(1.55,0.45){\pp{tl}{b}}
\put(1.5,1){\pp{}{\beta}}
\put(1.55,1.55){\pp{bl}{c}}
\put(1,1.5){\pp{}{\gamma}}
\put(0.45,1.55){\pp{br}{d}}
\put(0.5,1){\pp{}{\delta}}
\put(1,1){\pp{}{u}}
\end{picture}
                        \; = \;
\begin{picture}(2,0)(0,.85)
\multiput(0.5,0.5)(1,0){2}{\line(0,1){1}}
\multiput(0.5,0.5)(0,1){2}{\line(1,0){1}}
\put(.5,.52){\pp{bl}{\sss \searrow}}
\put(0.6,0.62){\pp{bl}{\sss pq}}
\put(0.45,0.45){\pp{tr}{\eta(a)}}
\put(1,.5){\pp{}{\alpha}}
\put(1.55,0.45){\pp{tl}{\eta(b)}}
\put(1.5,1){\pp{}{\beta}}
\put(1.55,1.55){\pp{bl}{\eta(c)}}
\put(1,1.5){\pp{}{\gamma}}
\put(0.45,1.55){\pp{br}{\eta(d)}}
\put(0.5,1){\pp{}{\delta}}
\put(1,1){\pp{}{u}}
\end{picture}
                        \; = \;
W^{pq}\W{
\eta(d)&\gamma&\eta(c)\\
\delta        &&\beta\\
\eta(a)&\alpha&\eta(b)}{|u}
\label{eq:zetacondition}
\end{equation}
and act through the special seam~\cite{CKP}
\begin{equation}
\setlength{\unitlength}{8mm}
      {\ds W^{q}_{(1,1,\eta)}}
      \W{d&c\\ \alpha&\beta\\ a & b}{.}\;=
      \delta_{b\,\eta(a)}\delta_{c\,\eta(d)}
\begin{picture}(2,0)(0,.85)
\multiput(0.5,0.5)(1,0){2}{\line(0,1){1}}
\multiput(0.5,0.5)(0,1){2}{\line(1,0){1}}
\put(0.5,0.52){\pp{bl}{\sss\searrow}}
\put(0.55,0.57){\pp{bl}{\sss q}}
\put(0.45,0.45){\pp{tr}{a}}\put(1.55,0.45){\pp{tl}{\eta(a)}}
\put(1.55,1.55){\pp{bl}{\eta(d)}}\put(0.45,1.55){\pp{br}{d}}
\put(1.5,1){\pp{}{\beta}}
\put(0.5,1){\pp{}{\alpha}}
\put(1,1){\pp{}{\eta}}
\end{picture}
\;=\;
\cases{1, &n_{q+1\,a}{}^{d}\neq 0,\ \alpha=\beta\\ & b=\eta(a),\ c=\eta(d),\cr 
       0, & \text{otherwise.}}
      \label{eq:seamWzeta}
\end{equation}
Notice that the $(r,s,\eta)=(1,1,1)$ seam, where $\eta=1$ denotes
the identity automorphism, is the empty seam
\begin{equation}
   W^{q}_{(1,1,1)}\W{d&c\\ \alpha&\beta\\ a & b}{.}\;=
\;\delta_{ab}\;\delta_{cd}\;\delta_{\alpha\beta}\;n_{q+1\,b}{}^{c}.
      \label{eq:sW111}
\end{equation}
The push-through property is also satisfied for an $\eta$-type seam.

\subsection{Construction of $a$-type seams for Type I theories}
\label{sec:Const-a-type}

In this section we construct a new type of fusion for Type I models labelled by
the nodes $a\in G$. The fused Boltzmann weights giving  the $(1,a)$-seams are obtained in the braid limits $u\to i\epsilon\infty$ with $\epsilon=\pm 1$ and are independent of the spectral parameter $u$. 
This new type of fusion is associated to the $\hat{N}_a$
graph fusion matrices in exactly the same way as the usual fusions are associated to the fused adjacency matrices $n_s$. For Type II models, the construction is applied to the parent Type I graphs in the next section. 
Previously, all known fusions were labelled by Young tableaux. In our case, $r$ and $s$ label $s\ell(2)$
tableaux with $r$ or $s$ boxes in a single row.
The construction of fusions labelled by nodes of graphs is an important step in understanding the graph
fusion algebras associated with integrable and conformal seams.


%
%
In constructing the $a$-type seams in this section we focus on vertically unfused face weights with $q=1$.
Recall that
the $s$-type seams are the braid limit of the $(s-1)$ fused
Boltzmann weights which are obtained by acting with the fusion projector $P_j^{s}$
on the face operators $X_j(u)$.
From \eqref{eq:RankP} we see that the admissible spins at the corners of the
$(s-1)$ fused face weights and the number of horizontal bond variables are given by the non-zero entries
in the fused adjacency matrices $n_s$ with $s\in A_{g-1}$.
From Section~\ref{sec:ADEfusion}, 
these fused adjacency matrices are linear combinations of the graph
fusion matrices
\begin{equation}
    n_{s}=\sum_{a\in G} n_{s\,1}{}^{a}\hat N_{a}
\end{equation}
This motivates us to define a new type of fusion projector
$\hat P_j^{(s)\,a}(b,c)$ associated with $\hat N_a$ and acting on paths between $b$ and $c$ by orthonormally decomposing the $s$-type projector $P_j^s(b,c)$
\begin{equation}
P_j^s(b,c) = \sum_{a\in G} n_{s\,1}{}^{a}\hat P_j^{(s)\,a}(b,c)
\end{equation}
where each of the $\hat{N}_a$ fusion projectors $\hat P_j^{(s)\,a}(b,c)$  separately satisfies the
{\em push-through property \eqref{eq:PushThru} in the braid limit}. 
These projectors $\hat P^{(s)\,a}(b,c)$ which may have complex entries are required to satisfy
\begin{eqnarray}
&&\mbox{1. Decomposition:}\qquad P^s(b,c) =\sum_{a\in G} n_{s\,1}{}^{a}\hat P^{(s)\,a}(b,c)
\label{eq:PhatDecomp}\\
&&\mbox{2. Orthogonality:}\qquad \hat P^{(s)\,a}(b,c)\ \hat P^{(s)\,f}(b,c)=0\, ,\qquad\quad a\neq f\, ,
\label{eq:PhatOrtho}\\
&&\mbox{3. Projection:}\qquad\quad\ \ \hat P^{(s)\,a}(b,c)^2 =\hat P^{(s)\,a}(b,c),\qquad
\hat P^{(s)\,a}(b,c)^\dag = \hat P^{(s)\,a}(b,c)\qquad\qquad
\label{eq:PhatHermit}\\
&&\mbox{4. Adjacency:}\qquad\quad\ \ \mbox{Rank}\big(\hat P^{(s)\,a}(b,c)\big)= \hat N_{a\,b}{}^c\, ,
\qquad\quad n_{s 1}{}^a\neq0.
\label{eq:PhatAdj}
\end{eqnarray}
%
%

By the above conditions  \eqref{eq:PhatDecomp}-\eqref{eq:PhatAdj},
$P^s(b,c)$ and $\{\hat P^{(s)\,a}(b,c)\}_{a\in G}$ are simultaneously diagonalizable
with a common set of eigenvectors. It follows that
$P^{(s)\,a}(b,c)$ can be decomposed as
\begin{equation}
\hat P^{(s)\,a}(b,c)=\sum_{\upsilon=1}^{\hat N_{ab}{}^c}
\vec{\hat U}^{(s)\,a}_{\upsilon}(b,c)\, \vec{\hat U}^{(s)\,a}_{\upsilon}(b,c)^{\dag}
\end{equation}
where $\vec{\hat U}^{(s)\,a}_{\upsilon}(b,c)$ are the simultaneous eigenvectors of
$P^s(b,c)$ and $\{\hat P^{(s)\,a}(b,c)\}_{a\in G}$ satisfying
\begin{equation}
\begin{array}{c}
P^s(b,c)\, \vec{\hat U}^{(s)\,a}_{\upsilon}(b,c)= \hat P^{(s)\,a}(b,c)\, \vec{\hat U}^{(s)\,a}_{\upsilon}(b,c)
=\vec{\hat U}^{(s)\,a}_{\upsilon}(b,c)\, ;\\
\hat P^{(s)\,f}(b,c)\, \vec{\hat U}^{(s)\,a}_{\upsilon}(b,c)=\vec 0,\qquad\quad a\neq f.
\end{array}
\end{equation}
Thus, the construction of the new type of fusion
at $u\rightarrow i\epsilon\infty$, with $\epsilon=\pm1$,
is equivalent to finding the appropriate orthonormal basis for the fusion
vectors $\{\vec{\hat U}^{(s)\,a}_{\epsilon\,\upsilon}(b,c)\}$ of $P^s(b,c)$
%
%
satisfying the conditions
\begin{gather}
    \setlength{\unitlength}{.8cm}
    \setlength{\unitlength}{.7cm}
\vtop to 1\unitlength{}
\begin{picture}(5,2.0)(-0.9,-0.42)
\put(0,-1){\framebox(3.6,1.5){\p{}{\mbox{\small $X^{(s-1)\,1}(i\epsilon\infty)$}}}}
\put(.1,-.9){\pp{bl}{\searrow}}
\put(-.25,-1.25){\pp{}{b}}
\put(1.85,-1.4){\pp{}{\vec {\hat U}^{(s)\,a}_{\epsilon\,\upsilon}(b,c)}}
\put(3.9,-1.25){\pp{}{c}}
\put(3.9,.75){\pp{}{d}}
\put(1.85,1.0){\pp{}{{{\vec {\hat U}^{(s)\,f}_{\epsilon\,\varsigma}(e,d)}}^{\dag}}}
\put(-.25,.75){\pp{}{e}}
\end{picture}
=0
\qquad
\mbox{for all $\upsilon,\varsigma$, $a\neq f$, $\hat N_{ab}{}^c,\ \hat N_{fe}{}^d\neq0$}
\label{eq:UhatOrtho}
\end{gather}
which follows from the push-through property and the orthogonality 
of $\hat P_j^{(s)\,a}$ \eqref{eq:PhatOrtho}.
Note that the $\hat N_a$ fusion vectors at $u\rightarrow i\epsilon\infty$ are
different for $\epsilon=+1$ and $\epsilon=-1$.
By \eqref{eq:braidpm},
$\vec {\hat U}^{(s)\,a}_{\epsilon\,\upsilon}(b,c)^{*}$ is
the $\hat N_a$ fusion vector at $u\rightarrow -i\epsilon\infty$.
Hence, the $\hat{N}_a$ fusion projectors at
$u\rightarrow\pm i\infty$ are complex conjugates
\begin{equation}
\hat P_{u\rightarrow -i\infty}^{(s)\,a}(b,c)= \hat P_{u\rightarrow +i\infty}^{(s)\,a}(b,c)^{*}.
\end{equation}
From now on, unless otherwise stated, the terms $\hat N_a$ fusion projector
and $\hat N_a$ fusion vector refer to the braid limit
$u\rightarrow+i\infty$.
We emphasize that $\hat P_j^{(s)\,a}$ does not satisfy the push-through property if
the face operator $X^{(s-1)1}(u)$ depends on $u$, that is, if we move away from the braid limit. 


The $\hat N_a$ fusion vectors 
can be obtained by solving \eqref{eq:UhatOrtho}.
However, there is a more convenient approach taking advantage of the
unitarity of the face operators in the braid limit
\eqref{eq:braid-inversion}. This provides a unitary transformation
between essential paths, so that the unknown $\hat N_a$
fusion vectors can be obtained from known ones by a unitary
transformation.

The fusion vectors $\vec U^s(e,d)$ with $\hat N_{ae}{}^d=1$ are
automatically the same as the $\hat N_a$ fusion vectors due to the
adjacency condition \eqref{eq:PhatAdj}
\begin{equation}
\vec U^s(e,d)\equiv \vec {\hat U}^{(s)\,a}(e,d),
\qquad\quad \hat N_{ae}{}^d=1
\end{equation}
So suppose that $\vec {\hat U}^{(s)\,a}_{\upsilon}(e,d)$ is known for some given $s,d,e$. 
Then for any $b,c\in G$ with $\hat N_{ab}{}^c\neq0$ satisfying
\begin{equation}
n_{2\,d}{}^c\neq0\qquad\mbox{and}\qquad
n_{2\,e}{}^{b'}\, n_{s\,b'}{}^c=n_{sb}{}^c\, \delta_{b',b}
\label{eq:req-unitary}
\end{equation}
the $\hat N_a$ fusion vector $\vec {\hat U}^{(s)\,a}_{\upsilon}(b,c)$
is given by the unitary transformation
\begin{equation}
\setlength{\unitlength}{0.55cm}
\begin{picture}(16,6)(-10,0)
\put(-12.5,3){
$\vec {\hat U}^{(s)\,a}_{\upsilon}(e,d)\mapsto \vec {\hat U}^{(s)\,a}_{\upsilon}(b,c):\qquad
\vec {\hat U}^{(s)\,a}_{\upsilon}(b,c)=\qquad$
}
\put(3,1.5){
\setlength{\unitlength}{0.55cm}
\begin{picture}(5,6)(0,0)
\put(1.2,0.05){
\pp{}{\bold {\hat U}^{(s)\,a}_{\upsilon}(e,d)}
}
\multiput(0,3)(1,1){2}{\line(1,-1){3}}
\multiput(0,3)(3,-3){2}{\line(1,1){1}}
\put(1.7,2.3){\pp{}{X^{1\,(\!s\!-\!1\!)}}}
\put(2.3,1.6){\pp{}{(-i\infty)}}
\put(3,0.3){\pp{}{\rightarrow}}
\multiput(-0.1,0.4)(0,0.4){8}{\pp{}{.}}
\multiput(4.1,0)(0,0.4){10}{\pp{}{.}}
\put(-0.4,3){\pp{}{e}}
\put(0.9,4.4){\pp{}{b}}
\put(3.3,-0.4){\pp{}{d}}
\put(4.5,1.1){\pp{}{c}}
\end{picture}
}
\end{picture}
\label{eq:Uhat-in-xfm}
\end{equation}
Clearly, it is a fusion vector of $P^s(b,c)$ by the push-through property of the
ordinary projector. It is also an $\hat N_a$ fusion vector because

\begin{equation}
\setlength{\unitlength}{0.4cm}
\begin{picture}(8,10)(3.7,-1)
\put(-1.6,0){
\begin{picture}(8,8)(-0.5,0)
\multiput(0,3)(1,1){2}{\line(1,-1){3}}
\multiput(0,3)(3,-3){2}{\line(1,1){1}}
\put(2.3,1.6){\pp{}{-i\infty}}
\put(3,0.3){\pp{}{\rightarrow}}
\multiput(0,5)(1,-1){2}{\line(1,1){3}}
\multiput(0,5)(3,3){2}{\line(1,-1){1}}
\put(1.4,5.3){\pp{}{+i\infty}}
\put(1,4.3){\pp{}{\rightarrow}}
\multiput(0,3)(0,0.4){6}{.}
\multiput(4,1)(0,0.4){16}{.}
\put(-0.4,3){\pp{}{e}}
\put(0.45,4){\pp{}{b}}
\put(3.4,-0.4){\pp{}{d}}
\put(4.5,1.1){\pp{}{c}}
\put(-0.4,5){\pp{}{e}}
\put(3.4,8.4){\pp{}{d'}}
\put(4.5,7.1){\pp{}{c}}
\put(0.5,0.1){
\pp{}{\bold {\hat U}^{(s)\,a}_{\upsilon}(e,d)}
}
\put(0.5,8){
\pp{}{\bold {\hat U}^{(s)\,f}_{\varsigma}(e,d')^{\dag}}
}
\end{picture}
}
\put(-12.2,3.8){\setlength{\unitlength}{.65cm}
\begin{picture}(5,2.5)(-0.9,-0.42)
\put(0,-1){\framebox(3.6,1.5){\p{}{\mbox{\small $X^{(s-1)\,1}(i\infty)$}}}}
\put(.07,-1){\pp{bl}{\searrow}}
\put(-.25,-1.25){\pp{}{b}}
\put(1.85,-1.4){\pp{}{\vec {\hat U}^{(s)\,a}_{\upsilon}(b,c)}}
\put(3.9,-1.25){\pp{}{c}}
\put(3.9,.75){\pp{}{d'}}
\put(1.85,1.0){\pp{}{{{\vec {\hat U}^{(s)\,f}_{\varsigma}(e,d')}}^{\dag}}}
\put(-.25,.75){\pp{}{e}}
\end{picture}
}
\put(-3,4){$=$}
%
\put(4.5,4){$=$}
\put(5.9,4){$
\vec {\hat U}^{(s)\,f}_{\varsigma}(e,d)^{\dag}\cdot
\vec {\hat U}^{(s)\,a}_{\upsilon}(e,d)\,\delta_{d',d}
$}
\put(18.9,4){$=$}
\put(20.2,4){$
\delta_{a,f}\ \delta_{\upsilon,\varsigma}\ \delta_{d',d}$}
\end{picture}
\end{equation}
by the inversion relation \eqref{eq:braid-inversion}, which is
guaranteed to be valid by \eqref{eq:req-unitary} since, for any $b'$
adjacent to $e$, we have
$\mbox{Rank}\big(P^{(s)\,a}(b',c)\big)=n_{sb}{}^c\delta_{b',b}$.
Thus, the $\hat N_a$ fusion vectors $\vec {\hat
U}^{(s)\,a}_{\upsilon}(b,c)$ are determined
uniquely up to a phase factor and unitary gauge transformations
within blocks given by $\hat N_{ab}{}^c>1$.
To be consistent, one should check that  \eqref{eq:Uhat-in-xfm}
satisfies \eqref{eq:UhatOrtho} for all the other admissible spins.
By the crossing symmetry \eqref{eq:CrossingSymmetry} and
\eqref{eq:braidpm}, \eqref{eq:Uhat-in-xfm} can be
expressed as the linear combination
of ordinary fusion vectors
\begin{equation}
\vec {\hat U}^{(s)\,a}_{\upsilon}(b,c)=
\sum_\gamma
\sqrt{\frac{\psi_d\psi_b}{\psi_e\psi_c}}
\ W_{(1,s,1)}\W{b&\gamma&c\\  e&\upsilon &d}{.}
\vec {U}^{s}_{\gamma}(b,c)\, .
\label{eq:Uhat-lin}
\end{equation}

We now illustrate the calculations of
$\hat N_a$ fusion vectors for the case $G=E_6$.
In $E_6$, $n_s=\hat N_s$ for $1\leq s\leq3$, so it remains to find
the $\hat N_a$ fusion vectors for $\hat N_4$, $\hat N_5$ and 
$\hat N_6$.
First, $n_4$ can be decomposed into $\hat N_4$ and $\hat N_6$
\begin{equation}
n_4=
{\tiny
\left(\tiny \begin{array}{cccccc}
0&0&0&1&0&1\\
0&0&2&0&1&0\\
0&2&0&2&0&1\\
1&0&2&0&0&0\\
0&1&0&0&0&1\\
1&0&1&0&1&0\\
\end{array} \right)
=
\left( \begin{array}{cccccc}
0&0&0&1&0&0\\
0&0&1&0&1&0\\
0&1&0&1&0&1\\
1&0&1&0&0&0\\
0&1&0&0&0&0\\
0&0&1&0&0&0\\
\end{array} \right)
\!+\!
\left( \begin{array}{cccccc}
0&0&0&0&0&1\\
0&0&1&0&0&0\\
0&1&0&1&0&0\\
0&0&1&0&0&0\\
0&0&0&0&0&1\\
1&0&0&0&1&0\\
\end{array} \right)
}=\hat N_4+\hat N_6
\end{equation}
and  only the $\hat N_a$ fusion vectors for $(b,c)=(2,3),(3,2),(3,4),(4,3)$
need to be determined.

Consider $(b,c)=(2,3)$. By  \eqref{eq:req-unitary},
$\vec {\hat U}^{(4)\,4}(1,4)
\mapsto \vec {\hat U}^{(4)\,4}(2,3)$
so it follows from
\eqref{eq:Uhat-lin} that
\begin{equation}
\vec {\hat U}^{(4)\,4}(2,3)=
\sum_\gamma
\sqrt{\frac{\psi_4\psi_2}{\psi_1\psi_3}}
\ W_{(1,4,1)}\W{2&\gamma&3\\  1&1&4}{.}
\vec {U}^{4}_{\gamma}(2,3)
\label{eq:E6-eg-23}
\end{equation}
and similarly $\vec {\hat U}^{(4)\,6}(1,6)\mapsto
\vec {\hat U}^{(4)\,6}(2,3)$.
For $(b,c)=(3,2)$,
$ \{\vec {\hat U}^{(4)\,4}(2,3), \vec {\hat U}^{(4)\,6}(2,3) \}
\mapsto
\{\vec {\hat U}^{(4)\,4}(3,2), \vec {\hat U}^{(4)\,6}(3,2) \} $,
so $\vec {\hat U}^{(4)\,4}(3,2)$ is obtained by  
applying \eqref{eq:Uhat-in-xfm} to \eqref{eq:E6-eg-23}
\begin{equation}
\vec {\hat U}^{(4)\,4}(3,2)=
\sum_\gamma
\frac{\psi_3}{\psi_2}
\ W_{(1,4,1)}\W{3&\gamma&2\\  2&\hat{4}&3}{.}
\vec {U}^{4}_{\gamma}(3,2)
\end{equation}
where
\begin{eqnarray}
\!\!\!
W_{(1,4,1)}\W{3&\gamma&2\\  2&\hat{4}&3}{.}\!\!&=&\!\!
\sum_\alpha
\vec {U}^{4}_{\alpha}(2,3)^{\dag}\cdot\vec {\hat U}^{(4)\,4}(2,3)
\,\ W_{(1,4,1)}\W{3&\gamma&2\\  2&\alpha&3}{.}
\\
{}\!\!&=&\!\!\sum_\alpha
\sqrt{\frac{\psi_4\psi_2}{\psi_1\psi_3}}
\ W_{(1,4,1)}\W{2&\alpha&3\\  1&1&4}{.}\,W_{(1,4,1)}\W{3&\gamma&2\\  2&\alpha&3}{.}
\end{eqnarray}
is the $s$-type seam with the bond variable $\hat 4$ specifically
chosen so that the corresponding fusion vector between $(2,3)$
is $\hat N_4$.
Similarly we obtain $\vec {\hat U}^{(4)\,6}(3,2)$.
For the other cases, we apply the same procedure
\begin{equation}
\begin{array}{c}
\{\vec {\hat U}^{(4)\,4}(6,3), \vec {\hat U}^{(4)\,6}(6,5) \}
\mapsto
\{\vec {\hat U}^{(4)\,4}(3,4), \vec {\hat U}^{(4)\,6}(3,4) \}\\
\{\vec {\hat U}^{(4)\,4}(5,2), \vec {\hat U}^{(4)\,6}(5,6) \}
\mapsto
\{\vec {\hat U}^{(4)\,4}(4,3), \vec {\hat U}^{(4)\,6}(4,3) \}
\end{array}
\end{equation}
Finally, we repeat the whole process for $n_5=\hat N_3+ \hat N_5$ to obtain
the $\hat N_5$ fusion vectors.

In summary, the $\hat N_a$ fusion vectors  for $E_6$ derived from $n_4$ and $n_5$ are listed below
in terms of the given path basis with $\lambda=\pi/12$.\\
\noindent
For $n_4=\hat N_4+\hat N_6$:
%
%
%
\begin{equation}\small
\begin{array}{ll}
\mbox{$(2,3)$ Basis:}&
\{(2,1,2,3),(2,3,2,3),(2,3,4,3),(2,3,6,3)\}\\
{\hat U}^{(4)\,4}(2,3)=&
\frac{\sqrt{3}-1}{2}\left(
(\sqrt{3}+1)^{1/2} e^{-\frac{7\lambda}{2}i} ,
-e^{-\frac{7\lambda}{2}i} ,
e^{\frac{5\lambda}{2}i} ,
-(\sqrt{3}+1)^{1/2} e^{\frac{11\lambda}{2}i} 
\right) \\
{\hat U}^{(4)\,6}(2,3)=&
\frac{\sqrt{3}-1}{2}\left(
\sqrt{2} e^{-\frac{7\lambda}{2}i} ,
-(\sqrt{3}-1)^{1/2} e^{-\frac{7\lambda}{2}i} ,
-(\sqrt{3}+1)^{1/2} e^{\frac{11\lambda}{2}i} ,
\sqrt{2} e^{\frac{7\lambda}{2}i} 
\right)
\end{array}
\end{equation}
%
%
\begin{equation}\small
\begin{array}{ll}
\mbox{$(3,2)$ Basis:}&
\{(3,2,1,2),(3,2,3,2),(3,4,3,2),(3,6,3,2)\}\\
{\hat U}^{(4)\,4}(3,2)=&
\frac{\sqrt{3}-1}{2}\left(
-(\sqrt{3}+1)^{1/2} e^{5\lambda i} ,
e^{5\lambda i} ,
-e^{-\lambda i} ,
(\sqrt{3}+1)^{1/2} e^{-4\lambda i} 
\right) \\
{\hat U}^{(4)\,6}(3,2)=&
\frac{\sqrt{3}-1}{2}\left(
-\sqrt{2} e^{5\lambda i} ,
(\sqrt{3}-1)^{1/2} e^{5\lambda i} ,
(\sqrt{3}+1)^{1/2} e^{-4\lambda i} ,
-\sqrt{2} e^{-2\lambda i}
\right)
\end{array}
\end{equation}
%
%
\begin{equation}\small
\begin{array}{ll}
\mbox{$(3,4)$ Basis:}&
\{(3,2,3,4),(3,4,3,4),(3,6,3,4),(3,4,5,4)\}\\
{\hat U}^{(4)\,4}(3,4)=&
\frac{\sqrt{3}-1}{2}\left(
e^{-\frac{\lambda}{2}i} ,
-e^{\frac{11\lambda}{2}i} ,
-(\sqrt{3}+1)^{1/2} e^{-\frac{7\lambda}{2}i} ,
(\sqrt{3}+1)^{1/2} e^{\frac{11\lambda}{2}i}
\right) \\
{\hat U}^{(4)\,6}(3,4)=&
\frac{\sqrt{3}-1}{2}\left(
-(\sqrt{3}+1)^{1/2} e^{-\frac{11\lambda}{2}i} ,
-(\sqrt{3}-1)^{1/2} e^{\frac{7\lambda}{2}i} ,
\sqrt{2} e^{-\frac{7\lambda}{2}i} ,
\sqrt{2} e^{\frac{7\lambda}{2}i}
\right)
\end{array}
\end{equation}
%
%
\begin{equation}\small
\begin{array}{ll}
\mbox{$(4,3)$ Basis:}&
\{(4,3,2,3),(4,3,4,3),(4,5,4,3),(4,3,6,3)\}\\
{\hat U}^{(4)\,4}(4,3)=&
\frac{\sqrt{3}-1}{2}\left(
e^{\frac{5\lambda}{2}i} ,
-e^{-\frac{7\lambda}{2}i} ,
(\sqrt{3}+1)^{1/2} e^{-\frac{7\lambda}{2}i} ,
-(\sqrt{3}+1)^{1/2} e^{\frac{11\lambda}{2}i}
\right) \\
{\hat U}^{(4)\,6}(4,3)=&
\frac{\sqrt{3}-1}{2}\left(
-(\sqrt{3}+1)^{1/2} e^{\frac{11\lambda}{2}i} ,
-(\sqrt{3}-1)^{1/2} e^{-\frac{7\lambda}{2}i} ,
\sqrt{2} e^{-\frac{7\lambda}{2}i} ,
\sqrt{2} e^{\frac{7\lambda}{2}i}
\right)
\end{array}
\end{equation}
For $n_5=\hat N_3+\hat N_5$:
%
%
\begin{equation}\small
\begin{array}{ll}
\mbox{$(2,4)$ Basis:}&
\{(2,1,2,3,4),(2,3,2,3,4),(2,3,4,3,4),(2,3,4,5,4),(2,3,6,3,4)\}\\
{\hat U}^{(5)\,5}(2,4)=&
\frac{\sqrt{3}-1}{2}\left(
-\sqrt{2} e^{3\lambda i} ,
(\sqrt{3}-1)^{1/2} e^{3\lambda i} ,
(\sqrt{3}-1)^{1/2} e^{-3\lambda i} ,
-\sqrt{2} e^{-3\lambda i} ,
-\sqrt{2} 
\right) \\
{\hat U}^{(5)\,3}(2,4)=&
(\frac{2}{\sqrt{3}}-1)^{1/2}
\left(
-\sqrt{2} e^{3\lambda i} ,
(\sqrt{3}-1)^{1/2} e^{3\lambda i} ,
-(\sqrt{3}-1)^{1/2} e^{-\lambda i} ,
\sqrt{2} e^{-\lambda i} ,
e^{-5\lambda i}  
\right)
\end{array}
\end{equation}
%
%
\begin{equation}\small
\begin{array}{ll}
\mbox{$(3,3)$ Basis:}&
\{(3,2,1,2,3),(3,2,3,2,3),(3,2,3,4,3),(3,2,3,6,3), (3,4,3,2,3),(3,4,3,4,3) \\
{}&\ \,(3,4,3,6,3),(3,6,3,2,3),(3,6,3,4,3),(3,6,3,6,3),(3,4,5,4,3)\}\\[6pt]
{\hat U}^{(5)\,5}(3,3)=&
(\frac{\sqrt{3}-1}{2})^{3/2}\left(
(\sqrt{3}+1)^{1/2} ,
-1, i,
(\sqrt{3}+1)^{1/2} e^{-3\lambda i} , 
-i, 1,
-(\sqrt{3}+1)^{1/2} e^{-3\lambda i} , \right.\\
{}&\left.\qquad\qquad\quad
(\sqrt{3}+1)^{1/2} e^{3\lambda i} ,
-(\sqrt{3}+1)^{1/2} e^{3\lambda i} ,
0, -(\sqrt{3}+1)^{1/2} 
\right) \\[6pt]
%
%
{\hat U}^{(5)\,3}_1(3,3)=&
\frac{1}{2}(1-\frac{1}{\sqrt{3}})^{1/2}
\left(
(\sqrt{3}+1)^{1/2} e^{-2\lambda i},
-e^{-2\lambda i},
e^{-2\lambda i},
0, e^{-2\lambda i},
-e^{-2\lambda i},
0,0,0, \right. \\
{}& \left. \qquad\qquad\qquad\quad 
0,
(\sqrt{3}+1)^{1/2} e^{-2\lambda i}
\right)  \\[6pt]
%
%
{\hat U}^{(5)\,3}_2(3,3)=&
3^{-\frac{1}{4}} (\frac{\sqrt{3}-1}{2})^{3/2}
\left(
(\sqrt{3}+1)^{1/2} e^{-5\lambda i},
-e^{-5\lambda i},
-\sqrt{3} e^{\lambda i},
(2(\sqrt{3}+1))^{1/2} e^{-\lambda i}, \right. \\
{}& \left. \qquad\qquad\qquad\  
\sqrt{3} e^{\lambda i},
e^{-5\lambda i},
-(2(\sqrt{3}+1))^{1/2} e^{-\lambda i},
-(2(\sqrt{3}+1))^{1/2} e^{3\lambda i}, \right. \\
{}& \left. \qquad\qquad\qquad\ 
(2(\sqrt{3}+1))^{1/2} e^{3\lambda i}, 0,
-(\sqrt{3}+1)^{1/2} e^{-5\lambda i} \right) \\
\end{array}
\end{equation}
or any other orthonormal basis spanned by ${\hat U}^{(5)\,3}_1(3,3)$ and 
${\hat U}^{(5)\,3}_2(3,3)$.
%
%
\begin{equation}\small
\begin{array}{ll}
\mbox{$(4,2)$ Basis:}&
\{(4,3,2,1,2),(4,3,2,3,2),(4,3,4,3,2),(4,5,4,3,2),(4,3,6,3,2)\}\\
{\hat U}^{(5)\,5}(4,2)=&
\frac{\sqrt{3}-1}{2}\left(
-\sqrt{2} e^{-3\lambda i} ,
(\sqrt{3}-1)^{1/2} e^{-3\lambda i} ,
(\sqrt{3}-1)^{1/2} e^{3\lambda i} ,
-\sqrt{2} e^{3\lambda i} ,
-\sqrt{2} 
\right) \\
{\hat U}^{(5)\,3}(4,2)=&
(\frac{2}{\sqrt{3}}-1)^{1/2}
\left(
\sqrt{2} e^{-\lambda i} ,
-(\sqrt{3}-1)^{1/2} e^{-\lambda i} ,
(\sqrt{3}-1)^{1/2} e^{3\lambda i} ,
-\sqrt{2} e^{3\lambda i} ,
e^{-5\lambda i}  
\right)
\end{array}
\end{equation}

The $a$-type seam weights, associated with the $\hat N_a$ fusion projectors, are given by
\begin{gather}
    \setlength{\unitlength}{.8cm}
   \mbox{}\hspace{-.7in}
 W_{(1,a,1)}\W{e&\varsigma&d\\  b&\upsilon &
  c}{.}=\!
  \setlength{\unitlength}{8mm}
\begin{picture}(2,0)(0,.85)
\multiput(0.5,0.5)(1,0){2}{\line(0,1){1}}
\multiput(0.5,0.5)(0,1){2}{\line(1,0){1}}
\put(.5,.52){\pp{bl}{\sss \searrow}}
\put(0.45,0.45){\pp{tr}{b}}
\put(1,.4){\pp{}{\upsilon}}
\put(1.55,0.45){\pp{tl}{c}}
\put(1.55,1.55){\pp{bl}{d}}
\put(1,1.6){\pp{}{\varsigma}}
\put(0.45,1.55){\pp{br}{e}}
\put(1,1){\pp{}{(1,a)}}
\end{picture}
  =
    \lim_{u\rightarrow i \infty}\,
       \frac{e^{-i\frac{(g+1)(s-1)}{2}\lambda}}{s(u\!+\!\xi)\,
       s^{s-2}_{-1}(u\!+\!\xi)}
    \setlength{\unitlength}{.9cm}
\vtop to 2\unitlength{}
\begin{picture}(3.6,1.25)(-0.5,-.25)
\put(0,-1){\framebox(3,1.30){\p{}{\sc \mbox{\small $X^{(s-1)1}(u\!+\!\xi)$}}}}
\put(.1,-.9){\pp{bl}{\searrow}}
\put(-.25,-1.25){\pp{}{b}}
\put(1.5,-1.4){\pp{}{\vec {\hat U}^{(s)\,a}_{\upsilon}(b,c)}}
\put(3.25,-1.25){\pp{}{c}}
\put(3.25,.65){\pp{}{d}}
\put(1.5,.75){\pp{}{{{\vec {\hat U}^{(s)\,a}_{\varsigma}(e,d)}}^{\dag}}}
\put(-.25,.65){\pp{}{e}}
\end{picture}           \label{eq:a-seam}
\end{gather}

\vspace{-.25in}\noindent
where
$s$ is chosen so that the corresponding fused adjacency matrix contains
$\hat N_a$ \eqref{eq:hatNn}. 
Note that the choice of $s$ to obtain a particular $a$-type seam is not unique.
In the $E_6$ case, for example, we can obtain the $\hat N_4$ seam from $n_4=\hat N_4+\hat N_6$ or from
$n_6=\hat N_2+\hat N_4$. Remarkably, the weights for such cases agree up to a unitary similarity transformation and give the same spectra for the transfer matrices. 

As an example we give the $a$-type seam weights explicitly for $E_6$: 
\setlength{\unitlength}{.25cm}
\begin{eqnarray}
\hat{N}_4:&\face 2341=\face 3232=\face 3436=\face 4325=1,&\\[12pt]
&\face 2543=\face 3632=e^{-2i\lambda},\qquad \face 2541=\face 4125=e^{-\frac{9}{2}i\lambda}&\\[12pt]
&\face 3236=e^{-i\frac{\lambda}{2}},\quad \face 3434=-e^{i\frac{\lambda}{2}},\quad 
\face 3636=-e^{-\frac{9}{2}i\lambda},\quad \face 4123=-e^{-\frac{3}{2}i\lambda}&\\[12pt]
&\face 4323=\frac{-1+\sqrt{3}}{2}\,e^{3i\lambda},\qquad\qquad \face 2343=\frac{1-\sqrt{3}}{2}\,e^{\frac{5}{2}i\lambda}&
\end{eqnarray}
\begin{eqnarray}
\hat{N}_6:&\face 1632=\face 3232=\face 3456=\face 4365=1,&\\[12pt]
&\face 3432=-e^{\frac{9}{2}i\lambda},\quad \face 3434=-e^{-\frac{3}{2}i\lambda},\quad 
\face 3216=-e^{\frac{3}{2}i\lambda},\quad \face 3234=i&
\\[18pt]
\hat{N}_5:&\face 1542=\face 5124=\face 3366=1,\qquad\qquad
\face 3342=\face 4233=-i&\\[-14pt]\notag
\end{eqnarray}
Note that we can make the weights of the $\hat{N}_5$ and $\hat{N}_6$ seams real by changing
the phase of the fusion vectors. Consequently, the corresponding partition functions are ambichiral since
the transfer matrices contain only real entries
and thus their eigenvalues must occur in complex conjugate pairs.
Note also that the weights of the $\hat{N}_5$ seam
can be transformed to be either 1 or 0.
Thus the $\hat{N}_5$ seam is seen to be 
identical to the automorphism seam \eqref{eq:seamWzeta} which implements
the $\Bbb Z_2$ involution on the $E_6$ graph.

For $D_{2\ell}$ cases, the ordinary fusion projectors are decomposed
according to
\begin{equation}
n_s=\hat N_s\,,\quad 1\leq s\leq 2\ell-2\, , \qquad\mbox{and}\qquad
n_{2\ell-1}=\hat N_{2\ell-1} + \hat N_{2\ell}
\label{eq:D2l_Nhatsplit}
\end{equation}
and the $a$-type seams for $\hat N_{2\ell-1}$
and $\hat N_{2\ell}$ can be obtained by the same process applied to $E_6$.
In the $D_4$ case, the $\hat{N}_3$ and $\hat{N}_4$ $a$-type seams 
\begin{eqnarray}
\hat{N}_3:\qquad&\face 2231=1,\qquad\quad \face 2214=\face 3422=e^{2\pi i/3}&\\[16pt]
\hat{N}_4:\qquad&\face 2241=1,\qquad\quad \face 2213=\face 4322=e^{2\pi i/3}&
\end{eqnarray}
are identical, up to the phase of the fusion vectors, to the $\Bbb Z_3$ automorphism seams
and yield ambichiral partition functions.

\subsection{Construction of $a$-type seams for Type II theories}

So far we have discussed only the $a$-type seam of Type I graphs.
On a Type II graph $G$, the seam cannot be associated with $\hat N_a$ since
these graph fusion matrices contain negative integers which fail to give
a meaningful description in terms of lattice paths. Instead,
for a Type II graph, an
$a$-type seam must be associated with a node $a\in H$ in the parent Type I graph 
rather than to a node $a\in G$. 
For the $A$-$D$-$E$ graphs, only $D_{2\ell+1}$ and $E_7$ are of Type II. For $D_{2\ell+1}$
we do not need $a$-type seams only the $r$- and $s$-type seams. So
here we consider only the case of $G=E_7$ with parent graph $H=D_{10}$. 

The $s$-type seams of $E_7$ constructed via \eqref{eq:sSeam} are labelled by
$s\in A_{17}$.  We decompose these according to \eqref{eq:D2l_Nhatsplit}
\begin{equation}
n_s= n^{(E_7\,D_{10})}_{s}\, ,\quad 1\leq s\leq 8\,, \qquad\mbox{and}\qquad
n_{9}= n^{(E_7\,D_{10})}_{9} + n^{(E_7\,D_{10})}_{10}
\end{equation}
where the intertwiners $\{n^{(E_7\,D_{10})}_{a}\}_{a\in D_{10}}$ of $E_7$ relative to $D_{10}$ (\ref{intGrelH})
form a representation of the $D_{10}$ graph fusion algebra.
Explicitly, we have
\begin{equation}
\begin{array}{lcccc}
n_9=
{\tiny
\left(\tiny \begin{array}{ccccccc}
1&0&1&0&0&0&1\\
0&2&0&2&0&0&0\\
1&0&3&0&2&0&1\\
0&2&0&4&0&2&0\\
0&0&2&0&2&0&2\\
0&0&0&2&0&0&0\\
1&0&1&0&2&0&1\\
\end{array} \right)}
&\!\!=\!\!&
{\tiny
\left( \begin{array}{ccccccc}
0&0&1&0&0&0&0\\
0&1&0&1&0&0&0\\
1&0&1&0&1&0&1\\
0&1&0&2&0&1&0\\
0&0&1&0&1&0&1\\
0&0&0&1&0&0&0\\
0&0&1&0&1&0&0\\
\end{array} \right) }
&\!\!+\!\!&
{\tiny
\left( \begin{array}{ccccccc}
1&0&0&0&0&0&1\\
0&1&0&1&0&0&0\\
0&0&2&0&1&0&0\\
0&1&0&2&0&1&0\\
0&0&1&0&1&0&1\\
0&0&0&1&0&0&0\\
1&0&0&0&1&0&1\\
\end{array} \right)  }  \\
{}&\!\!=\!\!&n^{(E_7\,D_{10})}_9 & \!\!+\!\! & n^{(E_7\,D_{10})}_{10}
\end{array}
\end{equation}

The $n^{(E_7\,D_{10})}_a$ fusion vectors for $E_7$ can be obtained
by unitary transformation. However, as $n_{9}$ is complicated, we need to introduce
another  unitary transformation formula
before we proceed. Since \eqref{eq:Uhat-in-xfm} is unitary, its inverse is also unitary.
By the same argument, 
suppose that $\vec {\hat U}^{(s)\,a}_{\upsilon}(e,d)$ is known for 
some given $s,d,e$. 
Then for any $b,c\in G$ with $\hat 
N_{ab}{}^c\neq0$ satisfying
\begin{equation}
n_{2\,e}{}^b\neq0\qquad\mbox{and}\qquad
n_{s\,b}{}^{c'}\,n_{2\,c'}{}^{d} =n_{s\,b}{}^c\, \delta_{c',c}
%
\label{eq:req-unitary-2}
\end{equation}
the $\hat N_a$ fusion vector $\vec {\hat U}^{(s)\,a}_{\upsilon}(b,c)$
is given by the unitary transformation
\begin{equation}
\setlength{\unitlength}{0.55cm}
\begin{picture}(16,6)(-10,0)
\put(-12.5,3){
$\vec {\hat U}^{(s)\,a}_{\upsilon}(e,d)\mapsto \vec {\hat 
U}^{(s)\,a}_{\upsilon}(b,c):\qquad
\vec {\hat 
U}^{(s)\,a}_{\upsilon}(b,c)=\qquad$
}
\put(3,1.5){
\setlength{\unitlength}{0.53cm}
\begin{picture}(5,6)(0,3.5)
\put(2.4,3.5){
\pp{}{\bold {\hat U}^{(s)\,a}_{\upsilon}(e,d)}
}
\multiput(0,5)(1,-1){2}{\line(1,1){3}}
\multiput(0,5)(3,3){2}{\line(1,-1){1}}
\put(2.3,6.2){\pp{}{X^{(\!s\!-\!1)\,1\!}}}
\put(1.4,5.3){\pp{}{(+i\infty)}}
\put(1,4.3){\pp{}{\rightarrow}}
\multiput(-0.1,4.1)(0,0.4){6}{.}
\multiput(4,4.2)(0,0.4){9}{.}
\put(0.5,3.8){\pp{}{e}}
\put(-0.4,5){\pp{}{b}}
\put(3.3,8.2){\pp{}{c}}
\put(4.5,7.1){\pp{}{d}}
\end{picture}
}
\end{picture}
\label{eq:Uhat-in-xfm2}
\end{equation}
and can be expressed as 
\begin{equation}
\vec {\hat U}^{(s)\,a}_{\upsilon}(b,c)=
\sum_\gamma
\ W_{(1,s,1)}\W{b&\gamma&c\\  e&\upsilon &d}{.}
\vec {U}^{s}_{\gamma}(b,c) \label{eq:Uhat-lin-inverse-1} 
\end{equation}
%
%
%
%
Comparing \eqref{eq:Uhat-lin-inverse-1} with \eqref{eq:Uhat-lin}
we note that it does not contain the crossing factor. 
Depending on the condition \eqref{eq:req-unitary} or \eqref{eq:req-unitary-2},
the unknown $\hat N_a$ fusion vectors can be obtained from known ones by either
\eqref{eq:Uhat-lin} or \eqref{eq:Uhat-lin-inverse-1}.

Note, however, that for $(b,c)=(3,3)$ in $n_{9}$, there does not exist $(e,d)$ which satisfy either
\eqref{eq:req-unitary} or \eqref{eq:req-unitary-2}. Thus, we need to extend the range of the
transformation in order to make it unitary
\begin{equation}
\begin{array}{c}
\{\vec {\hat U}^{(9)\,10}(2,2), \vec {\hat U}^{(9)\,10}(2,4) \}
\stackrel{\eqref{eq:Uhat-lin}}{\mapsto}
\{\vec {\hat U}^{(9)\,10}_1(3,3), \vec {\hat U}^{(9)\,10}_2(3,3) \}\\
\{\vec {\hat U}^{(9)\,9}(2,2), \vec {\hat U}^{(9)\,9}(2,4) \}
\stackrel{\eqref{eq:Uhat-lin}}{\mapsto}
\{\vec {\hat U}^{(9)\,9}(3,3), \vec U(1,3) \}
\end{array}
\end{equation}
and the transformations to $\vec {\hat U}^{(9)\,a}_\upsilon(3,3)$ for both $a=9,10$ must be in the
nullspace of $\vec U(1,3)$. Thus
\begin{equation}
\setlength{\unitlength}{0.55cm}
\begin{picture}(16,6)(0,0)
\put(-6.4,3){
$\vec {\hat 
U}^{(9)\,9}(3,3)=$
}
\put(-2,3){
$ 
\sqrt{\frac{\psi_1\psi_4}{\psi_3(\psi_2+\psi_4)}}
$}
\put(2.1,3){\large
$ ($  }
\put(20.5,3){\large
$ )$  }
\put(5.3,1.5){
\setlength{\unitlength}{0.5cm}
\begin{picture}(5,6)(0,0)
%
%
\put(0,0){
\begin{picture}(16,6)(0,0)
\put(1.1,0){
\pp{}{\bold {\hat U}^{(9)\,9}(2,2)}
}
\put(-3.4,1.5){
$
\sqrt{\frac{\psi_2\psi_3}{\psi_2\psi_1}}
$ }
\multiput(0,3)(1,1){2}{\line(1,-1){3}}
\multiput(0,3)(3,-3){2}{\line(1,1){1}}
\put(1.7,2.3){\pp{}{X^{1\,8}}}
\put(2.3,1.6){\pp{}{(-i\infty)}}
\put(3,0.3){\pp{}{\rightarrow}}
\multiput(-0.1,0.4)(0,0.4){8}{\pp{}{.}}
\multiput(4.1,0)(0,0.4){10}{\pp{}{.}}
\put(-0.4,3){\pp{}{2}}
\put(0.9,4.4){\pp{}{3}}
\put(3.3,-0.4){\pp{}{2}}
\put(4.5,1.1){\pp{}{3}}
\end{picture}
}
\put(5.5,1.5){$-$}
\put(9.5,0){
\begin{picture}(16,6)(0,0)
\put(1.1,0){
\pp{}{\bold {\hat U}^{(9)\,9}(2,4)}
}
\put(-3.4,1.5){
$
\sqrt{\frac{\psi_2\psi_3}{\psi_4\psi_1}}
$ }
\multiput(0,3)(1,1){2}{\line(1,-1){3}}
\multiput(0,3)(3,-3){2}{\line(1,1){1}}
\put(1.7,2.3){\pp{}{X^{1\,8}}}
\put(2.3,1.6){\pp{}{(-i\infty)}}
\put(3,0.3){\pp{}{\rightarrow}}
\multiput(-0.1,0.4)(0,0.4){8}{\pp{}{.}}
\multiput(4.1,0)(0,0.4){10}{\pp{}{.}}
\put(-0.4,3){\pp{}{2}}
\put(0.9,4.4){\pp{}{3}}
\put(3.3,-0.4){\pp{}{4}}
\put(4.5,1.1){\pp{}{3}}
\end{picture}
}
\end{picture}
}
\end{picture}
\label{eq:E7-33}
\end{equation}
where the scalar factors in \eqref{eq:E7-33} cancel the crossing factors
in \eqref{eq:Uhat-lin} when we take the
inverse of $\vec U (1,3)\mapsto \vec {\hat U}^{(9)\,9}(2,2)$
and $\vec U (1,3)\mapsto \vec {\hat U}^{(9)\,9}(2,4)$ respectively.

There are 58 possible spin configurations for non-zero $s$-type seam 
weights $W^{1}_{(1,9,1)}$. 
We constructed all the $a$-type seams for $n^{(E_7\,D_{10})}_9$ 
and $n^{(E_7\,D_{10})}_{10}$ symbolically and confirmed that
\eqref{eq:UhatOrtho} is satisfied for all possible spin configurations
at the four corners of the seam weight
and for all possible bond variables.
Note that, by the quantum symmetry, the partition functions of $x=(a,b)=(9,1)$ and 
$(3,1)$ are complex conjugates. Thus, the spectra of the transfer matrices of the 
respective $n^{(E_7\,D_{10})}_9$ and $n_3$ seams must be complex conjugates and 
this is verified numerically.

\section{Transfer Matrices}
\setcounter{equation}{0}
\label{sec:Transfer}
Given the fusion hierarchy, we build commuting transfer matrices for
different fusion levels and boundary conditions: on the torus and on
the cylinder, with or without seams.

\subsection{Torus transfer matrices}
\label{sec:TorusTransfer}
The transfer matrix for the $(p,q)$-fused \ade lattice model with an 
$(r,s,\eta)$-seam, on a toroidal square lattice is
given, in the basis of the cyclic paths in $N$ steps plus the seam, with bond
variables between adjacent spins, by the product of the corresponding
Boltzmann weights. 
The entries of the transfer matrix with an $(r,s,\eta)$ seam are
given by
\begin{gather}
\rule{0pt}{24pt}\< \vec{a},\vec{\alpha}|\;
\vec{T}_{(r,s,\eta)}^{pq}(u,\xi)\;
       |\vec{b},\vec{\beta}\>=
\setlength{\unitlength}{12mm}
\vtop to .7\unitlength {}
\begin{picture}(7.2,.8)(.4,.9)
     \multiput(0.5,0.5)(0,1){2}{\line(1,0){7}}
\multiput(0.5,0.5)(1,0){2}{\line(0,1){1}}
\multiput(3.5,0.5)(1,0){5}{\line(0,1){1}}
\put(0.5,0.52){\pp{bl}{\sss\searrow}}
\put(0,0){\put(0.5,0.55){\pp{bl}{\sss pq}}}
\multiput(3,0)(1,0){4}{
\put(0.5,0.52){\pp{bl}{\sss\searrow}}
}
\put(2.5,1){\pp{}{\cdots}}
\put(3,0){\put(0.5,0.55){\pp{bl}{\sss pq}}}
\multiput(4,0)(1,0){3}{\put(0.55,0.57){\pp{bl}{\sss q}}}
\put(0.5,0.35){\pp{c}{a_{1}}}
\put(1,0.5){\pp{c}{\alpha_{1}}}
\put(1.5,0.35){\pp{c}{a_{2}}}
\put(3.5,0.35){\pp{c}{a_{N}}}
\put(4,0.5){\pp{c}{\alpha_{N}}}
\put(4.5,0.35){\pp{c}{a_{N\plus1}}}
\put(5,0.5){\pp{c}{\,\alpha_{N\plus1}}}
\put(5.5,0.35){\pp{c}{\,a_{N\plus2}}}
\put(6,0.5){\pp{c}{\alpha_{N\plus2}}}
\put(6.5,0.35){\pp{c}{a_{N\plus3}}}
\put(7.5,0.35){\pp{c}{a_{1}}}
\put(0.5,1.7){\pp{c}{b_{1}}}
\put(1,1.5){\pp{c}{\beta_{1}}}
\put(1.5,1.7){\pp{c}{b_{2}}}
\put(3.5,1.7){\pp{c}{b_{N}}}
\put(4,1.5){\pp{c}{\beta_{N}}}
\put(4.5,1.7){\pp{c}{b_{N\plus1}}}
\put(5,1.5){\pp{c}{\beta_{N\plus1}}}
\put(5.5,1.7){\pp{c}{b_{N\plus2}}}
\put(6,1.5){\pp{c}{\beta_{N\plus2}}}
\put(6.5,1.7){\pp{c}{b_{N\plus3}}}
\put(7.5,1.7){\pp{c}{b_{1}}}
\put(1,1){\pp{}{u}}
\put(4,1){\pp{}{u}}
\put(5,1){\pp{}{{}^{r}\!(u,\xi)}}
\put(6,1){\pp{}{{(1,s)}}}
\put(7,1){\pp{}{\eta}}
\put(2.1,0.8){\p{}{}}\end{picture}
=
       \notag \\[16pt]
    \sum_{\vec\gamma}\prod_{i=1}^{N}W^{pq}\W{b_{i}&\beta_{i}&b_{i+1}\\
    \gamma_{i}&&\gamma_{i\plus 1}\\
    a_{i}&\alpha_{i}&a_{i+1}}{|u}
    W^{q}_{(r,1)}\W{b_{N}&\beta_{N}&b_{N+1}\\
    \gamma_{N}&&\gamma_{N\plus 1}\\
    a_{N}&\alpha_{N}&a_{N+1}}{|u,\xi}\times\qquad\notag\\ \qquad
    W^{q}_{(1,s)}\W{b_{N\plus 1}&\beta_{N\plus 1}&b_{N\plus 2}\\
    \gamma_{N\plus 1}&&\gamma_{N\plus 2}\\
    a_{N\plus 1}&\alpha_{N\plus 1}&a_{N\plus 2}}{.}
    W^{q}_{(1,1,\eta)}\W{b_{N\plus 2}&b_{1}\\
    \gamma_{N\plus 2}&\gamma_{1}\\
    a_{N\plus 2}&a_{1}}{.}
        \label{eq:TpqTorus}
\end{gather}
where the sum is over all allowed vertical bond variables.  The usual
periodic boundary condition is obtained for $(r,s,\eta)=(1,1,1)$. 
The $s$-type seam can be replaced with a single $a$-type seam or a pair of  $a$ and $b$ seams. Recall that the $a$ and $b$-type seams derive from the two braid limits $u\to\pm i\infty$ 
respectively of the $s$-type seams and are related by complex conjugation.  
Indeed the definition can be generalized to accommodate an arbitrary number of
seams.  Because the seam faces, other than the automorhism seams, are 
modified bulk faces they automatically satisfy
the GYBE. They can therefore be moved around freely with respect to each 
other  and the bulk
faces without effecting the spectrum of the transfer matrices.  
However, in the $D_{2\ell}$ cases when there are several
seams, their order in general cannot be interchanged because the automorphism 
seams do not commute with the $a$-type seams.

\subsection{Non-commutativity of seams}

To understand the origin of non-commutativity of seams let us begin by considering $D_4$. 
The graph $D_4$ exhibits an $\Bbb S_3$ symmetry on the external nodes $T=\{1,3,4\}$. This permutation group contains non-commuting two- and three-cycles
\begin{equation}
\Bbb S_3=\{();(1\,3),(1\,4),(3\,4);(1\,3\,4),(1\,4\,3)\}
\end{equation}
In the lattice model seams, these symmetries are realized by the $\Bbb Z_2$ and $\Bbb Z_3$
automorphism seams \eqref{eq:seamWzeta} with $\eta\in\Bbb S_3$. Notice however that
the graph fusion matrix  $\hat{N}_3$ implements the permutation $(1\,3\,4)$ and  
$\hat{N}_4$ implements the permutation $(1\,4\,3)$. Accordingly, on the lattice these automorphisms are implemented by the $a$-type seams with $a=3,4$. This means that the fused seams need only be supplemented by the $\Bbb Z_2$ transposition $\sigma=(3\,4)$ to generate all of $\Bbb S_3$. Moreover, $D_4$ is the only graph with an automorphism group other than $\Bbb Z_2$, so we can always restrict to automorphisms $\eta=\sigma^{\kappa-1}\in \Bbb Z_2$, $\kappa=1,2$.

The fused
seams, together with the bulk face weights, satisfy the generalized Yang-Baxter equation
and thus they commute with each other and can propagate through the bulk face 
weights along the row~\cite{BP01}.  Two transfer matrices differing by a propagation of fused seams differ only by a similarity transformation. Thus the spectrum of the torus transfer
matrices with a given set of regular fused seams does not depend on the order of these
seams nor on their positions. On the other hand, the automorphism seams commute with the bulk face weights and the $r$- and $s$-type fused seams but not, in general, with the $a$-type seams related to $\hat{N}$ fusions. This is manifest in the $D_4$ case since the transposition $(3\,4)$ does not commute with the three-cycles $(1\,3\,4)$ and  $(1\,4\,3)$ implemented by $\hat{N}_3$  and $\hat{N}_4$. 

Consider several such seams on the lattice.  The
non-commutativity of seams shows up by the fact that we get different
resultant seams when placing given seams in different order.
For instance, if $\sigma=(3\,4)$
\begin{equation}
(1\,4\,3)(3\,4)(1\,3\,4)(3\,4)=(1\,3\,4),\qquad (1\,4\,3)(1\,3\,4)(3\,4)(3\,4)=()
\end{equation}
and accordingly the product of four $(r,a,\eta)$ seams
\begin{equation}
W_{(1,{4},1)}W_{(1,1,\sigma)}W_{(1,{3},1)}W_{(1,1,\sigma)}\label{eq:4z3z}
\sim W_{(1,{3},1)}
\end{equation}
gives the same spectrum as the single seam $W_{(1,{3},1)}$ whereas the combination of seams
\begin{equation}
W_{(1,{4},1)}W_{(1,{3},1)}W_{(1,1,\sigma)}W_{(1,1,\sigma)}\label{eq:43zz}
\sim W_{(1,1,1)}
\end{equation}
yields the modular invariant partition function. Notice that 
$W_{(1,{3},1)}W_{(1,1,\sigma)}\sim W_{(1,1,\sigma)}W_{(1,{3},1)}$ gives the same spectra even though $(1\,3\,4)(3\,4)\ne (3\,4)(1\,3\,4)$ since the positions of the two seams can be interchanged by propagating one of them full cycle around the periodic row. Consequently, four or more seams are required to see the effects of non-commutativity in the spectra.

The same phenomenon is exhibited for the whole family $D_{2\ell}$ of Type I
models. The $a$-type seams $(1,2\ell-1,1)$,  $(1,2\ell,1)$ related to $\hat{N}_{2\ell-1}$ and $\hat{N}_{2\ell}$ do not commute with the automorphism seam
$(1,1,\sigma)$ where $\sigma=(2\ell\!-\!1,\,2\ell)$ is the $\Bbb Z_2$ transposition, although their sum does commute as indicated by the relation $\hat{N}_{2\ell-1}\sigma=\sigma\hat{N}_{2\ell}$. 

\subsection{Integrable seams on the cylinder}
\label{sec:BoundaryWeigths}
Although twisted partition functions occur on the torus, it is
striking to see that the Ocneanu algebra still plays a role on the cylinder.  
Indeed, Petkova and Zuber~\cite{PZ2001M} give the minimal conformal partition functions 
on the cylinder with a seam $(r,x)=(r,a,b,\kappa)$ and boundary conditions $(r',a)$ and $(r'',b)$ as
\begin{equation}
Z_{(r',a)|(r,x)|(r'',b)}(q)=\sum_{(r''',s)} (N^{A_{g-2}}_{r'''}\,N^{A_{g-2}}_{r})_{r'}{}^{r''}\,(n_s \tilde
n_x)_a{}^b\,\chi_{r''',s}(q)
\end{equation}
Another remarkable observation~\cite{PZ0011021} 
is that the twisted partition functions on the torus can be written as a bilinear 
combination of cylinder partition functions, summed over some boundary conditions.

In fact, the Ocneanu graph labelling the seams gives a complete
set of boundary conditions, not only on the torus but also on the
cylinder for the continuous conformal field theory as well as for the
integrable statistical mechanics model. In the latter context, a full set of
integrable boundary weights of type $(r,a)$ for Type I theories 
can be obtained by propagating a seam of type $(r,a)$ to the 
boundary and combining it with the simplest 
boundary condition $(r,a)=(1,1)$ called the vacuum. 
Similarly, for Type II theories, the seams $(r,c)$ with $c$ in the parent graph $H$
can be propagated to the boundary to produce an integrable boundary condition 
by combining it with the vacuum. However, in this case, only $|G|$ of the $|H|$ 
values for $c$ produce linearly independent boundary conditions. 
This is in accord with the fact that 
only $|G|$ of the $|H|$ intertwiners $n_c^{(G\,H)}$ of $G$ relative to $H$ are 
linearly independent
\begin{equation}
n_c^{(G\, H)}=\sum_{a\in G} \tilde n_{ca}{}^1\,\hat N_a^{(G)}
\end{equation}
In other words, although $G$-type boundary conditions are applied on the edge of the cylinder, only $H$-type seams propagate into the bulk. 
With this caveat, the algebra of defect lines (or seams) can 
be applied on the cylinder as well. 
 
The vacuum boundary condition corresponds to $(r,a)=(1,1)$. 
The $(1,a)$ boundary weights, for two $q$-adjacent nodes of
$G$, $c$ and $a$ (i.e. $n_{q+1,a}{}^c\neq 0$) are given by
\begin{gather}
    B^{q}_{(1,a)}\B{&&a\\
        &\gamma\\
       c\\
        &\alpha\\
                &&a}{.}=
\setlength{\unitlength}{8mm}
\begin{picture}(1.4,1)(-.25,.9)
\put(0,1){\line(1,-1){1}}
\put(0,1){\line(1,1){1}}
\multiput(1,0)(0,.205){10}{\line(0,1){.15}}
\put(.6,1){\pp{}{(1,a)}}
\put(1.1,2){\pp{l}{a}}
\put(.5,1.5){\pp{}{\gamma}}
\put(0,1){\pp{r}{c\,}}
\put(.5,.5){\pp{}{\alpha}}
\put(1.1,0){\pp{l}{a}}
\end{picture}
= {\frac{\psi_{c}^{\half}}{\psi_{a}^{\half}}} \;
\vec U^{q+1}_{\gamma}(c,a)^{\dag}\vec U^{q+1}_{\alpha}(c,a)=
{\frac{\psi_{c}^{\half}}{\psi_{a}^{\half}}}\;\delta_{\gamma\alpha}\;.
                \label{eq:B1a}
\end{gather}

\smallskip\noindent
In the Type I case, it is obtained by the action of an $a$-seam on
the vacuum boundary condition. The full $(r,a)$ boundary weights are
then given by the action of an $r$-type seam on the $(1,a)$-boundary
weight.  The double row seams are given by two regular $r$-seams sharing
the same extra spectral parameter $\xi$, placed one on top of the other,
with the same spectral parameters as the bulk faces appearing in the
double row transfer matrix (see \eqref{eq:TpqCylinder})
\begin{gather}
    B^{q}_{(r,a)}\B{&&d&\; \delta\\
        &\gamma\\ 
       c\\
        &\alpha\\
                &&b&\; \beta}{|u,\xi}=
\setlength{\unitlength}{9mm}
\begin{picture}(1.8,1.2)(-.25,.9)
\put(1,2){\line(1,0){.5}}
\put(0,1){\line(1,-1){1}}
\put(0,1){\line(1,1){1}}
\put(1,0){\line(1,0){.5}}
\multiput(1.5,0)(0,.205){10}{\line(0,1){.15}}
\put(1,1.2){\pp{}{(r,a)}}
\put(1,.8){\pp{}{(u,\xi)}}
\put(1.5,2){\pp{l}{\,a}}
\put(1.25,2){\pp{}{\delta}}
\put(1,2.1){\pp{b}{d}}
\put(.5,1.5){\pp{}{\gamma}}
\put(0,1){\pp{r}{c\;}}
\put(.5,.5){\pp{}{\alpha}}
\put(1,-.1){\pp{t}{b}}
\put(1.25,0){\pp{}{\beta}}
\put(1.5,0){\pp{l}{\,a}}
\end{picture}
\;= 
\;
\setlength{\unitlength}{9mm}
\begin{picture}(3.4,1)(-2.8,.9)
\put(-2.5,0){
\put(0,0){\framebox(2.5,2){}} 
\put(0,1){\line(1,0){2.5}}
\multiput(0,0)(0,1){2}{
\put(0.,.05){\pp{bl}{\sss\searrow}\pp{bl}{\,\sss q}}
}}
\put(0,1){\line(1,-1){1}}
\put(0,1){\line(1,1){1}}
\multiput(1,0)(0,.205){10}{\line(0,1){.15}}
\multiput(.2,2)(.2,0){4}{\pp{}{.}}
\multiput(.2,0)(.2,0){4}{\pp{}{.}}
\put(-1.25,1.6){\pp{}{{}^{r}\!(\mu\mi u-{\sss (q\mi 1)}\lambda,\, \xi)}}
\put(-1.25,.5){\pp{}{{}^{r}\!(u,\xi)}}
\put(.6,1){\pp{}{(1,a)\,}}
\put(1.1,2){\pp{l}{a}}
\put(-1.25,2){\pp{r}{\delta}}
\put(-2.5,2.1){\pp{b}{d}}
\put(-2.5,1.5){\pp{}{\gamma\,}}
\put(-2.5,1){\pp{r}{c\;\;{}}}
\put(-2.5,.5){\pp{}{\alpha}}
\put(-2.5,-.1){\pp{t}{b}}
\put(-1.25,0){\pp{r}{\beta}}
\put(1.1,0){\pp{l}{a}}
\end{picture}
        \label{eq:Bra}
\end{gather}

\medskip\noindent
and the left boundary weights are simply equal to the right boundary 
weights after applying crossing symmetry.

These boundary weights satisfy boundary analogs of the bulk local relations. 
The Generalized Boundary Yang-Baxter 
Equation or reflection equation is
    \newsavebox{\Bu}
    \newsavebox{\Bv}
\begin{equation}
\setlength{\unitlength}{17pt}
       \savebox{\Bu}(1.5,1){
       \begin{picture}(1.5,1)
    \put(1,1){\line(1,0){.5}}
    \put(0,0){\line(1,-1){1}}
    \put(0,0){\line(1,1){1}}
    \put(1,-1){\line(1,0){.5}}
    \multiput(1.5,-1)(0,.205){10}{\line(0,1){.15}}
    \put(1,.2){\pp{}{(r,a)}}
    \put(1,-.2){\pp{}{(u,\xi)}}
    \end{picture}}
       \savebox{\Bv}(2,1.5){
       \begin{picture}(2,1.5)
    \put(1.5,1,5){\line(1,0){.5}}
    \put(0,0){\line(1,-1){1.5}}
    \put(0,0){\line(1,1){1.5}}
    \put(1.5,-1.5){\line(1,0){.5}}
    \multiput(2,-1.5)(0,.205){15}{\line(0,1){.15}}
    \put(1.25,.2){\pp{}{(r,a)}}
    \put(1.25,-.2){\pp{}{(v,\xi)}}
    \end{picture}}
\vtop to 3\unitlength {}
\begin{picture}(3,2.5)(1,1.2)
    \put(-1,1){
\put(0,1){\line(1,1){1.5}}
\put(0,1){\line(1,-1){1}}
\put(2.5,1.5){\line(-1,1){1}}
\put(1.55,1.55){\pp{}{u-v+}}
\put(1.2,1.2){\pp{}{(q\mi p)\lambda}}
\put(.1,1){\pp{l}{\sss\uparrow}}
\put(.25,.9){\pp{b}{\sss qp}}
    }
    \put(1.4,2.5){\usebox{\Bu}}
    \put(0,0){
\put(0,1){\line(1,1){1.5}}
\put(0,1){\line(1,-1){1}}
\put(1.55,1.55){\pp{}{\mu-u-v}}
\put(1.2,1.2){\pp{}{-(p\mi 1)\lambda}}
\put(.1,1){\pp{l}{\sss\uparrow}}
\put(.25,.9){\pp{b}{\sss qp}}
    }
    \put(.9,0){\usebox{\Bv}}
\multiput(.7,3.5)(.2,0){9}{\pp{}{.}}
\put(3,3.5){\pp{l}{\;a}}
\put(3,1.5){\pp{l}{\;a}}
\put(3,-1.5){\pp{l}{\;a}}
\put(2.75,3.5){\pp{}{\alpha}}
\put(2.5,3.7){\pp{b}{b}}
\put(.5,3.7){\pp{b}{b}}
\put(-.25,2.75){\pp{}{\beta}}
\put(-1,2){\pp{r}{c\,}}
\put(-.5,1.5){\pp{}{\gamma}}
\put(0,1){\pp{r}{d\,}}
\put(.5,.5){\pp{}{\delta}}
\put(1,0){\pp{tr}{e\,}}
\put(1.75,-0.75){\pp{}{\epsilon}}
\put(2.5,-1.5){\pp{tr}{f\,}}
\put(2.75,-1.5){\pp{}{\phi}}
\end{picture}
=\;
\frac{\ss s_{1+q\mi p}^{q\,p}(u\mi v)\,s_{1-(p\mi 1)}^{q\,p}(\mu-u\mi v)
}{\ss s_{1}^{p\,q}(u\mi v)\,s_{1-(q\mi 1)}^{p\,q}(\mu-u\mi v)
}
\begin{picture}(4,2.5)(-.9,4.2)
    \put(.9,5){\usebox{\Bv}}
    \put(0,2.5){
\put(0,1.5){\line(1,1){1}}
\put(0,1.5){\line(1,-1){1.5}}
\put(1.25,1.3){\pp{}{\mu\mi{\sss (q\mi 1)}\lambda}}
\put(1.5,.95){\pp{}{-u-v}}
\put(.1,1.5){\pp{l}{\sss\uparrow}}
\put(.25,1.4){\pp{b}{\sss pq}}
    }
    \put(1.4,2.5){\usebox{\Bu}}
    \put(-1,1.5){
\put(0,1.5){\line(1,1){1}}
\put(0,1.5){\line(1,-1){1.5}}
\put(2.5,1){\line(-1,-1){1}}
\put(1.25,1.25){\pp{}{u\mi v}}
\put(.1,1.5){\pp{l}{\sss\uparrow}}
\put(.25,1.4){\pp{b}{\sss pq}}
    }
\multiput(.7,1.5)(.2,0){9}{\pp{}{.}}
\put(3,6.5){\pp{l}{\;a}}
\put(3,3.5){\pp{l}{\;a}}
\put(3,1.5){\pp{l}{\;a}}
\put(2.75,6.5){\pp{}{\alpha}}
\put(2.5,6.6){\pp{br}{b}}
\put(1.75,5.75){\pp{}{\beta}}
\put(1,5){\pp{br}{c\;}}
\put(.5,4.5){\pp{}{\gamma}}
\put(0,4){\pp{rb}{d\,}}
\put(-.5,3.5){\pp{}{\delta}}
\put(-1,3){\pp{r}{e\,}}
\put(-.25,2.25){\pp{}{\epsilon}}
\put(.5,1.3){\pp{t}{f}}
\put(2.5,1.3){\pp{t}{f\,}}
\put(2.75,1.5){\pp{}{\phi}}
\end{picture}
    \label{eq:BYBE}
\end{equation}
which is proved using the GYBE~\eqref{eq:GYBE} and the abelian 
property~\eqref{eq:Abelian}.
We refer to~\cite{BPO'B96,BP01} for the boundary crossing equation.

The double row transfer matrix is given by two rows similar to the one
appearing in the torus transfer matrix, with spectral parameters $u$
for the bottom one and $\mu-u-(q-1)\lambda$ for the top one, where
$\mu$ is a fixed parameter and $q$ is the vertical fusion level.  The
boundary condition is not cyclic but given by the boundary weights~\eqref{eq:Bra}.

\begin{gather}
        \< \vec{a},\vec{\alpha}|\;
\vec{T}_{
        (r_{L},a_{L})|
        (r,s,\eta)|
        (r_{R},a_{R})
        }^{pq}
        (u,\xi_{L},\xi,\xi_{R})\;
                |\vec{b},\vec{\beta}\>=\notag\\
\setlength{\unitlength}{14mm}
\vtop to 1.7\unitlength {}
\begin{picture}(9.5,1.7)(-.75,1.5)
     \multiput(0.5,0.5)(0,1){3}{\line(1,0){6.5}}
\multiput(0.5,0.5)(1,0){2}{\line(0,1){2}}
\multiput(3,0.5)(1,0){5}{\line(0,1){2}}
\multiput(0,0)(0,1){2}{
\put(0.5,.52){\pp{bl}{\sss\searrow}\pp{bl}{\sss pq}}
\multiput(2.5,0)(1,0){4}{\put(0.5,0.52){\pp{bl}{\sss\searrow}}}
\put(3,.52){\pp{bl}{\sss pq}}
\multiput(4.05,.525)(1,0){3}{\pp{bl}{\sss q}}
\put(2.25,1){\pp{}{\cdots}}
}
\put(-1,0.35){\pp{c}{a_{L}}}
\put(-.75,0.5){\pp{c}{\alpha_{L}}}
\put(0.5,0.35){\pp{c}{a_{1}}}
\put(1,0.5){\pp{c}{\alpha_{1}}}
\put(1.5,0.35){\pp{c}{a_{2}}}
\put(3,0.35){\pp{c}{a_{N}}}
\put(3.5,0.5){\pp{c}{\alpha_{N}}}
\put(4,0.35){\pp{c}{a_{N\plus1}}}
\put(4.5,0.5){\pp{c}{\alpha_{N\plus1}}}
\put(5,0.35){\pp{c}{a_{N\plus2}}}
\put(5.5,0.5){\pp{c}{\alpha_{N\plus2}}}
\put(6,0.35){\pp{c}{a_{N\plus3}}}
\put(7,0.35){\pp{c}{a_{N\plus4}}}
\put(8.25,0.5){\pp{c}{\alpha_{R}}}
\put(8.5,0.35){\pp{c}{a_{R}}}
\put(-1,2.7){\pp{c}{a_{L}}}
\put(-.75,2.5){\pp{c}{\beta_{L}}}
\put(0.5,2.7){\pp{c}{b_{1}}}
\put(1,2.5){\pp{c}{\beta_{1}}}
\put(1.5,2.7){\pp{c}{b_{2}}}
\put(3,2.7){\pp{c}{b_{N}}}
\put(3.5,2.5){\pp{c}{\beta_{N}}}
\put(4,2.7){\pp{c}{b_{N\plus1}}}
\put(4.5,2.5){\pp{c}{\beta_{N\plus1}}}
\put(5,2.7){\pp{c}{b_{N\plus2}}}
\put(5.5,2.5){\pp{c}{\beta_{N\plus2}}}
\put(6,2.7){\pp{c}{b_{N\plus3}}}
\put(7,2.7){\pp{c}{b_{N\plus4}}}
\put(8.25,2.5){\pp{c}{\beta_{R}}}
\put(8.5,2.7){\pp{c}{a_R}}
\put(1,1){\p{}{u}}
\put(3.5,1){\p{}{u}}
\put(1,2.15){\pp{r}{\bigl(\mu\mi u}}
\put(1,1.85){\pp{}{\; -\!{\sss (q\mi 1)}\lambda\bigr)}}
\put(3.5,2.15){\pp{r}{\bigl(\mu\mi u}}
\put(3.5,1.85){\pp{}{\; -\!{\sss (q\mi 1)}\lambda\bigr)}}
\put(4.5,1){\p{}{{}^{r}\! (u,\xi)}}
\put(4.75,2.15){\pp{r}{\bigl(^{\kern-.6em r\kern .5em}\mu - u}}
\put(4.5,1.85){\pp{}{-\!{\sss (q\mi 1)}\lambda,\, \xi\bigr)}}
\multiput(0,0)(0,1){2}{\put(5.5,1){\p{}{{(1,s)}}}
\put(6.5,1){\p{}{{\eta}}}}
\put(-1,2.5){\line(1,0){.5}}
\put(-.5,2.5){\line(1,-1){1}}
\put(-.5,.5){\line(1,1){1}}
\put(-1,.5){\line(1,0){.5}}
\multiput(-1,.5)(0,.205){10}{\line(0,1){.15}}
\put(-.5,1.7){\pp{}{(r_{L},a_{L})}}
\put(-.5,1.3){\pp{}{(\mu\mi u,\xi_{L})}}
\multiput(0,0)(0,2){2}{
        \multiput(0,0)(7.5,0){2}{
\multiput(.4,.5)(-.1,0){9}{\pp{}{.}}
        }}
\put(8.5,2.5){\line(-1,0){.5}}
\put(7,1.5){\line(1,-1){1}}
\put(7,1.5){\line(1,1){1}}
\put(8.5,.5){\line(-1,0){.5}}
\multiput(8.5,.5)(0,.205){10}{\line(0,1){.15}}
\put(8,1.7){\pp{}{(r_{R},a_{R})}}
\put(8,1.3){\pp{}{(u,\xi_{R})}}
\end{picture}
        \label{eq:TpqCylinder}
\end{gather}
The GYBE~\eqref{eq:GYBE} and other local relations imply that double row transfer matrices 
with the same boundary conditions and boundary fields commute
\begin{gather}
\vec{T}_{
        (r_{L},a_{L})|
        (r,s,\eta)|
        (r_{R},a_{R})
        }^{pq}
        (u,\xi_{L},\xi,\xi_{R})\;
\vec{T}_{
        (r_{L},a_{L})|
        (r,s,\eta)|
        (r_{R},a_{R})
        }^{pq'}
        (v,\xi_{L},\xi,\xi_{R})\;
        =\qquad\qquad
  \notag \\
\qquad\qquad
\vec{T}_{
        (r_{L},a_{L})|
        (r,s,\eta)|
        (r_{R},a_{R})
        }^{pq'}
        (v,\xi_{L},\xi,\xi_{R})\;
\vec{T}_{
        (r_{L},a_{L})|
        (r,s,\eta)|
        (r_{R},a_{R})
        }^{pq}
        (u,\xi_{L},\xi,\xi_{R})\;.
         \label{eq:Commutation}
\end{gather}

\section{Finite-Size Corrections and Numerical Spectra} 
\setcounter{equation}{0}
\label{sec:FiniteSize}

\subsection{Finite-size corrections and conformal spectra}
A critical \ade lattice model with a spectral parameter in the range
$0<u<\lambda$ gives rise to a conformal field theory in the continuum scaling
limit, namely, an $\hat{s\ell}(2)$ unitary minimal model.  The properties of
the \ade lattice model connect to the data of this conformal field
theory through the finite-size corrections to the eigenvalues of the
transfer matrices.  

Consider a periodic row transfer
matrix $\vec{T}(u,\xi)$ with a seam $x$ of type $(r,s,\eta)$ or $(r,a,\eta)$ and $N$ faces excluding the seams. If we write the eigenvalues of this transfer matrix  as
\begin{equation}
T_n(u)=\exp(-E_n(u)),\quad n=0,1,2,\ldots
\end{equation}
then the finite-size corrections to the energies $E_n$ take the form
\begin{eqnarray}
&&\hspace{1.7in}\mbox{} E_n(u)=Nf(u)+f_{r}(u,\xi) \label{eq:finiteSizeCorr}\\
&&\mbox{}\hspace{-.3in}\mbox{}
+\frac{2\pi}{N}\,\left(\big(-{\frac c{12}}+\Delta_n+\bar\Delta_n+k_n+\bar k_n\big)\sin\vartheta
+i(\Delta_n-\bar\Delta_n+k_n-\bar k_n)\cos\vartheta\right)
+o\left(\frac{1}{N}\right)\quad\mbox{}\notag
\end{eqnarray}
where $f(u)$ is the bulk free energy, $f_{r}(u,\xi)$ is the boundary
free energy (independent of $s$, $a$ and $\eta$), $c$ is the central
charge, $\Delta_n$ and $\bar\Delta_n$ are the conformal weights,
$ k_n,\bar k_n\in{\Bbb N}$ label descendent levels and
the anisotropy angle $\vartheta$ is given by
\begin{equation}
\vartheta=g u
\end{equation}
where $g$ is the Coxeter number.

On a finite $M\times N$ periodic lattice, the partition
function can be written as
\begin{eqnarray}
Z_x^{M,N}&=&\exp(-MNf(u)-Mf_r(u,\xi))\, Z_x(q) \notag\\
{}&=&\mbox{Tr\,}\vec{T}(u,\xi)^M=\sum_n T_n(u)^M=\sum_{n\ge 0} \exp(-ME_n(u))
\end{eqnarray}
where $Z_x(q)$ is the conformal partition function and $q=\exp(2\pi i \tau)$ is
the modular parameter with $\tau=\frac{M}{N}\exp[i(\pi-\vartheta)]$.
Removing the bulk and boundary contributions to the partition function on a
torus leads to the twisted partition functions 
$Z_x(q)$~\cite{PZ0011021} described in Sections~\ref{sec:OcneanuGraph} and \ref{sec:Twisted}.

\subsection{Bulk and seam free energies} 
\label{sec:SeamFree}

In~\cite{CMP}, we showed that the row transfer matrix with an 
$(r,s,\eta)$-seam satisfies the inversion identity hierarchy
\begin{equation}\label{eq:InversionIdentity}
\vec{T}_0^1\vec{T}_1^1\;=\;s_{-1}^{r-1}s_{1}^{r-1}f_{-1}^1f_{1}^1\,\vec{I}
+s_0^{r-1}f_0^1\,\vec{T}_0^2
\end{equation}
where $\vec{T}_k^q\;=\;\vec{T}_{r,s}^q(u+k\lambda)$ are vertically 
$q$-fused transfer matrices,  
\begin{equation}
f_q^p=\big[s_q^p(u)\big]^N\, ,
\qquad
  s_q^p=\left\{\begin{array}{cl}
1&r=1\, ,\\
s_q^p(u+\xi)&r\geq2
\end{array}\right.
\end{equation}
In the thermodynamic limit, the second term
in \eqref{eq:InversionIdentity} vanishes and the resulting formula
is called an \emph{Inversion Relation}.  This equation can be solved, using 
the structure of zeros and poles, first at order $N$ and then at order $1$ to find the 
bulk and seam free energies as we explain in this section.

We calculate the bulk and  seam free energies, $f(u)$
and  $f_r(u,\xi)$, or equivalently the partition function per
face $\kappa(u)=\exp(-f(u))$ and partition function per length 
$\kappa_r(u,\xi)=\exp(-f_r(u,\xi))$.  Two \ade models
sharing the same Coxeter number are related by intertwiners so their
bulk and seam free energies are the same.  Thus we only need to find the free 
energies for the $A_{L}$
or ABF models~\cite{BaxterBook}.

The bulk free energy $f(u)=-\log\kappa(u)$ for the ABF models was computed by 
Baxter
\begin{equation}
       \kappa (u)=\exp\int_{-\infty}^{+\infty}
        \frac{
        \cosh (\pi-2\lambda)t\;
        \sinh ut\;
        \sinh (\lambda-u)t}
        {t\sinh\pi t\;
        \cosh \lambda t}\;
        dt.
        \label{eq:fLodd}
\end{equation}
This integral has a closed form when $L$ is even
\begin{equation}
       \kappa (u)=
       \frac{
        \sin (u+\lambda)}
        {\sin\lambda}
        \prod_{k=1}^{\frac{L-2}{2}}
        \frac{
        \sin (u+(2k\plus 1)\lambda)}
        {\sin(u+2k\lambda)}.
        \label{eq:fLeven}
\end{equation}
The partition function per face of the $A_L$
model satisfies  the {\em crossing symmetry}
\begin{equation}
    \kappa (u)  =  \kappa (\lambda -u)        \label{eq:crossing}
\end{equation}
and the {\em inversion relation}
\begin{equation}
\kappa (u)\; \kappa
(u+\lambda)=\frac{\sin(u+\lambda)\sin(u-\lambda)}{\sin^2\lambda}.
        \label{eq:bulkinversion}
\end{equation}
This solution is the unique solution of the inversion
relation, crossing symmetry and height reversal symmetry which is
analytic and non-zero in the analyticity strip $\Re\!u\in(0,\lambda)$.

Likewise,  the seam inversion relation for the order one term gives
\begin{eqnarray}
\kappa_r (u)\; \kappa_r
(u+\lambda)&=&\frac{\sin(u+\xi+\lambda)\sin(u+\xi-(r\mi1)\lambda)}{\sin^2\lambda}.
        \label{eq:BoundaryInversion}
\end{eqnarray}
The range of validity for the parameter $\xi$ is
\begin{equation}
   -\lambda-\frac{\pi}{2}<\Re\!(u+\xi)<-\frac{\lambda}{2}
    \label{eq:xiStrip}
\end{equation}

 Let
$q$ be the RHS of (\ref{eq:BoundaryInversion}).  It
is Analytic and Non-Zero in the strip $\Re\!u\in(0,\lambda)$. 
Furthermore the derivative $q'$ approaches
a Constant when $\Im\!u\to\pm\infty$ (ANZC).  Hence we can introduce
the Fourier transforms of the logarithmic derivatives
\begin{eqnarray}
\mathcal{F}(k)&:=&{\frac 1{2 
\pi i}}\int\limits_{\hidewidth 0<\Re\!u<\lambda \hidewidth}f'_r(u)\,e^{-ku}du \\
{\frac d{du}}f_r(u)&=&
\int\limits_{-\infty}^{+\infty}\mathcal{F}(k)e^{ku}dk
\end{eqnarray}
so that (\ref{eq:BoundaryInversion}) becomes
\begin{equation}
\mathcal{F}(k)(1+e^{k\lambda})=
{\frac 1{2 i\pi}}
\int\limits_{\hidewidth 0<\Re(u)<\lambda \hidewidth}\frac{q'(u)}{q(u)}e^{-ku}du
      \label{eq:Ffebound}
\end{equation}
The solution by inverse Fourier transforms gives
\begin{eqnarray}
{\frac d{du}}f_r(u)&=&{\frac 1{2 i\lambda}}
\int\limits_{\hidewidth0<\Re\!w<\lambda\hidewidth}d
w\left({\frac d{du}}\;
\log q(u-w)\right)\frac{1}{\sin { (L+1)}w}
      \label{eq:soldbound}
\end{eqnarray}
Integrating with respect to $u$ and taking the $w$ integration along
the vertical line $\Re\!w=\epsilon>0$ we obtain in the limit
$\epsilon\to 0$
\begin{eqnarray}
f_r(u)
&=&\frac{\log q(u)}{2}
+\frac{1}{\lambda}\int_{-\infty}^\infty
\frac {\log q(u-iw)}{\sinh { (L+1)}w}\,d w
      \label{eq:solbound}
\end{eqnarray}

This integral formula admits a closed form under certain conditions
\begin{equation}
\kappa_r(u)   =
\cases{
\displaystyle {(u+\xi)\over \tan {\ts\frac{L+1}{2}}(u+\xi)}  \prod_{k=1}^{\frac{r\mi 1}{2}}
 \frac{(u+\xi-2k\lambda)}{(u+\xi-(2k\mi1)\lambda)}, & \text{$r$ odd}\\[12pt]
\displaystyle{(u+\xi+\lambda)\over \tan {\ts\frac{L+1}{2}}(u+\xi+\lambda)}
\prod_{k=1}^{\frac{L\mi r}{2}}
\frac{(u+\xi+(2k\plus1)\lambda)}{(u+\xi+2k\lambda)} , &\text{$L+r$ even.}}
\end{equation}
The tangent parts are solutions of the homogeneous functional equation 
and fix up the zeros and poles of the sine parts which 
are solutions of the functional equation~(\ref{eq:BoundaryInversion}) 
but have an unwanted zero in the analyticity strip.

The seam free energy for $s$-type seams are given by the braid limit of the $r$-type seams and are constants. The seam free energies of the $a$-type seams are the same as the $s$-type seam from which they originate. We remove these seam free energies by the normalization of the transfer matrices. Lastly, the seam free energy of an $\eta$ seam is zero.

\subsection{Numerical determination of conformal spectra}
\label{sec:Conf}

The twisted conformal partition functions are obtained numerically from finite-size spectra.
Since integrable lattice realizations of the $s\ell(2)$ \ade Wess-Zumino-Witten conformal field theories are not known we can only obtain numerically the twisted partition functions of the unitary minimal \ade models labelled by a pair of graphs $(A,G)$.
We use Mathematica~\cite{Wolfram} to construct and diagonalize numerically the finite-size transfer matrices $\vec{T}(u,\xi)$ with specified seams for
different numbers of faces $N$. For simplicity, we restrict ourselves to the isotropic conformal point given by $u=\frac\lambda2$ and $\xi=-3\lambda/2$ . For the first ten or so eigenvalues, we extrapolate the conformal corrections to $N=\infty$ using a combination of polynomial fits in the
inverse number of faces and van den Broeck-Schwartz~\cite{vBS} sequence extrapolation. 
The resulting sequences give approximations to the rational exponents that appear in the
$q$-expansion of the twisted partition functions in increasing powers of the
modular parameter $q$.

In subsequent subsections, we analyse the numerical data for the
$D_4, D_5, D_6$ cases of the \ade lattice models. The $A_L$ cases for $L=3,4,5,6,7,9$ were reported in \cite{CMOP2}. All the numerical results confirm the quantum symmetries and twisted partition functions stated in Sections~\ref{sec:OcneanuGraph} and \ref{sec:Twisted}.
Given the coincidence of the construction labels of our integrable seams and the conformal labels of Petkova and Zuber, we expect that our list of integrable seams will also exhaust the twisted conformal boundary conditions for the exceptional $E_6$, $E_7$ and $E_8$ cases as well. These cases, however, are too large to convincingly confirm numerically.

Consideration of a seam of type $(r,a)$ gives access only to a chiral half of the Ocneanu graph. In general, to obtain the complete Ocneanu graph, one needs to consider the composition of two seams $(r,a,\eta)$ and $(r',b,\eta')$. Since the \ade models are labelled by pairs $(A,G)$, with the first member always for type-$A$, it is sufficient to take $r'=r$. Also, as we have seen, we can restrict the automorphisms to a $\Bbb Z_2$ subgroup of the full automorphism group given by $\eta=\sigma^{\kappa-1}$. Indeed, the only graph $G$ with an automorphism group larger than $\Bbb Z_2$ is $D_4$ and, in this case, the three-cycles of $\Bbb S_3$ are reproduced within the $\hat{N}$ graph fusions. We conclude that it suffices to consider integrable seams of the form $x=(r,a,b,\sigma^{\kappa-1})$.

For simplicity, we take the absolute values of the eigenvalues which is equivalent to taking the modular parameter $q$ real
\begin{equation}
|T_n^{(N)}|=e^{-Nf-f_r+\frac{2\pi}{N}\frac{c}{12}}\, \exp\left(-\frac{2\pi}{N}\, x_n +\mbox{o}\left(
\frac{1}{N}\right)\right)\label{absFS}
\end{equation}
We thus numerically estimate the conformal dimensions or exponents
\begin{equation}
x_n=\Delta+\bar\Delta+k_n+\bar k_n
\end{equation}
and ignore the spins $s_n=\Delta-\bar\Delta+k_n-\bar k_n$.
Since we are at the isotropic point the geometric factor $\sin\vartheta=1$. 
To obtain the bulk free energy, we extrapolate the sequence $-\frac{1}{N}\log
|T_n^{(N)}|$. 
The seam free energy is obtained, in a similar way, after removing the bulk contribution in 
\eqref{absFS}.
An $s$- or $a$-type seam contributes a constant to the seam free energy which is
removed by our choice of normalization.
The extrapolated numerical values for the bulk and seam free energies
agree with the analytic results \eqref{eq:fLodd}
and \eqref{eq:solbound} within an accuracy of $\pm0.3\%$.

To estimate the exponents $x_n$, we extrapolate the sequences 
\begin{equation}
x_n^{(N)} = -\frac{N}{2\pi}\left( \log|T_n^{(N)}|+Nf+f_r-\frac{2\pi}{N}\frac{c}{12}
\right)
\end{equation}
and compare values and degeneracies with $q$-series of the twisted partition functions
$Z_x(q)$ \cite{PZ0011021}.

The accuracy of our numerical results is restricted by the data for different system sizes $N$ which, in turn, is limited by available computer memory. A typical maximum size matrix that we can
construct and diagonalize is around $4500\times 4500$. The dimension of a transfer matrix
with $N$ faces and an $(r,a)$ seam is $\mbox{Tr\,}(n_2^N n_r \hat{N}_a)$.
This grows rapidly as the number of nodes in $G$ increases and grows exponentially with $N$.  
In practice, this means we are typically restricted to $|G|\le 6$ and to system sizes $N\le 12$. Furthermore, because of parity constraints, we are either restricted to odd or even system sizes $N$ so we are typically extrapolating sequences of length six. Nevertheless, because of the quantized values of the conformal weights, the integer spacing of conformal towers and recognizable degeneracies of the characters, we are able to identify the various twisted partition functions with considerable confidence.

\subsection{Numerical spectra of $(A_4,D_4)$}

\begin{figure}[htbp]
\setlength{\unitlength}{6.05mm}
\begin{picture}(8,14)
\put(0,0){
\begin{picture}(8,14)
\put(2,0){
\includegraphics[width=.8\linewidth]{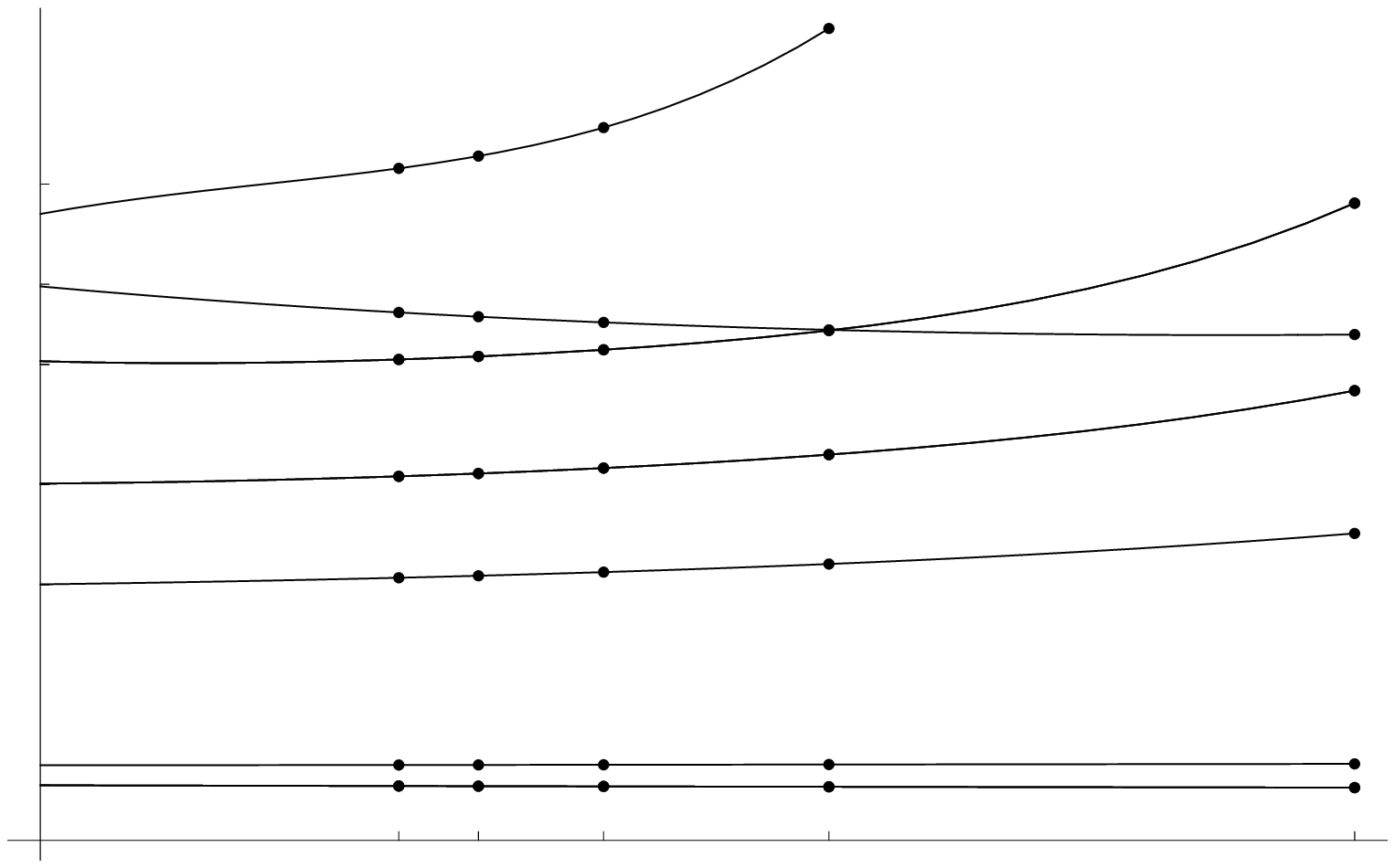}
}
\put(2.2,0.8){\small\pp{c}{\frac{11}{120}}}
\put(2.2,1.7){\small\pp{c}{\frac{1}{8}}}
\put(2.2,4.3){\small\pp{c}{\frac{17}{40}}}
\put(2.2,6){\small\pp{c}{\frac{71}{120}}}
\put(2.2,7.8){\small\pp{c}{\frac{19}{24}}}
\put(2.2,9.1){\small\pp{c}{\frac{37}{40}}}
\put(2.2,10.8){\small\pp{c}{\frac{131}{120}}}
\put(8.5,-0.4){\small\pp{c}{\frac{1}{11}}}
\put(9.8,-0.4){\small\pp{c}{\frac{1}{9}}}
\put(11.8,-0.4){\small\pp{c}{\frac{1}{7}}}
\put(15.3,-0.4){\small\pp{c}{\frac{1}{5}}}
\put(23.5,-0.4){\small\pp{c}{\frac{1}{3}}}
\end{picture} }
\end{picture}
\vspace{0.4cm}
     \caption{Energy levels of the minimal $(A_4,D_4)$ model with $(1,2,1)$ seam. The numerical values of $x_n^{(N)}$ are plotted against $1/N$. The polynomial fits intersect the $y$-axis at the
extrapolated values $x_n$. Notice that eigenvalues can cross for small values of $N$.}
\label{D4-1-2-1}
\end{figure}

The $(A_4,D_4)$ lattice model corresponds to the critical 3-state Potts model and is special in that the $D_4$ graph admits an $\Bbb S_3$ automorphism group.

However, we only need seams of the form $(r,a,\kappa)\equiv(r,a,\sigma^{\kappa-1})$ with $\kappa=1,2$ and
$\sigma=(3\,4)$ the $\Bbb Z_2$ transposition represented by
\begin{equation}
\sigma=
{\tiny
\left(
\begin{array}{cccc}
1&0&0&0\\
0&1&0&0\\
0&0&0&1\\
0&0&1&0\\
\end{array}
\right)   }
\end{equation}

To illustrate our numerical procedure consider the twisted partition function $Z_{(1,2,1)}(q)$ 
with seam $(r,a,\kappa)=(1,2,1)$. The $q$-series at the isotropic conformal point with real $q$ is 
\begin{eqnarray}
\!\!\!\!Z_{(1,2,1)}(q)\!\!\!&=&\!\!\!
\hat\chi_{1,2}^{ }(\hat\chi_{1,1}^{ }+2\hat\chi_{1,3}^{ })
+\hat\chi_{3,2}^{ }(\hat\chi_{3,1}^{ }+2\hat\chi_{3,3}^{ }) \notag\\
{}&=& \!\!\!(\chi_{1,2}^{ }+\chi_{1,4}^{ })(\chi_{1,1}^{ }
+\chi_{1,5}^{ }+2\chi_{1,3}^{ }) 
+(\chi_{3,2}^{ }+\chi_{3,4}^{ })
(\chi_{3,1}^{ }+\chi_{3,5}^{ }+2\chi_{3,3}^{ }) \\
{}&=&\!\!\! q^{-1/15}(
2q^{11/120}\!+\!q^{1/8}\!+\!q^{17/40}\!+\!2q^{71/120}\!+\!2q^{19/24}
\!+\!q^{37/40}
+4q^{131/120}\!+\!o(q^{{131}/{120}})
)\notag
\end{eqnarray}
so that the exponents, counting degeneracies, are given by the sequence
\begin{equation}
\{x_n\}=\mbox{\small 
$\{\frac{11}{120},\frac{11}{120},\frac{1}{8},\frac{17}{40},\frac{71}{120},\frac{71}{120},
\frac{19}{24},\frac{19}{24},\frac{37}{40},\frac{131}{120},\frac{131}{120},\ldots\}$}
\label{theorexps}
\end{equation}

A plot of our numerical data is shown in Fig.~\ref{D4-1-2-1}. 
Notice that some eigenvalues cross so that it is not possible to simply order the eigenvalues according to their magnitudes at a given value of $N$ and this complicates the extrapolation procedure. Although we have not systematically done so, it is possible to remove this ambiguity in the identification of each eigenvalue at a given $N$ by examining the patterns of the zeros of the eigenvalue  $T_n^{(N)}(u)$ in the complex $u$-plane as explained in \cite{KP91}. 
The theoretical data (\ref{theorexps}) is to be compared with our numerical values of $\{x_n\}$ for the $(1,2,1)$ seam shown in Table~\ref{D4numericdata}.
The agreement over the first 10 levels is certainly good enough to unequivocally identify the $(1,2,1)$ integrable seam as giving rise to the $(1,2,1)$ twisted partition function.

Typical numerical data for two other seams is also shown in Table~\ref{D4numericdata}. In this way we can identify all of the integrable seams with the corresponding twisted boundary conditions in Table~\ref{tab:A4D4} given by Petkova and Zuber~\cite{PZ0011021}. In particular, we confirm that our integrable seams give a realization of the complete set of conformal twisted boundary conditions for the 3-state Potts model.

\begin{table}[htb]
{\small
\begin{equation}
\begin{array}{|c|cccccccccc|}\hline
\multicolumn{11}{|c|}{(1,2,1)}\cr\hline
n& 0 & 1 & 2 & 3 & 4 & 5 & 6 & 7 & 8 & 9 \cr\hline
\mbox{Exact}& \frac{11}{120} & \frac{11}{120} & \frac{1}{8} & \frac{17}{40} & \frac{71}
   {120} & \frac{71}{120} & \frac{19}{24} & \frac{19}{24} & \frac{37}{40} & \frac{131}
   {120} \cr\hline\mbox{Num.}& 0.0915 & 0.0915 & 0.1250 & 0.4258 & 0.5934 & 0.5934 & 0.7969 & 
   0.7969 & 0.9215 & 1.0421 \cr\hline|\mbox{diff.}|& 10^{-4} & 10^{-4} & 3\!\cdot\!10^{-5} & 
   8\!\cdot\!10^{-4} & 0.0018 & 0.0018 & 0.0053 & 0.0053 & 
 0.0035 & 0.0496\cr\hline\hline
\multicolumn{11}{|c|}{(1,3,1)}\cr\hline
\mbox{Exact}& \frac{2}{15} & \frac{7}{15} & \frac{7}{15} & \frac{2}{3} & \frac{2}{3} & 
    \frac{17}{15} & \frac{17}{15} & \frac{4}{3} & \frac{22}{15} & \frac{22}
   {15} \cr\hline\mbox{Num.}& 0.1335 & 0.4658 & 0.4658 & 0.6681 & 0.6681 & 1.1640 & 1.1640 & 
   1.3364 & 1.4445 & 1.4445 \cr\hline|\mbox{diff.}| & 10^{-4} & 8\!\cdot\!10^{-4} & 8\!\cdot\!10^{-4} & 
   0.0014 & 0.0014 & 0.0306 & 0.0306 & 0.0030 & 0.0222 & 
    0.0222 \cr\hline\hline
\multicolumn{11}{|c|}{(3,1,2)}\cr\hline
\mbox{Exact}& \frac{1}{20} & \frac{3}{20} & \frac{3}{20} & \frac{11}
   {20} & \frac{11}{20} & \frac{13}{20} & \frac{13}{20} & \frac{21}
   {20} & \frac{21}{20} & \frac{21}{20} \cr\hline\mbox{Num.}& 0.0492 & 0.1501 & 
   0.1501 & 0.5479 & 0.5479 & 0.6503 & 0.6503 & 1.0577 & 1.0577 & 
   1.0629 \cr\hline|\mbox{diff.}| & 8\!\cdot\!10^{-4} & 10^{-4} &10^{-4} & 0.0021 & 
   0.0021 & 3\!\cdot\!10^{-4} & 3\!\cdot\!10^{-4} & 0.0077 & 
   0.0077 & 0.0129 \cr \hline
\end{array}\nonumber
\end{equation}}
\caption{Numerical exponents $x_n$ for the $(A_4,D_4)$ minimal model with seams
of type $(r,a,\kappa)=(1,2,1)$, $(1,3,1)$ and $(3,1,2)$.}
\label{D4numericdata}
\end{table}

\subsection{Numerical spectra of $(A_6,D_5)$}

The numerical spectra of the minimal $(A_6,D_5)$ model can be obtained similarly to the $(A_4,D_4)$ model. 
The graph $D_5$, however, is a Type II graph with parent graph $A_7$. The integrable seams are thus labelled by $(r,s)\in(A_6,A_7)$
where the $s$-type fusion seams are subject to the fused adjacency matrices $n_s$ of $D_5$. 
Note that, although $n_2$=$n_6$, the eigenvalue spectra associated with these seams off the isotropic point are complex conjugates
of each other in
agreement with the quantum symmetry of the Oceneau graph. 
Also, note that $n_7$ and the corresponding seam $(1,7,1)=(1,1,2)$ implement the $\Bbb Z_2$ involution. Thus,
the labels
$(r,s)$ suffice to label all the distinct nodes on the Oceneau graph for $D_5$. In fact, this is a general feature for $D_{2\ell+1}$
for which the $\Bbb Z_2$ involution is implemented by $n_{4\ell-1}$.
Our data for seams of type $(r,a,\kappa)=(1,2,1)$, $(3,3,1)$ and $(1,1,2)$ are shown in Table~\ref{D5Table}.

\begin{table}[htbp]
{\small
\begin{equation}
\begin{array}{|c|cccccccccc|}\hline
\multicolumn{11}{|c|}{(1,2,1)}\cr\hline
n& 0 & 1 & 2 & 3 & 4 & 5 & 6 & 7 & 8 & 9 \cr\hline
\mbox{Exact}&  \frac{23}{224} & \frac{23}{224} & \frac{27}{224} & \frac{5}
   {32} & \frac{83}{224} & \frac{107}{224} & \frac{135}{224} & 
    \frac{135}{224} & \frac{167}{224} & \frac{167}
   {224} \cr\hline\mbox{Num.}&  0.102 & 0.102 & 0.1201 & 0.1562 & 0.3728 & 0.481 & 
   0.6074 & 0.6074 & 0.7527 & 0.7527 \cr\hline|\mbox{diff.}|& 7\!\cdot\!10^{-4} & 
   7\!\cdot\!10^{-4}  & 5\!\cdot\!10^{-4}  & 5\!\cdot\!10^{-5}  & 0.0023 & 
   0.0033 & 0.0047 & 0.0047 & 0.0071 & 
   0.00714\cr\hline\hline
\multicolumn{11}{|c|}{(3,3,1)}\cr\hline
\mbox{Exact}& \frac{1}{28} & \frac{1}{28} & \frac{1}{14} & \frac{9}
   {112} & \frac{9}{112} & \frac{15}{112} & \frac{1}{7} & \frac{1}
   {7} & \frac{3}{14} & \frac{25}{112} \cr\hline\mbox{Num.}& 0.036 & 0.036 & 
   0.0722 & 0.0813 & 0.0813 & 0.1354 & 0.1459 & 0.1459 & 0.215 & 
   0.2275 \cr\hline|\mbox{diff.}|& 2\!\cdot\!10^{-4} & 2\!\cdot\!10^{-4} & 7\!\cdot\!10^{-4} & 
   9\!\cdot\!10^{-4} & 9\!\cdot\!10^{-4} & 0.0015 & 0.0031 & 
   0.0031 & 0.0007 & 0.0043 \cr\hline\hline
\multicolumn{11}{|c|}{(1,1,2)}\cr\hline
\mbox{Exact}& \frac{3}{112} & \frac{15}{112} & \frac{5}{16} & \frac{4}
   {7} & \frac{4}{7} & \frac{99}{112} & \frac{115}{112} & \frac{115}
   {112} & \frac{127}{112} & \frac{127}{112} \cr\hline\mbox{Num.}& 0.027 & 0.1347 & 
   0.3126 & 0.569 & 0.569 & 0.8785 & 1.0310 & 1.0310 & 1.1673 & 
   1.1673 \cr\hline|\mbox{diff.}|&  2\!\cdot\!10^{-4} & 8\!\cdot\!10^{-4} & 1\!\cdot\!10^{-4} & 
    0.0026 & 0.0026 & 0.0054 & 0.0042 & 
    0.0042 & 0.033 & 0.033 \cr \hline
\end{array}\nonumber
\end{equation}}
\caption{Numerical exponents $x_n$ of the $(A_6,D_5)$ minimal model with seams of type $(r,a,\kappa)=(1,2,1)$, $(3,3,1)$ and $(1,1,2)$.}
\label{D5Table}
\end{table}

\renewcommand{\textfraction}{0}
\begin{table}[htbp]
{\small
\begin{equation}
\begin{array}{|c|cccccccccc|}\hline
\multicolumn{11}{|c|}{(1,5,1)}\cr\hline
n& 0 & 1 & 2 & 3 & 4 & 5 & 6 & 7 & 8 & 9 \cr\hline
\mbox{Exact}&  \frac{2}{45} & \frac{2}{15} & \frac{4}{15} & \frac{8}
   {15} & \frac{8}{15} & \frac{29}{45} & \frac{29}{45} & \frac{14}
   {15} & \frac{47}{45} & \frac{47}{45} \cr\hline\mbox{Num.}&  0.0447 & 0.1345 & 
   0.2688 & 0.5267 & 0.5267 & 0.6438 & 0.6438 & 0.9625 & 1.0256 & 
   1.0256 \cr\hline|\mbox{diff.}|& 2\!\cdot\!10^{-4} & 0.0011 & 0.0021 & 
    0.0066 & 0.0066 & 7\!\cdot\!10^{-4} & 7\!\cdot\!10^{-4} & 
   0.0292 & 0.0188 & 0.0188 \cr \hline\hline
\multicolumn{11}{|c|}{(8,2,1)}\cr\hline
\mbox{Exact}& \frac{13}{120} & \frac{13}{120} & \frac{43}{360} & \frac{17}
   {120} & \frac{7}{40} & \frac{41}{120} & \frac{151}{360} & \frac{61}
   {120} & \frac{73}{120} & \frac{73}{120} \cr\hline\mbox{Num.}&  0.1081 & 0.1081 & 
   0.1193 & 0.1408 & 0.1749 & 0.3461 & 0.4214 & 0.5061 & 0.6252 & 
   0.6252 \cr\hline|\mbox{diff.}|& 2\!\cdot\!10^{-4} & 2\!\cdot\!10^{-4} & 1\!\cdot\!10^{-4} & 
    8\!\cdot\!10^{-4} & 8\!\cdot\!10^{-4} & 0.0044 & 0.0019 & 
   0.0022 & 0.0169 & 0.0169 \cr \hline\hline
\multicolumn{11}{|c|}{(3,5,2)}\cr\hline
\mbox{Exact}& \frac{1}{20} & \frac{1}{20} & \frac{1}{12} & \frac{5}
   {36} & \frac{5}{36} & \frac{7}{36} & \frac{49}{180} & \frac{49}
   {180} & \frac{79}{180} & \frac{79}{180} \cr\hline\mbox{Num.}& 0.0503 & 0.0503 & 
   0.0854 & 0.1433 & 0.1433 & 0.1941 & 0.2741 & 0.2741 & 0.4276 & 
   0.4276 \cr\hline|\mbox{diff.}|& 3\!\cdot\!10^{-4} & 3\!\cdot\!10^{-4} & 0.0020 & 0.0044 & 
   0.0044 & 0.0004 & 0.0018 & 0.0018 & -0.0112 & 
    0.0112 \cr\hline
\end{array} \nonumber
\end{equation}}
\caption{Numerical exponents $x_n$ of the $(A_8,D_6)$ minimal model with seams of type $(r,a,\kappa)=(1,5,1)$, $(8,2,1)$ and $(3,5,2)$.}
\label{D6Table}
\end{table}

\subsection{Numerical spectra of $(A_8,D_6)$}

For $D_6$ we have fewer data points than for $D_4$ so the precision of the extrapolated values of the exponents is not as good. Nevertheless, the characteristic degeneracies of the exponents in the twisted partition functions are faithfully reproduced. Our data for seams of type $(r,a,\kappa)=(1,5,1)$, $(8,2,1)$ and $(3,5,2)$ are shown in Table~\ref{D6Table}.

As dictated by symmetry we observe generally, for $(A_{4\ell-4},D_{2\ell})$, that the $a$-type seams $(1,2\ell-1,1)$ and $(1,2\ell,1)$ give the same spectra for the transfer matrices. This is in accord with the symmetry of the graph fusion matrices. 
Indeed, if $\sigma$ implements the $\Bbb Z_2$ transposition for $(A_{4\ell-4},D_{2\ell})$ then
\begin{equation}
\sigma\,\hat{N}_{2\ell-1}=\hat{N}_{2\ell}\,\sigma
\end{equation}
so that $\hat{N}_{2\ell-1}$ is similar to $\hat{N}_{2\ell}$ under
interchange of the nodes $2\ell-1$ and $2\ell$.

\section{Discussion}
\setcounter{equation}{0}
\label{sec:Discussion}

In this paper we have constructed integrable
seams for each twisted conformal boundary
condition $x=(r,a,b,\kappa)$ of the \ade $s\ell(2)$ minimal models. 
Our construction labels are identical to the conformal labels of Petkova and Zuber. 
For $E_7$, we have not succeeded in constructing single integrable seams for the special seams denoted $(r,X)$, even though we have constructed integrable seams of type $(r,6,2,1)$ and $(r,4,2,1)$   which suffice to give the partition functions, so these seams remain somewhat mysterious. 

The construction of our integrable seams involves a new type of fusion, related to the graph fusion matrices $\hat{N}_a$ and labelled by the nodes $a$ of the graph $G$, rather than the usual Young tableaux. We find that the left-and right-chiral halves of the Ocneanu fusion algebra are related to the braid limits $u\to\pm i\infty$ of the seam weights.
The quantum symmetries and twisted conformal partition functions have been verified numerically for the $A_L$, $L=3,4,5,6,7,8,9$ and $D_L$, $L=4,5,6$ cases.
In general, the numerics reproduce the first 10 exponents to about a 1 or 2\% accuracy. 
Together with the agreement of the exact degeneracies, the numerics give strong evidence of the identification of our seams with the corresponding twisted conformal boundary conditions. The origin of non-commutativity in the Ocneanu graph fusion algebra is traced to the existence of graph automorphisms which do not commute with the fusions.

Lastly, we point out that the parameter $\xi$ appearing in the integrable seams is an arbitrary complex parameter. We have fixed its value to ensure conformal boundary conditions in the continuum scaling limit. By choosing it to have an imaginary part that scales appropriately with $N$ (the number of columns in the transfer matrix), it is possible~\cite{FPR} to perturb the boundary away from the conformal fixed point. This will induce renormalization group flows between the fixed points representing the various twisted conformal boundary conditions.

\section*{Acknowledgements}                     
\label{sec:Acknowldegments}
This research is supported by the Australian Research Council. Part of this work was done while CHOC and PAP were visiting IPAM, UCLA. We thank Valya Petkova and Jean-Bernard Zuber for their continued interest in this work and for comments on the manuscript.

\bibliographystyle{unsrt}
 \bibliography{seam}

\end{document}